\documentclass[12pt]{article}
\pdfoutput=1
\usepackage{amssymb}
\usepackage{amsmath}
\usepackage{amstext}
\usepackage{graphicx,epsfig}
\usepackage{epsfig}
\usepackage{verbatim} 
\usepackage{fancybox}
\usepackage{color}
\usepackage{ulem}
\usepackage{enumitem}
\usepackage{subfigure}
\usepackage{bbm}
\usepackage{parskip}
\usepackage{cite}
\usepackage{youngtab}
\usepackage{graphicx}
\usepackage{jheppub}
\usepackage{setspace}
\usepackage{longtable}
\usepackage{tabularx}
\usepackage{amsmath,amssymb,color,mathrsfs,verbatim,bbm,wasysym,pstricks,epsfig,colortbl}
\usepackage{slashed,mathtools,youngtab,xcolor,rotating,color,multirow,placeins}
\usepackage{ytableau}

\newcommand{\Comment}[1]{{}}
\definecolor{MyDarkBlue}{rgb}{0.15,0.15,0.45}
\usepackage[linktocpage=true]{hyperref}
\hypersetup{
colorlinks=true,
citecolor=MyDarkBlue,
linkcolor=MyDarkBlue,
urlcolor=MyDarkBlue,
pdfauthor={Kurt Hinterbichler},
pdftitle={},
pdfsubject={hep-th}
}

\newcommand{\be}{\begin{equation}}
\newcommand{\ee}{\end{equation}}
\newcommand{\bea}{\begin{eqnarray}}
\newcommand{\eea}{\end{eqnarray}}
\newcommand{\beas}{\begin{eqnarray*}}
\newcommand{\eeas}{\end{eqnarray*}}
\newcommand{\nn}{\nonumber}

\newcommand{\cD}{\mathcal{D}}

\newcommand{\tr}{\mathrm{Tr}}

\newcommand{\dcft}{d}
\newcommand{\dads}{D}
\newcommand{\dso}{\mathbb{D}}
\newcommand{\cB}{\mathcal{B}}
\newcommand{\cT}{\mathcal{T}}
\newcommand{\bW}{\hat{W}}
\newcommand{\PibW}{\bW^\Pi}
\newcommand{\PidbW}{\bW_\Pi}
\newcommand{\caA}{\mathcal{A}}
\newcommand{\caB}{\mathcal{B}}
\newcommand{\caC}{\mathcal{C}}
\newcommand{\cabW}{\hat{\mathcal{W}}}
\newcommand{\caPibW}{\cabW^\Pi}
\newcommand{\caV}{\mathcal{V}}
\newcommand{\cadC}{d\caC}
\newcommand{\caW}{\mathcal{W}}
\newcommand{\Epsilon}{\epsilon}
\definecolor{Gray}{gray}{0.9}

\def\({\left(}
\def\){\right)}
\newcommand{\la}{\langle}
\newcommand{\ra}{\rangle}

\newcommand{\Tr}{\text{Tr}}

\numberwithin{equation}{section}

\begin{document}

\begin{center}
{\LARGE {Partially Massless Higher-Spin Theory}}
\end{center} 
 \vspace{1truecm}
\thispagestyle{empty} \centerline{
{\large  { Christopher Brust$^{a,}$}}\footnote{E-mail: \Comment{\href{mailto:cbrust@perimeterinstitute.ca}}{\tt cbrust@perimeterinstitute.ca}},
{\large  { Kurt Hinterbichler$^{b,}$}}\footnote{E-mail: \Comment{\href{mailto:kurt.hinterbichler@case.edu}}{\tt kurt.hinterbichler@case.edu}}
                                                          }

\vspace{1cm}

\centerline{{\it 
${}^a$Perimeter Institute for Theoretical Physics,}}
 \centerline{{\it 31 Caroline St. N, Waterloo, Ontario, Canada, N2L 2Y5 }} 
 
 \vspace{.3cm}

\centerline{\it ${}^{\rm b}$CERCA, Department of Physics, Case Western Reserve University, }
\centerline{\it 10900 Euclid Ave, Cleveland, OH 44106, USA}

\begin{abstract}
~We study a generalization of the $D$-dimensional Vasiliev theory to include a tower of partially massless fields.  This theory is obtained by replacing the usual higher-spin algebra of Killing tensors on (A)dS with a generalization that includes ``third-order'' Killing tensors.   Gauging this algebra with the Vasiliev formalism leads to a fully non-linear theory which is expected to be UV complete, includes gravity, and can live on dS as well as AdS.  The linearized spectrum includes three massive particles and an infinite tower of partially massless particles, in addition to the usual spectrum of particles present in the Vasiliev theory, in agreement with predictions from a putative dual CFT with the same symmetry algebra.  We compute the masses of the particles which are not fixed by the massless or partially massless gauge symmetry, finding precise agreement with the CFT predictions.  This involves computing several dozen of the lowest-lying terms in the expansion of the trilinear form of the enlarged higher-spin algebra.  We also discuss nuances in the theory that occur in specific dimensions; in particular, the theory dramatically truncates in bulk dimensions $D=3,5$ and has non-diagonalizable mixings which occur in $D=4,7$.

\end{abstract}

\newpage

\tableofcontents
\newpage

\section{Introduction\label{sec:intro}}
\parskip=5pt
\normalsize

In this paper, we explore an explicit description of a partially massless (PM) higher-spin (HS) theory, discussed previously in \cite{Bekaert:2013zya,Basile:2014wua,Grigoriev:2014kpa,Alkalaev:2014nsa,Joung:2015jza}.  This is a fully interacting theory which can live on either anti-de Sitter (AdS) or de Sitter (dS), and is expected to be a UV complete and predictive quantum theory which includes gravity.  Like the original Vasiliev theory\footnote{Throughout this work, we refer only to the bosonic CP-even Vasiliev theory.} \cite{Vasiliev:1990en, Vasiliev:1992av, Vasiliev:1999ba, Vasiliev:2003ev} (see \cite{Vasiliev:1995dn, Vasiliev:1999ba, Bekaert:2005vh, Iazeolla:2008bp, Didenko:2014dwa,Vasiliev:2014vwa,Giombi:2016ejx} for reviews), it contains an infinite tower of massless fields of all spins, but in addition it contains a second infinite tower of particles, all but three of which are partially massless, carrying degrees of freedom intermediate between those of massless and massive particles.  This tower may be thought of as a partially Higgsed version of the tower in the Vasiliev theory.

The theory on AdS is expected to be the holographic dual to the singlet sector of the bosonic $U(N)$ $\square^2$ free conformal field theory (CFT) studied in \cite{Brust:2016gjy} (see also \cite{Karananas:2015ioa,Osborn:2016bev,Guerrieri:2016whh,Nakayama:2016dby,Peli:2016gio,Gwak:2016sma,Gliozzi:2016ysv,Gliozzi:2017hni}), and on dS is expected to be dual to the Grassmann counterpart CFT, just as the original Vasiliev theory is expected to be dual to an ordinary free scalar \cite{Sezgin:2002rt,Klebanov:2002ja,Anninos:2011ui}.  We define the bulk theory as the Vasiliev-type gauging of the CFT's underlying global symmetry algebra, which we refer to here as $hs_2$.  It is a part of a family of theories based on the $\square^k$ field theory which contain $k$ towers of partially massless states.  We study this theory for several reasons:

In our universe, we've confirmed the existence of seemingly fundamental particles with spins $0$, $\frac{1}{2}$ and $1$, and we have good reason to believe that gravity is described by a particle with spin $2$. It is an interesting field-theoretic question to ask, {\it even in principle}, what spins we are allowed to have in our universe. Famous arguments, such as those reviewed in \cite{Bekaert:2010hw,Porrati:2012rd}, would n\"aively seem to indicate that we should not expect particles with spin greater than $2$ to be relevant to an understanding of our universe, but these no-go theorems are evaded by specific counterexamples in the form of theories such as string theory and the Vasiliev theory, both of which contain higher-spin states and are thought to be complete.  
Of particular interest is the question of whether partially massless fields fall into the allowed class.  Partially massless fields are of interest due to a possible connection between partially massless spin-2 field and cosmology (see e.g.,~\cite{deRham:2013wv} and the review \cite{Schmidt-May:2015vnx}), which has led to many studies of the properties of the linear theory and possible nonlinear extensions~\cite{Zinoviev:2006im,Hassan:2012gz,Hassan:2012rq,Hassan:2013pca,Deser:2013uy,deRham:2013wv,Zinoviev:2014zka,Garcia-Saenz:2014cwa,Hinterbichler:2014xga,Joung:2014aba,Alexandrov:2014oda,Hassan:2015tba,Hinterbichler:2015nua,Cherney:2015jxp,Gwak:2015vfb,Gwak:2015jdo,Garcia-Saenz:2015mqi,Hinterbichler:2016fgl,Apolo:2016ort,Apolo:2016vkn, Gwak:2016sma}.  No examples (other than non-unitary conformal gravity \cite{Maldacena:2011mk,Deser:2012qg,Deser:2013bs}) of UV-complete theories in four dimensions containing an interacting partially massless field and a finite number of other fields are known, and so it has remained an open question whether these particles could even exist.  The theory we describe in this paper contains an infinite tower of partially massless higher-spin particles. Thus, the mere existence of this theory promotes further studies into partially massless gravity.

Although the past twenty years have seen great progress in our understanding of quantum gravity in spaces with negative cosmological constant, a grasp of the nature of quantum gravity in spaces with a positive cosmological constant such as our own remains elusive. There have been proposals inspired by AdS/CFT for a dS/CFT correspondence, which would relate quantum gravity on de Sitter to conformal theories at at least one of the past and future boundaries \cite{Strominger:2001pn, Hull:1998vg, Witten:2001kn, Strominger:2001gp, Balasubramanian:2002zh, Maldacena:2002vr}. It was argued in \cite{Anninos:2011ui} that the future boundary correlators of the non-minimal and minimal Vasiliev higher-spin theories on dS should match the correlators of the singlet sector of free ``$U(-N)$'' or $Sp(N)$ Grassmann scalar field theories, respectively. However, a lack of other examples has been an obstacle preventing us from answering deep questions we would like to understand in dS/CFT, such as how details of unitarity of the dS theory emerge from the CFT.  To that end, it seems a very exciting prospect to develop new, sensible theories on dS as well as their CFT duals to learn more about a putative correspondence.

Another interesting puzzle in the same vein is what the connection between the Vasiliev theory and string theory is. It is well-known that the leading Regge trajectory of string theory develops an enlarged symmetry algebra in the tensionless limit, (see, e.g. \cite{PhysRevLett.60.1229}), generally becoming a higher-spin theory. In particular, the tensionless limit of the superstring on AdS and the Vasiliev theory appear to be connected, and supersymmetrizing both \cite{Chang:2012kt} appears to relate the $\mathcal{N}=6$ super-Vasiliev theory and IIA superstring theory on ${\rm AdS}_4 \times \mathbb{C}\mathrm{P}^3$. However, the question of how to include in the Vasiliev theory the additional massive states which are present in the string spectrum is still a challenge. From the point of view of the Vasiliev theory, there are drastically too few degrees of freedom to describe string theory in full; string theory contains an infinite set of Vasiliev-like towers of ever increasing masses, and one would require an infinite number of copies of the fields in the Vasiliev theory in order to construct a fully Higgsed string spectrum. Without the aid of the $hs$ algebra underlying the Vasiliev construction, it is not clear how to proceed and add massive states to the Vasiliev theory to make it more closely resemble that of string theory. The theory we describe here contains partially massless states, which represent a sort of ``middle ground'' in the process of turning a theory with only massless degrees of freedom into one which contains massive (or partially massless) degrees of freedom as well by adding various St\"uckelberg fields.

It is natural to suspect that there should be a smooth Higgsing process by which an infinite set of massless Vasiliev towers eat each other and become the massive spectrum of string theory \cite{Girardello:2002pp,Bianchi:2003wx,Bianchi:2005ze}.  On AdS, there seems to be no obstruction to this, but on dS the situation is different.  As we review in section \eqref{sec:intropm}, there is a unitarity bound $m^2\geq H^2\left(s-1\right)\left(D+s-4\right)$ for a mass $m$, spin $s$ particle in $D$ dimensional dS space.  Below this bound, particles are non-unitary and so any smooth Higgs mechanism starting from $m=0$ would necessarily be doomed to pass through this non-unitary region before becoming fully massive.  The PM fields, however, are exceptions to this unitarity bound.  They form a discrete set of points below this bound where extra gauge symmetries come in to render the non-unitary parts of the fields unphysical (just as massless high-spin particles are unitary on dS despite lying below the unitarity bound).  Thus, one might suspect a discrete Higgs-like mechanism by which the massless theory steps up along the partially massless points on the way to full massiveness.  These intermediate theories should be Vasiliev-like theories with towers of partially massless modes (however, the theory we consider here continues to have a massless tower and we do not know any example of PM theory with no massless fields).

The partially massless higher-spin theory we describe in this paper is constructed in a similar fashion to the Vasiliev theory.  It is constructed at the level of classical equations of motion, although just as in the case of the Vasiliev theory, we believe the dual CFT defines the theory quantum-mechanically and in a UV-complete fashion.  There's no universally agreed-upon action for this theory or for the original Vasiliev theory (see \cite{Vasiliev:1988sa,Boulanger:2011dd,Doroud:2011xs,Boulanger:2012bj,Boulanger:2015kfa,Bekaert:2015tva,Bonezzi:2016ttk,Sleight:2016dba} for efforts in this direction), but this is believed to be a technical issue rather than a fundamental issue, and an action is expected to exist.
The theory can be defined on both AdS and dS, and is essentially nonlocal on the scale of the curvature radius $L$, though it has a local expansion in which derivatives are suppressed by the scale $L$. Nevertheless, this theory admits a weakly-coupled description and so can be studied perturbatively in $AdS_{\dads \geq 3}$; in particular it can be linearized, which we do in this paper. 

Our primary technical tool and handle on the theory is its symmetry algebra. The original Vasiliev theory in $AdS_{\dads \geq 4}$ is the gauge theory of the so-called $hs$ algebra, an infinite-dimensional extension of the diffeomorphism algebra which gauges all Killing tensors as well as Killing vectors on AdS. This algebra is equivalent to the global symmetry algebra of free scalar field theory in one fewer dimension, which consists of all conformal Killing tensors as well as conformal Killing vectors. The algebra we employ in this paper is the symmetry algebra of the $\square^2$ free field theory, which includes all of the generators of the $hs$ algebra, and in addition ``higher-order Killing tensors'', studied in \cite{2006math.....10610E}.  The representations and the bilinear form of this algebra were studied by Joung and Mkrtchyan \cite{Joung:2015jza}, and we make use of many of their results\footnote{They referred to this algebra as $\mathfrak{p}_2$; however as this algebra arises from a $\square^2$ dual CFT, we refer to this algebra in this paper simply as the $hs_2$ algebra.}. The structure of this algebra is very rigid, and its gauging completely fixes the structure of the corresponding theory on AdS, giving rise to the PM HS theory.

One crucial distinction between this PM HS theory and the original Vasiliev theory is that the PM theory on AdS is non-unitary/ghostly. This follows from the non-unitarity of the dual CFT, as well as the fact that the PM fields themselves are individually non-unitary on AdS.   Nevertheless, despite being nonunitary, our CFT is completely free, so there cannot be any issue of instability usually associated with nonunitary/ghostly theories.  We may compute its correlators with no issues, seemingly defining an interacting nonunitary theory.  The bulk theory should somehow not be unstable, since it is dual to a free theory.  Thus we believe that this theory exists in AdS and is stable despite its nonunitarity, and we believe that the infinite-dimensional underlying gauge algebra $hs_2$ is so constraining as to prevent any sort of instability from arising, though we will not attempt here to study interactions in detail in this theory, deferring such questions instead to future work.

We might suspect that the PM theory on dS is nonunitary as well, but without a Lagrangian description of the theory, and without the clearcut link between boundary and bulk unitarity enjoyed by AdS/CFT, we do not have a clear-cut answer as to whether the PM theory is unitary on dS.  The individual particles, including the PM particles, are all unitary on de Sitter, but unitarity could sill be spoiled if there are relative minus signs between kinetic term of different particles, and without a Lagrangian we cannot directly check whether this is the case.

In the $\square^2$ CFT, we demonstrated in \cite{Brust:2016gjy} that certain dimensions were special; in $d=2,4$ there existed what we dubbed the ``finite theories''; we will show here that the PM HS theory in $D=3,5$ mimics the structure of these finite CFTs. Furthermore, in $d=3,6$ there was module mixing that took place in the CFT.  We will see that this manifests as non-diagonalizability of the dual free PM HS action in $D=4,7$.  The fact that these Verma module structures mimic each other comes as no surprise, but does offer evidence that the PM HS theory is truly the AdS dual of the $\square^2$ CFT. Furthermore, the details of the duality in these cases are new, and are not specific to the Vasiliev formalism; this constitutes new evidence that the AdS/CFT duality continues to hold at the non-unitary level.

One interesting and powerful check of the duality between the Vasiliev theory and free field theory was the one-loop matching of the partition functions of the boundary and bulk theories \cite{Giombi:2013fka, Giombi:2014iua}. It has been argued that unitary higher-spin theories where the symmetry is preserved as we approach the boundary should have quantized inverse coupling constant \cite{Maldacena:2011jn}. Therefore, when computing the one-loop correction to the inverse Newton's constant in the Vasiliev theory, one was forced to obtain an integer multiple of the dual theory's $a$-type conformal anomaly (even $d$) or sphere free energy $F$ (odd $d$), which was precisely what happened. Despite the fact that the $\square^2$ CFT is non-unitary, its $N$ is nevertheless quantized, and so we continue to expect that the one-loop correction to the inverse Newton's constant is consistent with its quantization. In the companion paper \cite{Brust:2016xif} we do this computation in several dimensions and find a positive result (see also \cite{Gunaydin:2016amv}); we obtain integer multiples of the $a$-type conformal anomaly or sphere free energy $F$ of a single real conformally coupled $\square^2$ scalar in one dimension fewer. In particular, we obtain identical results to the Vasiliev case \cite{Giombi:2014iua}, namely $G_N^{-1} = N$ for the non-minimal/$U(N)$ duality and $G_N^{-1}=N-1$ for the minimal/$O(N)$ duality.

The outline of this paper is as follows: we begin by introducing and reviewing the properties of partially massless higher-spin free particles in AdS and dS in section \ref{sec:intropm}. We then turn to reviewing properties and the relevant representation of the algebra $hs_2$ in section \ref{sec:hs2}, as it is so central to all of the discussions in the paper, and discuss how to compute trilinear forms in the algebra, which are necessary for later calculations.  We gauge this algebra in section \ref{sec:pm}, linearize the theory, and discuss how the linearized master fields break up into unfolding fields for the physical particles. In section \ref{sec:masscomp}, we compute the masses of the four particles whose masses are not fixed by gauge invariance. We discuss which boundary conditions are necessary on the various fields so as to reproduce CFT expectations. In section \ref{sec:nuances}, we explore what happens to the PM HS spectrum in $D=3,4,5,7$, demonstrating agreement with expectations from the dual CFT. Finally, in section \ref{sec:conc}, we discuss various future directions for research, as well as implications for dS/CFT. We discuss the one-loop renormalization of the inverse Newton's constant in the companion paper \cite{Brust:2016xif}.

\textbf{Conventions:} We use the mostly plus metric signature, and the curvature conventions of \cite{Carroll:2004st}.  We (anti) symmetrize tensors with unit weight, e.g., $S_{(\mu\nu)} = \frac{1}{2}(S_{\mu\nu}+S_{\nu\mu})$.  The notation $(\cdots)_T$ indicates that the enclosed indices are to be symmetrized and made completely traceless.  Throughout this work, we unfortunately must reference three different spacetime dimensions; the dimension of the dual CFT is denoted $\dcft$, the dimension of the bulk (A)dS is denoted $\dads$, and the dimension of the ambient or embedding space in which the symmetry algebra is defined is denoted $\dso$.  They are related by $\dcft+2=\dads+1=\dso$. Embedding space coordinates are indexed by $A,B,C,\ldots$, and moved with the flat ambient metric $\eta_{AB}$.  (A)dS spacetime coordinates are indexed by $\mu,\nu,\rho,\ldots$, and moved with the (A)dS metric $g_{\mu\nu}$.  (A)dS tangent space indices are indexed by $a,b,c,\ldots$, and moved with the tangent space flat metric $\eta_{ab}$.  The boundary CFT indices are $i,j,k,\ldots$, and are moved with the flat boundary metric $\delta_{ij}$.  The background (A)dS space has a vielbein $\hat{e}_\mu^{\ a}$ which relates AdS spacetime and AdS tangent space indices.  $L$ refers to the AdS length scale, and $H$ refers to Hubble in dS (see section \ref{sec:intropm}).

All Young tableaux are in the manifestly antisymmetric convention, and on tensors we use commas to delineate the anti-symmetric groups of indices corresponding to columns (except on the metrics $\eta_{ab}$, $g_{\mu\nu}$, $\eta_{AB}$, $\delta_{ij}$.). We use the shorthand $[r_1,r_2,\ldots]$ to denote a Young tableau with $r_1$ boxes in the first row, $r_2$ boxes in the second row, etc.  All of the Young tableaux we work with will also be completely traceless, so we do not indicate tracelessness explicitly. The projector onto a tableau with row lengths $r_1, r_2,\cdots$ is denoted $P_{[r_1,r_2,\cdots]}$ where the indices to be projected should be clear from the context.  The action of the projector is to first symmetrize indices in each row, and then anti-symmetrize indices in each column, with the overall normalization chosen so that $P_{[r_1,r_2,\cdots]}^2=P_{[r_1,r_2,\cdots]}$.  This projector does not include the subtraction of traces.  Introductions to Young tableaux can be found in section 4 of~\cite{Bekaert:2006py} or the book \cite{Tung:1985na}.

%%%%%%%%%%%%%%%%%%%%%%%%%%%%%%%%%%%%%%%%%%%%%%%%%%%%%%%%%%%%%%%%%%%%%%%%%%%%%%%%

\section{Review of Partially Massless Fields}
\label{sec:intropm}

We begin by reviewing some properties of partially massless higher-spin fields in AdS or dS \cite{Deser:1983tm,Deser:1983mm,Higuchi:1986py,Brink:2000ag,Deser:2001pe,Deser:2001us,Deser:2001wx,Deser:2001xr,Zinoviev:2001dt,Skvortsov:2006at,Skvortsov:2009zu}, and how they behave as we take them to the boundary, i.e. the properties of the dual CFT operators.  Partially massless fields are fields with more degrees of freedom, and correspondingly less gauge symmetry, than a massless field, but fewer degrees of freedom, and correspondingly more gauge symmetry, than a fully massive field.  For a given spin, the amount of gauge symmetry fixes the mass on both AdS and dS.  Partially massless fields are necessarily below the unitarity bound in AdS, but are unitary in dS.  

\subsection{Free Massive Fields}

A spin-$s\geq1 $ field on $D$ dimensional (A)dS with mass $m$ is described by a symmetric $s$-index field $\phi_{\mu_1\ldots \mu_s}$ which satisfies the equations of motion
\bea && \left[\square +H^2\left( 6+D(s-2)+s(s-6)\right)-m^2\right]\phi_{\mu_1\ldots \mu_s}=0, \label{eomexp} \\
&&  \nabla^{\nu}\phi_{\nu\mu_2\ldots \mu_s}=0, \label{transverseexp} \\
&& \phi^\nu_{\phantom{\nu} \nu\mu_3\ldots\mu_s}=0,\label{massivesys}\eea
i.e. it is transverse, traceless, and satisfies a Klein-Gordon equation.  $\square\equiv\nabla^\mu \nabla_\mu$ is the curved space Laplacian.

Here $H$ is the (A)dS curvature scale, i.e. $H^2>0$ for dS, in which case $H$ is the Hubble constant, and $H^2<0$ for AdS (in which case we usually write $H^2=-{1/L^2}$ with $L$ the usual AdS radius).  The scalar curvature $R$ and cosmological constant $\Lambda$ are related to the Hubble constant as
\be R={D(D-1) {H}^2},\ \ \ \Lambda={(D-1)(D-2){H}^2\over 2}.\ee

In the AdS case, $H^2=-{1\over L^2}$, the high spin fields are dual to symmetric tensor ``single-trace'' primaries ${\cal O}_{i_1\ldots i_s}$.  For generic $m$, these satisfy no particular conservation conditions.   Their scaling dimensions are given in terms of the mass by
\be \Delta=   {d\over 2}\pm\sqrt{{\left(d+2(s-2)\right)^2\over 4}+m^2L^2},\ \ \ s\geq 1.\label{dimeqo}\ee
Here $d=D-1$ is the dimension of the dual CFT.   The positive root corresponds to the ``ordinary quantization" of AdS/CFT, and the negative root corresponds to the ``alternate quantization" of \cite{Klebanov:1999tb}.  

The unitarity bound \cite{Mack:1975je} for symmetric traceless tensor operators is
\be \Delta\geq d+s-2,\ \ \ s\geq 1,\label{unitboundd}\ee
For scalars, $s=0$, we have 
\be \Delta=   {d\over 2}\pm\sqrt{{d^2\over 4}+m^2L^2},\ \ \ s= 0,\label{dimeqosc}\ee
and the unitarity bound is
\be \Delta\geq {d\over 2}-1,\ \ \ s=0,\label{unitbounddsc}\ee
so for $s=0$ both ordinary and alternate quantizations are possible in a unitary theory.  For $s>1$, only the ordinary quantization is compatible with unitarity.  However, in the (non-unitary) partially massless theory, we will see that we do indeed need to use the alternate quantization for certain particles with $s\geq 1$. 

Solving for $m$ gives
\bea && m^2L^2 = \Delta(\Delta-d),\ \ \ s=0\, ,\nn\\
&& m^2L^2=\left(\Delta+s-2\right)\left(\Delta-s+2-d\right),\ \ s\geq 1.\label{m2dimdeadscfte}
\eea
For $s\geq 1$, we have $m^2\geq 0$ in the bulk, and there is no analog of the Breitenlohner-Freedman bound \cite{Breitenlohner:1982bm,Breitenlohner:1982jf}\footnote{The Breitenlohner-Freedman bound for scalars is $m^2\geq -{(D-1)^2\over 4 {L}^2}.$} allowing for slightly tachyonic but stable scalars. For $s\geq 1$, as soon as the mass is negative, we generically expect instabilities owing to the theory becoming ghostly/non-unitary.  

In the dS case, the unitarity bound for massive particles is {\it not} at $m=0$. Instead, the bound below which the particle is generically non-unitary is the Higuchi bound \cite{Higuchi:1986py,Higuchi:1986wu,Higuchi:1989gz},
\be m^2\geq H^2\left(s-1\right)\left(D+s-4\right).\label{unitaritybounddse}\ee
Below this bound, the kinetic term for one of the St\"uckelberg fields is generically of the wrong sign, indicating that some of the propagating degrees of freedom are ghostly. However, at special values of the mass between zero and the Higuchi bound, the particle develops a gauge symmetry which eliminates the ghostly degrees of freedom, and the field is unitary at these special points. These points are the partially massless fields, and we turn to them next.

\subsection{Free Partially Massless Fields}

Partially massless fields occur at the special mass values
\be m_{s,t}^2=H^2\left(s-t-1\right)\left(D+s+t-4\right),\ \ \ t=0,1,2,\ldots,s-1.\label{pmvalues}\ee
Here, $t$ is called the {\it depth} of partial masslessness.  At these mass values, the system of equations \eqref{massivesys} becomes invariant under a gauge symmetry,
\be \delta \phi_{\mu_1\ldots\mu_s}=\nabla_{(\mu_{t+1}\ldots \mu_{s}}\xi_{\mu_1\ldots \mu_t)}+\ldots \label{PMgaugesym}
\ee
and so $t$ counts the number of indices on the gauge parameter $\xi_{\mu_1\ldots \mu_t}$.  Here $\ldots$ stands for lower-derivative terms proportional to $H^2$.  On shell, the gauge parameter is transverse and traceless and satisfies a Klein-Gordon equation
\bea &&\square \xi_{\mu_1\ldots \mu_t}+\ldots=0,\nn\\
&&\nabla^{\nu}\xi_{\nu\mu_2\ldots \mu_t}=0,\nn\\ 
&&\xi^\nu_{\phantom{\nu} \nu\mu_3\ldots\mu_t}=0. \label{3eqpm}
\eea
The $\ldots$ terms in \eqref{3eqpm} and \eqref{PMgaugesym} as well as the mass values \eqref{pmvalues} are completely fixed by demanding invariance of the on-shell equations of motion \eqref{massivesys} under the on-shell gauge transformation \eqref{PMgaugesym}.

Just as massive and massless fields carry irreducible representations of the dS group, partially massless fields also carry irreducible representations, albeit ones which have no flat space counterpart.
A generic massive field has, in the massless limit, the degrees of freedom of massless fields of spin $s,s-1,\ldots,0$ (usually called, with some abuse of terminology, helicity components). The gauge symmetry of a PM field removes some of the lower helicity components; a depth $t$ PM field has helicity components \be s,s-1,\ldots, t+1.\ee

\begin{figure}[h]
\centering
\includegraphics[width=\textwidth]{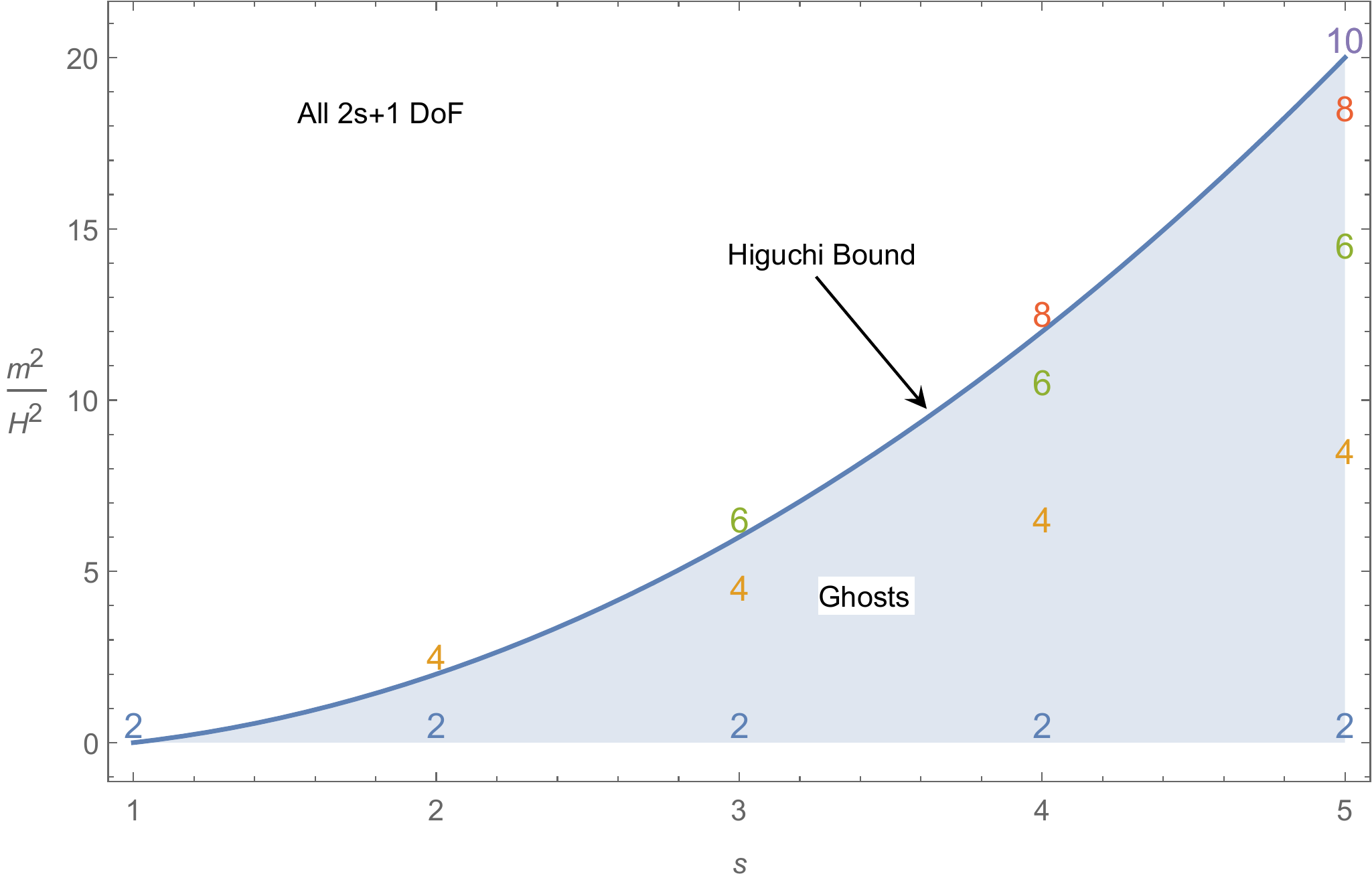}
\caption{\label{fig:PMpoints}Here we show the Higuchi bound on ${\rm dS}_4$ as well as the first few partially massless particles' masses and spins. Above the Higuchi bound, a massive particle is unitary. Below the bound, generically the kinetic term for at least one of the helicity components is of the wrong sign, indicating that some degrees of freedom are ghostly and the particle is non-unitary.  However, at the specific partially massless points (represented by the location of the numbers in the figure), the particle develops a gauge symmetry which eliminates the ghostly degrees of freedom, making the particle unitary. This comes at the expense of reducing the number of propagating degrees of freedom; how many degrees of freedom propagate is represented by the number at each partially massless location.}
\end{figure}

The highest depth is $t=s-1$, which corresponds to the usual massless field $m^2=0$ containing only helicity components $s$.  We see that on AdS, all but the highest depth PM fields have negative masses, and are non-unitary.  On dS, the masses are positive, and the PM fields are unitary (despite sitting below the Higuchi bound).  The lowest depth is $t=0$.  This saturates the unitarity/Higuchi bound on dS. Fields with masses below this bound are ghostly and therefore non-unitary, {\it unless }they are at one of the higher depth PM points.  As an illustration of this structure, see figure \ref{fig:PMpoints}, which shows the Higuchi bound on ${\rm dS}_4$ as well as the first few partially massless particles' masses and spins. 

PM fields are dual to {\it multiply}-conserved symmetric tensor single-trace primaries ${\cal O}_{i_1\ldots i_s}$, i.e. they satisfy a conservation condition involving multiple derivatives \cite{Dolan:2001ih},
\be \partial_{i_1}\ldots\partial_{i_{s-t}}{\cal O}^{i_1\ldots i_s}=0.\ee
For the massless case, $t=s-1$, this is the usual single-derivative conservation law. More generally, $s-t=c$, where $c$ is the degree of ``conservedness" of the operator (a notation we introduced in \cite{Brust:2016gjy}), i.e. the number of derivatives you need to dot into the operator to kill it.

On AdS, the mass-scaling dimension relation \eqref{dimeqo} (with the positive root) tells us that the dimension of these partially conserved currents should be\footnote{As a check, one can see that the general form for the two-point correlation functions, 
\begin{equation}\la {\cal O}_{i_1\ldots i_s}(x) {\cal O}_{j_1\ldots j_s} (0)\ra\sim {1\over x^{2\Delta }}\left[\ldots\right] \end{equation}
become conserved, doubly conserved, etc. precisely at these values, e.g. the $s=2$ expression satisfies
\begin{align} \partial^{i_1} \la {\cal O}_{i_1 i_2}(x) {\cal O}_{j_1 j_2} (0) \ra \sim ~&(\Delta-d)\left[\ldots\right], \nonumber \\
\partial^{i_1} \partial^{i_2} \la {\cal O}_{i_1 i_2}(x) {\cal O}_{j_1 j_2} (0)\ra \sim ~&(\Delta-d)(\Delta-(d-1))\left[\ldots\right]. 
\end{align}}
\be \Delta_{s,t}=d+t-1=d+s-2-(s-1-t).\ee
The second equality shows that these operators violate the CFT unitarity bound \eqref{unitboundd} except for the conserved operator with $t=s-1$, which saturates it.

\FloatBarrier

%%%%%%%%%%%%%%%%%%%%%%%%%%%%%%%%%%%%%%%%%%%%%%%%%%%%%%%%%%%%%%%%%%%%%%%%%%%%%%%%

\section{The $hs_2$ Algebra}
\label{sec:hs2}

We now discuss the symmetry algebra, $hs_2$, which we will ultimately gauge in order to obtain a partially massless higher spin theory.  In the linearized partially massless higher-spin theory, there will be two ``master'' fields (a gauge field and a field strength), and a ``master" gauge parameter, which are valued in the $hs_2$ algebra.  The Vasiliev equations themselves are also valued in the algebra.  

There is a multi-linear form which is defined on the algebra.  This will be used to extract component equations from the general Vasiliev equations, which in turn will allow us to calculate the masses of the four particles in the linearized PM HS theory without any gauge symmetry. 
Our ultimate goal will be to compute the multilinear forms we will need to compute the masses.

  The reader who is interested purely in the physics of the theory may familiarize themselves with the generators of the $hs_2$ algebra in subsection \ref{sec:hs2generalities}, and then move on to section \ref{sec:pm}, skipping the intermediate details of the computation.
The content of this section is mostly a review of, or slight extensions of, previous work \cite{Vasiliev:2003ev, Bekaert:2005vh, Joung:2014qya, Joung:2015jza}.  Our main contribution is the explicit calculation of several of the lowest-lying terms in the expansion of the trilinear form of this algebra, which are given in appendix \ref{sec:multilinearforms}.

First we describe the construction of the algebra abstractly, without reference to any particular realization. Then, we introduce oscillators with a natural star product which form a realization of the algebra which is useful for computations. Finally, we implement the technology of coadjoint orbits which can be used as a bookkeeping device for the different tensor structures which emerge and greatly simplifies calculations.

\subsection{Generalities About the $hs_2$ Algebra}
\label{sec:hs2generalities}

The $hs_2$ algebra is realized as the algebra of global symmetries of a conformal field theory \cite{2006math.....10610E,Bekaert:2013zya,Basile:2014wua,Grigoriev:2014kpa,Alkalaev:2014nsa,Joung:2015jza,Brust:2016gjy}, the $\square^2$ CFT described by the action
\begin{equation}S \propto \int d^\dcft x ~\phi^\dagger_a\square^2 \phi^a \, .\label{dualcft2}
\end{equation}
The $\square^2$ CFT contains as its underlying linearly realized\footnote{These are not to be confused with the non-linearly realized higher shift symmetries of \cite{Hinterbichler:2014cwa,Griffin:2014bta}, which are also present.} symmetry algebra precisely the algebra $hs_2$. The spectrum of operators and conserved currents form a representation of this algebra.

We first discuss this algebra abstractly.  $hs_2$ can be abstractly defined as a quotient of the universal enveloping algebra (UEA), $U(so(\dso))$, of the $\dso=d+2$ dimensional\footnote{This construction is independent of the signature.} embedding space Lorentz algebra $so(\dso)$, by a particular ideal.  The abstract generators $T_{AB}$ of $so(\dso)$ transform in the ${\tiny \Yvcentermath1 \yng(1,1)}$ adjoint representation of the $so(\dso)$ algebra,
\be T_{AB}\in {\tiny \Yvcentermath1 \yng(1,1)}\, .\ee
The commutation relations for $so(\dso)$ are
\be \left[ T_{AB},T_{CD}\right]=-\eta_{AC}T_{BD}+\eta_{BC}T_{AD}-\eta_{BD}T_{AC}+\eta_{AD}T_{BC},\label{sodcommr}\ee
where $\eta_{AB}$ is the $so(\dso)$ invariant metric tensor.

The universal enveloping algebra, and then $hs_2$, will be described as successive quotients of the algebra of all formal products of the $T$'s. First, we consider the tensor product algebra formed from the $T$'s. We can label the elements of the tensor product algebra by the irrep under $so(\dso)$, which we display as a tableau, as well as by the number of powers of $T$ they came from, which we indicate using a subscript $n$ on the tableau, and which we'll refer to as the ``level''. For example, we may decompose the product of two $T$'s as
\begin{align}T_{AB} T_{CD} = {\tiny \Yvcentermath1 \yng(1,1)} \otimes {\tiny \Yvcentermath1 \yng(1,1)} = {\tiny \Yvcentermath1 \yng(2,2)}_2 &\oplus {\tiny \Yvcentermath1 \yng(2)}_2 \oplus \bullet_2 \oplus {\tiny \Yvcentermath1 \yng(1,1,1,1)}_2 \nonumber \\
 &\oplus {\tiny \Yvcentermath1 \yng(2,1,1)}_2 \oplus {\tiny \Yvcentermath1 \yng(1,1)}_2 \label{2by2teqn}\end{align}
The scalar $\bullet_2$ is the quadratic Casimir $C_2=T_{AB}T^{AB}$, and the antisymmetric tensor ${\tiny \Yvcentermath1 \yng(1,1)}_2$ is the commutator.  

In the top line of \eqref{2by2teqn} are terms which are symmetric in the interchange of the two $T$s, whereas the bottom line contains terms which are antisymmetric in the interchange of the two $T$s.  To pass to the UEA, we use the commutation relations \eqref{sodcommr} to eliminate all anti-symmetric parts in terms of parts with a lower number of $T$'s, leaving only the symmetric parts in the top line (see the Poincar\'e-Birkhoff-Witt theorem).

To pass to the $hs_2$ algebra, we quotient by a further ideal.  The generators ${\tiny \Yvcentermath1 \yng(1,1,1,1)}_2$, ${\tiny \Yvcentermath1 \yng(4)}_4$, and finally $C_2 - \frac{1}{2}(\dso-6)(\dso+2)$ generate an ideal of the UEA (here ${\tiny \Yvcentermath1 \yng(4)}_4$ comes in at level 4, e.g. from the tensor product ${\tiny \Yvcentermath1 \yng(2)}_2 \otimes {\tiny \Yvcentermath1 \yng(2)}_2$). We quotient the UEA to the $hs_2$ algebra by replacing ${\tiny \Yvcentermath1 \yng(1,1,1,1)}_2 \rightarrow 0$, $\bullet_2 \rightarrow \frac{1}{2}(\dso-6)(\dso+2) $, and ${\tiny \Yvcentermath1 \yng(4)}_4 \rightarrow 0$.

Those generators which remain in the resulting quotient define the generators of the $hs_2$ algebra, and consist of the representations:
\begin{align} hs_2 \subset \bullet_0 \oplus {\tiny \Yvcentermath1 \yng(1,1)}_1 & \oplus {\tiny \Yvcentermath1 \yng(2,2)}_2 \oplus {\tiny \Yvcentermath1 \yng(3,3)}_3 \oplus {\tiny \Yvcentermath1 \yng(4,4)}_4 \oplus \ldots \nonumber \\
& \oplus {\tiny \Yvcentermath1 \yng(2)}_2 \oplus {\tiny \Yvcentermath1 \yng(3,1)}_3  \oplus {\tiny \Yvcentermath1 \yng(4,2)}_4 \oplus \ldots \label{eqn:hs2gens}\end{align}
The first line are generators which are in the same representations as the generators of the massless $hs$ algebra, which we will call $T_{(r)}$ at level $r$, whereas the second line are generators new to $hs_2$, which we will call $\tilde{T}_{(r)}$ at level $r$. The old generators  $T_{(r)}$ correspond to Killing tensors of AdS and conformal Killing tensors of the CFT, whereas the new generators $\tilde{T}_{(r)}$ correspond to so-called {\it order three Killing tensors} in AdS and {\it order three conformal Killing tensors} in the CFT, as reviewed in \cite{Joung:2015jza}, \cite{Brust:2016gjy}, and in the appendix. They are associated with multiply conserved currents in the CFT, and are gauged by partially massless fields in AdS. It is noteworthy that $hs_2$ contains $hs$ as a sub-vector space. However, as the values of the Casimirs do not match, it is not, strictly speaking, a subalgebra\footnote{We thank Evgeny Skvortsov for discussions of this point.}. 

In the process of taking the quotient, all of the Casimirs are fixed to specific values:
\bea 
 C_2&=&T_{AB}T^{AB}=-T^{A}_{\ B} T^{B}_{\ A}\sim\frac{1}{2}(\dso-6)(\dso+2) \nn\\
 C_4&=&T^{A}_{\ B} T^{B}_{\ C} T^C_{\ D} T^D_{\ A}\sim \frac{1}{8}(\dso-6)^2(\dso+2)^2 \nn\\
&\vdots &
\eea

All of the generators in the $hs_2$ algebra are traceless two-row Young tableaux, which we generically call $[r,s]$ with $s \leq r$. These generators can be written as elements of the UEA in the appropriate representations
\be T_{A_1B_1,\ldots,A_sB_s,A_{s+1},\ldots ,A_r}=P_{[r,s]} \left(T_{A_1B_1}\ldots  T_{A_sB_s} T_{A_{s+1}B_{s+1}} \ldots T_{A_rB_r} \eta^{B_{s+1}B_{s+2}}\ldots \eta^{B_{r-1}B_r}\right)-{\rm traces},\ee
where $P_{[r,s]}$ is the normalized projector onto the $[r,s]$ tableau (this definition fixes the normalization of the generators). These generators carry indices in the fully traceless tableau of shape $[r,s]$.  We use the anti-symmetric convention, which means that they are anti-symmetric in any $A,B$ pair, vanishes if we try to anti-symmetrize any $A,B$ pair with any third index to the right of the pair.  For $r=0$ we have only the constants, and for $r=s=1$ the original generators $T_{AB}$.
  In the original $hs$ algebra, all the generators have $s=r$.  In $hs_2$, as shown schematically in equation \eqref{eqn:hs2gens}, we have generators with $ s = r$, which we referred to as $T_r$, as well as generators with $s = r-2$, which we referred to as $\tilde{T}_r$.

A general algebra element is a linear combination of the above generators,
\begin{align} A=~~~&\sum_{r=0}^\infty {1\over 2^r r!} A_{(r)}^{A_1B_1,\ldots,A_rB_r}T_{A_1B_1,\ldots,A_rB_r} \nonumber \\
+&\sum_{r=2}^\infty {1\over 2^r r!}\tilde{A}_{(r)}^{A_1B_1,\ldots A_{r-2}B_{r-2}, A_{r-1},A_r} \tilde{T}_{A_1B_1,\ldots,A_{r-2}B_{r-2},A_{r-1},A_r} \label{alggenelA}\end{align}
where the coefficient tensors $A_{(r)}$ have the symmetry of a traceless $[r,r]$ tableau and the coefficient tensors $\tilde{A}_{(r)}$ have the symmetry of a traceless $[r,r-2]$ tableau.

The product on the $hs_2$ algebra is the product in the UEA mod the ideal, and we denote it by $\star$.   It takes the schematic form
\begin{align} T_r \star T_{r'} &=  T_{r+{r'}}+\ldots+T_{|r-{r'}|}+\tilde{T}_{r+{r'}}+\ldots+\tilde{T}_{|r-{r'}|}, \nonumber \\
T_r \star \tilde{T}_{r'} &= T_{r+{r'}}+\ldots+T_{|r-{r'}|}+\tilde{T}_{r+{r'}}+\ldots+\tilde{T}_{|r-{r'}|}, \nonumber \\
\tilde{T}_r \star \tilde{T}_{r'} &= T_{r+{r'}}+\ldots+T_{|r-{r'}|}+\tilde{T}_{r+{r'}}+\ldots+\tilde{T}_{|r-{r'}|}.\label{eqn:schematicstarproduct}\end{align}  
The product is bilinear and associative but not commutative.
The commutator of the star product, for any two algebra elements $A$ and $B$, is
\be \left[A,B\right]_\star\equiv A\star B-B\star A,\ee
and it gives the $hs_2$ algebra the structure of a Lie algebra which is isomorphic to the Lie algebra of linearly realized global symmetries of the $\square^2$ CFT.

There is a natural trace on the algebra which projects onto a singlet, defined simply as
\be \tr(A)=A_{(0)},\label{tracedefn}\ee
and a multi-linear form can be defined using this trace as
\be \mathcal{M}(A,B,\ldots) \equiv \tr(A\star B\star \ldots).\ee
Note that the bilinear form is diagonal in the degree $r$, because the product of a rank $r$ generator and a rank $r'$ generator only contains a zero component if $r=r'$.  But there can be mixing between algebra elements with the same degree but corresponding to different Young diagrams, which we will have to worry about later. 

\subsection{Oscillators and Star Products}

Although in principle the previous subsection contains all of the ingredients necessary to define the $hs_2$ algebra, it is incredibly cumbersome to use those definitions directly to compute anything in the algebra.  In this section we review an oscillator construction of the algebra, as introduced in \cite{Vasiliev:2003ev, Bekaert:2005vh, Joung:2015jza}.  The oscillator construction comes with its own natural star product, which is very convenient for computations, and ultimately reproduces the results of the computations in the ideal described in the previous section. One reason for the simplification is the introduction of a ``quasiprojector" which greatly assists with the step of modding out by the ideal, and makes it possible to compute the bilinear and trilinear forms of the algebra to a high enough order to extract what we need from the Vasiliev equations.

We introduce bosonic variables $Y_\alpha^A$, called oscillator variables, which carry an $sl(2)$ index\footnote{This $sl(2)$ is the Howe dual algebra to the $so(\dso)$, see e.g. the review \cite{Bekaert:2005vh}.} $\alpha=+,-$ in addition to an $so(\dso)$ index $A$.  (For us, this $sl(2)$ is a completely auxiliary structure useful for defining the representation and we do not think of it as being physical or related to any spacetime.)  At the end of the day, all physical quantities will be singlets under this $sl(2)$. The invariant tensor for $sl(2)$ is $\epsilon_{\alpha\beta}$ which is anti-symmetric,
\begin{equation}{
\epsilon_{\alpha\beta}= \epsilon^{\alpha\beta}=\left(\begin{array}{cc}0 & 1 \\-1 & 0  \end{array}\right),\ \ \ \ \epsilon^{\alpha\gamma}\epsilon_{\beta \gamma}=\delta_{\ \beta}^\alpha.}
\end{equation}

Suppose we have two arbitrary polynomials in the $Y_\alpha^A$ variables, $F(Y)$ and $G(Y)$. We may define an oscillator star product, $\ast$, between them. (Note that the oscillator star product $\ast$ is a priori different from the $hs_2$ product $\star$ which we defined in the previous subsection; we will discuss how to relate the two further below.  We will refer to both as ``the star product'' in this paper, leaving the distinction clear from context.) The (oscillator) star product between them is defined to be
\be  F\ast G= F\, {\rm exp}\left[{-{1\over 2}\eta^{AB}\epsilon_{\alpha\beta} {\overset{\leftarrow}{\partial}\over \partial Y_\alpha^A}{\overset{\rightarrow}{\partial}\over \partial Y_\beta^B}}\right]G.\ee
Like $\star$, $\ast$ is bi-linear and associative. Our goal is to understand how we can use this easy-to-evaluate product $\ast$ to evaluate the desired product $\star$. 

With the star product we define the star commutator
\be \left[F,G \right]_\ast=F\ast G-G\ast F.\ee
The star products and commutators among the basic variables are
\be Y_\alpha^A \ast Y_\beta^B=Y_\alpha^A  Y_\beta^B-{1\over 2}\epsilon_{\alpha\beta}\eta^{AB},\ \ \  \left[ Y_\alpha^A , Y_\beta^B\right]_\ast=-\epsilon_{\alpha\beta}\eta^{AB}.\ee

In addition, there is an integral version of this same star product \cite{Vasiliev:2003ev, Vasiliev:2004cm}:
\begin{equation}F \ast G = \frac{1}{\pi^{2\dso}} \int d^{2\dso}S d^{2\dso}T~F(Y+S)G(Y+T) e^{2\eta_{AB}\epsilon^{\alpha\beta}S^A_\alpha T^B_\beta}\, .\end{equation}

It should be noted that there are consequently two products available to the $Y$; an ordinary product and a star product. The $Y$'s commute as ordinary products, despite not commuting as star products. When we write polynomials in $Y$, we mean that they are polynomials in the ordinary product sense.

We define antisymmetric $so(\dso)$ and symmetric $sl(2)$ generators as
\be T^{AB}_{(Y)}=Y_\alpha^A  Y_\beta^B\epsilon^{\alpha\beta}, \ \ \ \ k_{\alpha\beta}=Y_\alpha^A  Y_\beta^B\eta_{AB}.\label{sodsl2yrep}\ee
We may use the above star product to evaluate the star commutators of \eqref{sodsl2yrep}, and these reproduce the commutation relations of decoupled $so(\dso)$ and $sl(2)$ algebras,
\bea&& \left[ T^{AB}_{(Y)},T^{CD}_{(Y)}\right]_\ast=-\eta^{AC}T^{BD}_{(Y)}+\eta^{BC}T^{AD}_{(Y)}-\eta^{BD}T^{AC}_{(Y)}+\eta^{AD}T^{BC}_{(Y)}, \\
&& \left[k_{\alpha\beta},k_{\gamma\delta}\right]_\ast=-\epsilon_{\alpha\gamma}k_{\beta\delta}-\epsilon_{\beta\gamma}k_{\alpha\delta}-\epsilon_{\beta\delta}k_{\alpha\gamma}-\epsilon_{\alpha\delta}k_{\beta\gamma}, \\
&& \left[k_{\alpha\beta},T^{AB}_{(Y)}\right]_\ast=0\, .
\eea 

To each element of the algebra $A$, we may associate a polynomial $A_{(Y)}$ in the $Y$'s by replacing the generators with a product of $Y$'s 
\begin{align} T^{A_1B_1,\ldots,A_rB_r} &\rightarrow T^{A_1B_1}_{(Y)}T^{A_2B_2}_{(Y)}\ldots T^{A_rB_r}_{(Y)}, \nonumber \\
\tilde{T}^{A_1B_1,\ldots,A_{r-2}B_{r-2},A_{r-1},A_r} &\rightarrow T^{A_1B_1}_{(Y)}T^{A_2B_2}_{(Y)}\ldots T^{A_rB_r}_{(Y)}\eta_{B_{r-1}B_r}\, .
\end{align}

We would like to be able to use the $\ast$ product on $A_{(Y)}$ in place of the $\star$ product on $A$, but there is an obstruction in that, in general, we still have nontrivial Casimir elements in the polynomial $A_{(Y)}$, which must be fixed to particular numbers. We may force all of the Casimir-type elements to be set to the values required by the $hs_2$ algebra by introducing a quasiprojector\footnote{This is referred to as a quasiprojector rather than a projector because the explicit form does not satisfy $\Delta_{hs_2}^2 = \Delta_{hs_2}$; rather, its square doesn't converge \cite{Vasiliev:2004cm}. This is not a problem at the level of working to any fixed order in the algebra, as we do.}, $\Delta_{hs_2}$, which will be useful for setting the Casimirs to their proper values, and extracting from a general polynomial $F$ an element of $hs_2$ when working within a trace:
\begin{equation}\Delta_{hs_2} \ast F(Y) \equiv F_{hs_2}(Y)\, .\end{equation} 
To extract the $hs_2$ trace, we merely take the $r=0$ component of $ F_{hs_2}(Y)$. This can be formally obtained by simply setting $Y\rightarrow 0$. Therefore
\begin{equation}\tr(F_{hs_2}) = \Delta_{hs_2} \ast F(Y) \bigg|_{Y=0}\, .\end{equation}
Once we have the quasiprojector, we can compute multi-linear forms using the $\ast$ product:
\be \mathcal{M}(A,B,\ldots)\equiv \tr(A \star B \star \ldots)= \Delta_{hs_2} \ast A_{(Y)} \ast B_{(Y)}\ast\ldots \bigg|_{Y=0} \,. 
\ee

We now need to know what $\Delta_{hs_2}$ is.  It should implement the modding out by the ideals, including replacing the Casimir $T_{AB}^{(Y)}\ast T^{AB}_{(Y)}$ with the appropriate number,
\begin{equation}\Delta_{hs_2} \ast T_{AB}^{(Y)}\ast T^{AB}_{(Y)}\bigg|_{Y=0} = \frac{1}{2}(\dso-6)(\dso+2)\, ,\end{equation}
 and likewise with all higher powers.

A useful form for the quasi-projector was found in \cite{Joung:2015jza},
\begin{equation}\Delta_{hs_2}^{\mathrm{}} = N \int_0^1 dx~ \sqrt{x}(1-x)^{\frac{\dso-6}{2}}~\phantom{}_2F_1\left(3,-1;\frac{3}{2};\frac{1}{1-x}\right)e^{-2\sqrt{x}Y_+\cdot Y_-}\label{eqn:altquasiproj},\end{equation}
with $N$ a normalization factor,
\begin{equation}N=-\frac{2^{3-\dso} \Gamma (\dso-1)}{\Gamma \left(\frac{\dso}{2}-3\right) \Gamma \left(\frac{\dso}{2}+1\right)}.\end{equation}

\subsection{Coadjoint Orbits}
\label{sec:formsexplanation}

In order to conveniently deal with the tensor structures which emerge, it is useful to introduce, following \cite{Joung:2015jza}, the technology of {\it coadjoint orbits}.  The coadjoint orbit method allows us to replace the coefficient tensors $A_{(r)}^{A_1B_1,\ldots,A_rB_r}$, $\tilde{A}_{(r)}^{A_1B_1,\ldots A_{r-2}B_{r-2}, A_{r-1},A_r}$ of a general algebra element \eqref{alggenelA} with products of a single antisymmetric tensor $\mathcal{A}_{AB}$, called a coadjoint orbit, which we write in a script font,   
\begin{align} A_{(r)}^{A_1B_1,\ldots,A_rB_r} &\rightarrow \mathcal{A}^{A_1B_1}\mathcal{A}^{A_2B_2}\ldots \mathcal{A}^{A_rB_r}-\mathrm{trace},\nonumber \\
\tilde{A}_{(r)}^{A_1B_1,\ldots,A_{r-2}B_{r-2},A_{r-1},A_r} &\rightarrow \mathcal{A}^{A_1B_1}\ldots \mathcal{A}^{A_{r-1}B_{r-1}}\mathcal{A}^{A_rB_r}\eta_{B_{r-1}B_r}.\label{wstringe1}\end{align}
These coadjoint orbits will serve as placeholders or bookkeeping devices.  Expressions for our multi-linear forms will be written in terms of products of matrix traces of products of these coadjoint orbits for various $hs_2$ valued fields. These are in one-to-one correspondence with the different tensor structures or ways of contracting the indices. Once we have obtained the multi-linear form with the coadjoint orbits, we may reconstruct the tensor structure in question by passing back to spacetime fields.

The coadjoint orbits $\mathcal{A}_{AB}$ satisfy what we will call here the coadjoint orbit conditions:
\be \caA^{(A}_{\phantom{A}B}\caA^{C}_{\phantom{C}D}\caA^{E}_{\phantom{E}F}\caA^{G)}_{\phantom{G}H}\eta^{BD}\eta^{FH}=0,\ \ \ \caA^{[AB}\caA^{C]D}=0.\ee
These two together serve to enforce that products of $r$ copies of $\caA_{AB}$ in \eqref{wstringe1} have the symmetry properties of either a trace-{\it ful} $[r,r]$ or trace-{\it less} $[r,r-2]$ tableau. (We often view $\caA$ as a matrix in what follows, and use $\langle \ldots \rangle$ to denote a matrix trace.) To see this, the first can be shown to imply the conditions $\langle \caA^2\rangle = 0$, and the second can be contracted with a second coadjoint orbit $\caB$ to show
\be \caA \caB \caA=\frac{1}{2} \caA \la \caB \caA\ra.\ee
Note that this identity also implies that $\caA^3=0$. Therefore, if we consider the quantity $\caA_{A_1 B_1}\ldots \caA_{A_r B_r}$, then it is in the $[r,r]$ representation, but it is not traceless (which is why the trace has to be explicitly subtracted in \eqref{wstringe1}), and taking a single trace of, say, any two $B$ indices puts the resulting tensor in the $[r,r-2]$ representation, which is automatically traceless.

In the computations we will do, we will have several different fields present in each multi-linear form, so we'll introduce several different, independent coadjoint orbits, one for each field, each satisfying their own coadjoint orbit conditions (and each with the script version of the letter associated to the particular field).

As mentioned, we must subtract the single traces manually from the $[r,r]$ fields.   There are no traces to subtract at level 0 or 1 in the algebra; we must first subtract traces at level 2, and (as we will see) we'll need trace-free replacements up to level 4. The explicit form of the traces can be worked out by adding all possible trace terms with arbitrary coefficients, and demanding that the resulting tensor is in the $[r,r]$ representation and is totally traceless given the coadjoint orbit conditions.  The results of this procedure for $r=2,3,4$ are:
\bea   A^{A_1B_1,A_2B_2}_{(2)} &&\rightarrow\caA^{A_1B_1}\caA^{A_2B_2} +{3\over \dads-1}P_{[2,2]} \left[\eta^{B_1B_2}\caA^{A_1C}\caA^{\phantom{C}A_2}_{C}\right] \, , \nn \\
A^{A_1B_1,A_2B_2,A_3B_3}_{(3)}  &&\rightarrow \caA^{A_1B_1}\caA^{A_2B_2}\caA^{A_3B_3}+{9\over \dads+1}P_{[3,3]} \left[\eta^{B_2B_3}\caA^{A_1B_1}\caA^{A_2C}\caA^{\phantom{C}A_3}_{C}\right] \, , \nn\\
  A^{A_1B_1,A_2B_2,A_3B_3,A_4B_4}_{(4)} &&\rightarrow \caA^{A_1B_1}\caA^{A_2B_2}\caA^{A_3B_3}\caA^{A_4B_4} +{18\over \dads+3}P_{[4,4]} \left[\eta^{B_3B_4} \caA^{A_1B_1}\caA^{A_2B_2}\caA^{A_3C}\caA^{\phantom{C}A_4}_{C}\right] \, .\nn \\
\eea
There are no such subtleties with the $[r,r-2]$ tensors, which are already traceless given the coadjoint orbit conditions, so we may simply replace
\bea
\tilde A^{A_1,A_2}_{(2)}  &&\rightarrow \caA^{A_1C}\caA^{\ \ A_2}_{C}\, , \nn\\
 \tilde A^{A_1B_1,A_2,A_3}_{(3)}   &&\rightarrow   \caA^{A_1B_1}\caA^{A_2C}\caA^{\ \ A_3}_{C}\, ,\nn \\
  \tilde A^{A_1B_1,A_2B_2,A_3,A_4}_{(4)}  &&\rightarrow    \caA^{A_1B_1}\caA^{A_2B_2}\caA^{A_3C}\caA^{\ \ A_4}_{C}\, .
\eea
With these replacements we may pass to coadjoint orbits, perform our computations of the multilinear form, and then pass back by the inverse operation:
\bea  \caA^{A_1B_1}\caA^{A_2B_2} &&\rightarrow A^{A_1B_1,A_2B_2}_{(2)} -{3\over \dads-1}P_{[2,2]} \left[\eta^{B_1B_2}\tilde A_{(2)}^{A_1,A_2}\right] \, ,\nn \\
 \caA^{A_1B_1}\caA^{A_2B_2}\caA^{A_3B_3} &&\rightarrow A^{A_1B_1,A_2B_2,A_3B_3}_{(3)}-{9\over \dads+1}P_{[3,3]} \left[\eta^{B_2B_3}\tilde A_{(3)}^{A_1B_1,A_2,A_3}\right] \, ,\nn\\
 \caA^{A_1B_1}\caA^{A_2B_2}\caA^{A_3B_3}\caA^{A_4B_4} &&\rightarrow  A^{A_1B_1,A_2B_2,A_3B_3,A_4B_4}_{{(4)}} -{18\over \dads+3}P_{[4,4]} \left[\eta^{B_3B_4}\tilde A_{(4)}^{A_1B_1,A_2B_2,A_3,A_4}\right]\, ,\nn \\ \label{eqn:orbitToTensor1}
&&
\eea
\bea
 \caA^{A_1C}\caA^{\ \ A_2}_{C} &&\rightarrow \tilde A^{A_1,A_2}_{2}\, , \nn\\
 \caA^{A_1B_1}\caA^{A_2C}\caA^{\ \ A_3}_{C} &&\rightarrow \tilde A^{A_1B_1,A_2,A_3}_{3}\, ,\nn \\
 \caA^{A_1B_1}\caA^{A_2B_2}\caA^{A_3C}\caA^{\ \ A_4}_{C} &&\rightarrow \tilde A^{A_1B_1,A_2B_2,A_3,A_4}_{4}\, .\label{eqn:orbitToTensor2}
\eea
In practice we will not need the initial replacement \eqref{eqn:orbitToTensor1}.  We will instead compute the multilinear form of particular elements in the algebra directly in terms of the coadjoint orbits, and then reconstruct the fields with the inverse operation \eqref{eqn:orbitToTensor2}.

\subsection{Computation of Multi-linear Forms}

Now we move onto the computation of the multilinear form. As stated above, it is convenient to use the quasiprojector \eqref{eqn:altquasiproj} for computations. The strategy for evaluating the $n^\mathrm{th}$ multi-linear form $\mathcal{M}(W_1,\ldots,W_n)$ is detailed at length in \cite{Joung:2015jza}, which we review here for completeness' sake. For each of the $n$ algebra elements in the argument of multi-linear form, we associate a different coadjoint orbit $\caW_i$, $ i =1,\ldots, n$. We then form a particular Gaussian $e^{Y_+ \cdot \caW_i \cdot Y_-}$ for each $i$. Finally, we may evaluate the trace by using the integral version of the star product to star together the $n$ Gaussians as well as the Gaussian from the alternate quasiprojector \eqref{eqn:altquasiproj}.  In total, we have:
\begin{equation}\mathcal{M}(W_1,\ldots,W_n)\equiv \Delta_{hs_2}^{\mathrm{}} \ast e^{Y_+\cdot \caW_1 \cdot Y_-}\ast \ldots \ast e^{Y_+ \cdot \caW_n \cdot Y_-}\Bigg|_{Y=0} \, .\end{equation}
After this has been evaluated, it may be series-expanded in each $\caW$ to extract the relevant part of the multilinear form, which can be written in terms of products of traces of products of $\caW$.

Now we describe the process of evaluating the multilinear form to obtain a series expansion for the answer in the desired form, products of traces. First, we need to star in the Gaussian form of the quasiprojector. Then we evaluate the star products with the integral version of the star product. This returns a determinant to be evaluated on the matrix of $\caW$'s. We do this by using $\det(I+M) = e^{\tr(\ln M)}$, and then finally we can expand in powers of $\caW$ and carry out the resulting $x$-integrals term-by-term. In all, we have:
\begin{align}\mathcal{M}&= \Delta_{hs_2}^{\mathrm{}} \ast e^{Y_+\cdot \caW_1 \cdot Y_-}\ast \ldots \ast e^{Y_+ \cdot \caW_n \cdot Y_-}\Bigg|_{Y=0} \nonumber \\
&= N\int_0^1 dx~ \sqrt{x}(1-x)^{\frac{\dso-6}{2}}~\phantom{}_2F_1\left(3,-1;\frac{3}{2};\frac{1}{1-x}\right) \nonumber \\
&\qquad \qquad \qquad \times e^{-2\sqrt{x}Y_+ \cdot Y_-} \ast e^{Y_+\cdot \caW_1 \cdot Y_-}\ast \ldots \ast e^{Y_+ \cdot \caW_n \cdot Y_-}\Bigg|_{Y=0} \nonumber \\
&= N \int_0^1 dx~\frac{ \sqrt{x}(1-x)^{\frac{\dso-6}{2}}~\phantom{}_2F_1\left(3,-1;\frac{3}{2};\frac{1}{1-x}\right)}{\det\left[{1\over 2^{n+1}}\left(\prod_{j=1}^n\left(2-\caW_j\right)\right) \left(\left(1-\sqrt{x}\right)\left(\prod_{j=1}^n{2+\caW_j\over 2-\caW_j}\right)+1+\sqrt{x}\right)\right]} \nonumber \\
&= N \int_0^1 dx~\frac{ \sqrt{x}(1-x)^{\frac{\dso-6}{2}}~\phantom{}_2F_1\left(3,-1;\frac{3}{2};\frac{1}{1-x}\right)}{\exp\left[\tr\left(\ln\left[{1\over 2^{n+1}}\left(\prod_{j=1}^n\left(2-\caW_j\right)\right) \left(\left(1-\sqrt{x}\right)\left(\prod_{j=1}^n{2+\caW_j\over 2-\caW_j}\right)+1+\sqrt{x}\right)-1\right]\right)\right]}\, .\label{eqn:multilinearform}\end{align}

From here it is conceptually straightforward but computationally quite intensive to Taylor expand the log, perform the trace, series expand the exponential, then finally expand the $\frac{1}{1-\ldots}$, all the while exploiting the coadjoint conditions satisfied by $\caW$. The only other piece of information we need are the values of the integrals.  (Although a $\sqrt{x}$ appears in the determinant in \eqref{eqn:multilinearform}, only integer powers of $x$ come out at the end of the day.) We are then able to do the integrals over $x$ finding
\begin{equation}N\int_0^1 dx~ \sqrt{x}(1-x)^{\frac{\dso-6}{2}}~\phantom{}_2F_1\left(3,-1;\frac{3}{2};\frac{1}{1-x}\right) x^m=\frac{(-2)^{-m} (2 m+1)\text{!!} (\dso+4 m)}{\dso \left(-\frac{\dso}{2}-m+\frac{3}{2}\right)_m}\, .\end{equation}
We may collect the forms by trace structures and powers of each $\caW$ and read off the coefficients. 

%%%%%%%%%%%%%%%%%%%%%%%%%%%%%%%%%%%%%%%%%%%%%%%%%%%%%%%%%%%%%%%%%%%%%%%%%%%%%%%%

In section \ref{sec:masscomp}, we will see that we need only the bilinear form, which we'll call $\mathcal{B}$, and trilinear form, which we'll call $\mathcal{T}$, for the mass computations we're interested in doing,
\begin{align}\mathcal{B}(W_1,W_2) &\equiv \Delta_{hs_2} \ast W_{1(Y)} \ast W_{2(Y)} \bigg|_{Y=0} \, ,\nn\\
\mathcal{T}(W_1,W_2,W_3)&\equiv \Delta_{hs_2} \ast W_{1(Y)} \ast W_{2(Y)} \ast W_{3(Y)} \bigg|_{Y=0}\, .\label{bitrideff}\end{align}
Suppose that $W_{1,2,3}$ are $hs_2$ fields valued in only particular levels of the algebra; call the levels $n_1$, $n_2$, and $n_3$. We denote the corresponding bilinear and trilinear forms $\mathcal{B}_{(n_1,n_2)}$ and $\mathcal{T}_{(n_1,n_2,n_3)}$, respectively.
We have computed the bilinear form up to fifth order in both $\caW_1$ and $\caW_2$, as well as the trilinear form up to fifth order in $\caW_1$, first order in $\caW_2$, and fifth order in $\caW_3$ which include all the cases we will need to compute the linearized mass spectrum of the $hs_2$ theory. The results are rather lengthy, and so we list them in appendix \ref{sec:multilinearforms}.

\section{A Partially Massless Higher-Spin Theory}
\label{sec:pm}

Just as the original Vasiliev theory can be thought of, in a sense, as a Yang-Mills-like gauge theory with gauge algebra $hs$ on AdS, so too can the partially massless Vasiliev theory be thought of as a Yang-Mills-like gauge theory based on the $hs_2$ algebra on AdS.  We now turn to providing a description of the degrees of freedom of the partially massless higher-spin theory and the way in which they are embedded into $hs_2$ valued fields.

The full non-linear theory can be constructed using the generalized formalism of \cite{Alkalaev:2014nsa}, and  should also be reconstructible from the dual CFT \eqref{dualcft2}, along the lines of e.g. \cite{Boulanger:2015ova,Sleight:2016dba,Sleight:2016xqq}.  We are interested here in studying the linear theory and subtleties of the spectrum, and matching to the dual CFT.   Rather than linearize the full theory, it will be easier for us to directly construct the linear theory from the $hs_2$ algebra.

\subsection{Expectation for the Spectrum}

AdS/CFT tells us that the spectrum of physical fields in the bulk should match the spectrum of single trace primary operators in the CFT.  The spectrum of single trace primaries for the $U(N)$ $\square^2$ CFT has been worked out in \cite{Basile:2014wua,Brust:2016gjy}.  There is a tower of conserved higher spin currents with spins $s=1,2,3,\ldots$ and dimensions $\Delta=d+s-2$.  These should correspond to massless bulk fields with full massless gauge symmetry.
On top of this, there is a tower of ``triply-conserved" currents with spins $s=3,4,5,\ldots$ and dimensions $\Delta=d+s-4$.  These should correspond to $t=s-3$ partially massless bulk fields.  In addition, there are four operators which do not satisfy any conservation condition: two $s=0$ operators of dimension $\Delta=d-4,d-2$, an $s=1$ of dimension $\Delta=d-3$, and an $s=2$ of dimension $\Delta=d-2$.

In the case of the theory on AdS, there is a straightforward map between unitarity of the boundary CFT and unitarity of the bulk theory.  In particular, the sign of the kinetic term of a field in the bulk theory is the same as the sign of the coefficient of the two point function of the field's dual operator.  We may therefore deduce the signs of the kinetic terms of the fields from the calculations of the two-point functions in \cite{Brust:2016gjy}, and we can see precisely which fields are non-unitary due to a wrong sign kinetic term (in addition to the already non-unitary nature of the PM fields).  Unfortunately, as there's no universally agreed-upon action for the Vasiliev theory, we cannot directly check the signs of the bulk kinetic terms to verify this correspondence.  Note that only the relative sign between fields is relevant, as the overall sign can be changed by multiplying the entire bulk action (and CFT action) by $-1$. 

For the theory on dS, however, there is not a straightforward connection between unitarity in the bulk and unitarity of the boundary CFT.  We know that the partially massless fields are themselves unitary on dS, but because we lack an action or a clean link to boundary unitarity, we cannot say whether the relative kinetic signs between the PM fields and other fields of the theory on dS is positive, and thus we cannot make any definitive claim about unitarity of the bulk dS theory.

In tables \ref{tab:spectrum3} through \ref{tab:spectrumD}, we display the expected spectrum on ${\rm AdS}_D$ derived from the $U(N)$ version of the CFT dual, for all dimensions $D\geq 3$ (the masses of the de Sitter version of the theory may be obtained by simply replacing $L^2 \rightarrow -\frac{1}{H^2}$).  In the lower dimensions, various subtleties and truncations occur; in $D=3,5$ the spectrum dramatically truncates, and in $D=4,7$, extended Verma modules appear.  In $D=3,5$, we would not know a priori whether the AdS theory is the dual of the ``finite'' or ``log'' CFTs discussed in \cite{Brust:2016gjy}, however, we will see below that indeed the PM HS theories, as based on the $hs_2$ algebra, are the duals of the finite theories and {\it not} the log theories. Furthermore, as we'll see, the particles in AdS present in $D=3,5$ carry a finite number of modes, rather than the infinite number of modes expected from a full propagating degree of freedom.  This corresponds to the fact that the primaries in the finite CFT have a finite number of descendants.  

As we will discuss below, there is a consistent truncation of this theory where we keep only the even-spin particles, just as in the Vasiliev theory, and this is the dual of the $O(N)$ CFT. The resulting spectra may be read off from the tables below by simply dropping all odd-spin particles.

{
\renewcommand{\arraystretch}{1.3} 
\begin{table}[h]
\centering
\resizebox{5.5in}{.45in}{
\begin{tabular}{|c|ccccc|}\hline
Bulk Field & Spin & Mass & Quantization & Dual Operator   & Kinetic Term Relative Sign \\ \hline
\rowcolor{Gray}  Scalar  & $s=0 $ & $m^2L^2= 8$ & Alternate & $\Delta=-2$ & $+$\\ 
 Scalar  & $s=0$ & $m^2L^2= 0$ & Alternate & $\Delta=0$ & $+$ \\ 
\rowcolor{Gray} Massive Vector  & $s=1 $ & $m^2L^2=4$ & Alternate & $\Delta=-1$  & $-$ \\ 
\hline
\end{tabular}}
\caption{Spectrum of the partially massless higher-spin theory on $AdS_3$.}
\label{tab:spectrum3}
\end{table}
}

{
\renewcommand{\arraystretch}{1.3} 
\begin{table}[h]
\centering
\resizebox{6.6in}{.7in}{
\begin{tabular}{|c|ccccc|}\hline
Bulk Field & Spin & Mass & Quantization & Dual Operator   & Kinetic Term Relative Sign \\ \hline
\rowcolor{Gray} Massless Tower $(t=s-1)$ & $s=1,2,3,\cdots $ & $m^2L^2=0$ & Standard & $\Delta=s+1$ & $\begin{cases} + & s=1 \\ - & s>1 \end{cases}$ \\ 
Partially Massless Tower $(t=s-3)$ & $s=3,4,5,\cdots $ & $m^2L^2=-2(2s-3)$ & Standard & $\Delta=s-1$ & $+$\\ 
\rowcolor{Gray}  Scalar  & $s=0$ & $m^2L^2= -2$ & Alternate & $\Delta=1$ & $+$ \\ 
 Massive Vector  & $s=1 $ & $m^2L^2=2$ & Alternate & $\Delta=0$  & $-$ \\ 
\rowcolor{Gray} Partially Massless Graviton + Scalar Mixture & $s=2,0 $ & $m^2L^2=-2,4$ & Alternate & $\Delta=1,-1$  & $+$\\ 
\hline
\end{tabular}}
\caption{Spectrum of the partially massless higher-spin theory on $AdS_4$.}
\label{tab:spectrum4}
\end{table}
}

{
\renewcommand{\arraystretch}{1.3} 
\begin{table}[h]
\centering
\resizebox{5.5in}{.25in}{
\begin{tabular}{|c|ccccc|}\hline
Bulk Field & Spin & Mass & Quantization & Dual Operator   & Kinetic Term Relative Sign \\ \hline
\rowcolor{Gray}  Scalar  & $s=0 $ & $m^2L^2= 0$ & Alternate & $\Delta=0$ & $+$\\ 
\hline
\end{tabular}}
\caption{Spectrum of the partially massless higher-spin theory on $AdS_5$.}
\label{tab:spectrum5}
\end{table}
}

{
\renewcommand{\arraystretch}{1.3} 
\begin{table}[h]
\centering
\resizebox{6.5in}{.75in}{
\begin{tabular}{|c|ccccc|}\hline
Bulk Field & Spin & Mass & Quantization & Dual Operator   & Kinetic Term Relative Sign \\ \hline
\rowcolor{Gray} Massless Tower $(t=s-1)$ & $s=1,2,3,\cdots $ & $m^2L^2=0$ & Standard & $\Delta=s+3$ & $-$ \\ 
Partially Massless Tower $(t=s-3)$ & $s=3,4,5,\cdots $ & $m^2L^2=-2(2s-1)$ & Standard & $\Delta=s+1$ & $+$\\ 
\rowcolor{Gray}  Scalar  & $s=0 $ & $m^2L^2= -4$ & Alternate & $\Delta=1$ & $+$\\ 
 Scalar  & $s=0$ & $m^2L^2= -6$ & Alternate & $\Delta=3$ & $+$ \\ 
\rowcolor{Gray} Massive Vector  & $s=1 $ & $m^2L^2=-2$ & Alternate & $\Delta=2$  & $+$ \\ 
Massive Graviton & $s=2 $ & $m^2L^2=-6$ & Alternate & $\Delta=3$  & $+$\\ 
\hline
\end{tabular}}
\caption{Spectrum of the partially massless higher-spin theory on $AdS_6$}
\label{tab:spectrum6}
\end{table}
}

{
\renewcommand{\arraystretch}{1.3} 
\begin{table}[h]
\centering
\resizebox{6.5in}{.65in}{
\begin{tabular}{|c|ccccc|}\hline
Bulk Field & Spin & Mass & Quantization & Dual Operator   & Kinetic Term Relative Sign \\ \hline
\rowcolor{Gray} Massless Tower $(t=s-1)$ & $s=1,2,3,\cdots $ & $m^2L^2=0$ & Standard & $\Delta=s+4$ & $-$ \\ 
Partially Massless Tower $(t=s-3)$ & $s=3,4,5,\cdots $ & $m^2L^2=-4s$ & Standard & $\Delta=s+2$ & $+$\\ 
\rowcolor{Gray} Two Scalar Mixture  & $s=0 $ & $m^2L^2= -8,-8$ & Alternate & $\Delta=2,4$ & $+$\\ 
 Massive Vector  & $s=1 $ & $m^2L^2=-4$ & Alternate & $\Delta=3$  & $+$ \\ 
\rowcolor{Gray} Massive Graviton & $s=2 $ & $m^2L^2=-8$ & Alternate & $\Delta=4$  & $+$\\ 
\hline
\end{tabular}}
\caption{Spectrum of the partially massless higher-spin theory on $AdS_7$.}
\label{tab:spectrum7}
\end{table}
}

{
\renewcommand{\arraystretch}{1.3} 
\begin{table}[h]
\centering
\resizebox{6.5in}{.75in}{
\begin{tabular}{|c|ccccc|}\hline
Bulk Field & Spin & Mass & Quantization & Dual Operator   & Kinetic Term Relative Sign \\ \hline
\rowcolor{Gray} Massless Tower $(t=s-1)$ & $s=1,2,3,\cdots $ & $m^2L^2=0$ & Standard & $\Delta=d+s-2$ & $-$ \\ 
Partially Massless Tower $(t=s-3)$ & $s=3,4,5,\cdots $ & $m^2L^2=-2(D+2s-7)$ & Standard & $\Delta=d+s-4$ & $+$\\ 
\rowcolor{Gray}  Scalar  & $s=0 $ & $m^2L^2= -4(D-5)$ & Alternate & $\Delta=d-4$ & $+$\\ 
 Scalar  & $s=0$ & $m^2L^2= -2(D-3)$ & Alternate & $\Delta=d-2$ & $-$ \\ 
\rowcolor{Gray} Massive Vector  & $s=1 $ & $m^2L^2=-2(D-5 )$ & Alternate & $\Delta=d-3$  & $+$ \\ 
Massive Graviton & $s=2 $ & $m^2L^2=-2(D-3)$ & Alternate & $\Delta=d-2$  & $+$\\ 
\hline
\end{tabular}}
\caption{Spectrum of the partially massless higher-spin theory on $AdS_D$ for $D>7$.}
\label{tab:spectrumD}
\end{table}
}

In $D=4$, the massive spin-2 mass value becomes $m^2L^2=-2$, which is the value for a $t=0$ partially massless graviton on $AdS$.  However, as we'll see, no gauge symmetry associated with the $s=2,t=0$ partially massless gauge transformation appears in the $hs_2$ algebra.  This is because in $D=4$, as we'll see in section \ref{sec:ads4}, instead of becoming a partially massless graviton, the graviton pairs up with the $m^2L^2=4$ spin-0 and becomes a field theoretic realization of the extended module, with a total of six propagating degrees of freedom on ${\rm AdS}_4$.

\subsection{Fields}

We now turn to describing the fields of the $hs_2$ theory and how they encode the spectrum discussed above.
The linear partially massless higher-spin theory will be of the same form as the linearized Vasiliev theory, but the linearized master fields, a one-form $W$ and a zero-form $C$, will take values in the $hs_2$ algebra rather than the original $hs$ algebra.  Thus the dynamical fields of the theory are an $hs_2$ valued one-form,
\be  W=\sum_{r=0}^\infty {1\over 2^r r!} W_{(r)}^{A_1B_1,\ldots,A_rB_r}T_{A_1B_1,\ldots,A_rB_r} 
+\sum_{r=2}^\infty {1\over 2^r r!}\tilde{W}_{(r)}^{A_1B_1,\ldots A_{r-2}B_{r-2}, A_{r-1},A_r} \tilde{T}_{A_1B_1,\ldots,A_{r-2}B_{r-2},A_{r-1},A_r}\, ,\ee
and an $hs_2$ valued zero-form
\be C=\sum_{r=0}^\infty {1\over 2^r r!} C_{(r)}^{A_1B_1,\ldots,A_rB_r}T_{A_1B_1,\ldots,A_rB_r} 
+\sum_{r=2}^\infty {1\over 2^r r!}\tilde{C}_{(r)}^{A_1B_1,\ldots A_{r-2}B_{r-2}, A_{r-1},A_r} \tilde{T}_{A_1B_1,\ldots,A_{r-2}B_{r-2},A_{r-1},A_r} \,.\label{Cexpansion}\ee
The components of $W$ will encode the gauge fields, i.e. the massless and partially massless fields and their associated generalized spin connections.  The components of $C$ will encode the field strengths of the gauge fields, as well as the additional massive fields, neither of which will transform under any gauge symmetry at the linear level.

The gauge symmetry will be described in terms of a gauge parameter zero-form, also valued in $hs_2$,
\be \Epsilon=\sum_{r=0}^\infty {1\over 2^r r!} \Epsilon_{(r)}^{A_1B_1,\ldots,A_rB_r}T_{A_1B_1,\ldots,A_rB_r} 
+\sum_{r=2}^\infty {1\over 2^r r!}\tilde{\Epsilon}_{(r)}^{A_1B_1,\ldots A_{r-2}B_{r-2}, A_{r-1},A_r} \tilde{T}_{A_1B_1,\ldots,A_{r-2}B_{r-2},A_{r-1},A_r}\,.\ee
The components of $\Epsilon$ will encode the gauge parameters for all the massless and partially massless fields, as well as St\"uckelberg symmetries associated to generalized local Lorentz transformations.

The linear theory also uses a zeroth-order ``background" one-form $\bW$, which has non-trivial values only at the first level of the algebra,
\be \bW={1\over 2}\bW_{}^{AB} T_{AB}.\ee
The components of the background one form are the spin connection one-form, $\omega^{ab}$, and the vielbein one-form, $e^a$, of the background AdS space,
\be \bW_{}^{AB}=\begin{cases} \bW_{}^{aD}=-{1\over L}e^a \, ,\\ \bW_{}^{ab}=\omega^{ab}\, .\end{cases}
\ee

At this point, we have restricted to $AdS$, so that the embedding space metric becomes
\be \eta_{AB}=\left(\begin{array}{cc}\eta_{ab} & 0 \\0 & -1\end{array}\right)\, ,\ee
with the $A,B,\ldots$ indices ranging from $0,1,\cdots,D$.  Raising or lowering a $D$ index costs a minus sign.

The $AdS_D$ background is a solution of the fully nonlinear equations, and it satisfies a covariant flatness condition
\begin{equation} d\bW +  \bW \star \bW = 0\, .\end{equation}
(Note that this is a two-form equation where we've left the wedge product between the two $\bW$'s implicit; we do the same in many equations below.)

\subsection{Equations of Motion}

The linearized equations of motion and gauge symmetries are
\bea && \cD W = \mathcal{C}(C)\, ,\label{fulllineq1}\\
&& \tilde{\cD}C =0,  \label{fulllineq2}\\
&& \delta W = \cD \Epsilon\, , \ \ \delta C=0\, . \label{fulllineq3}
\eea
Several explanations are in order:
First, several covariant derivatives have been invoked, which are defined as follows:
\bea && \cD W \equiv dW + \frac{1}{2}[\bW,W]_\star\, , \label{fulllineq4}\\
&& \tilde{\cD}C \equiv dC + [\bW , C]_\star^\Pi ,  \label{fulllineq5}\\
&& \cD \Epsilon \equiv d\Epsilon + [\bW, \Epsilon]_\star \,  . \label{fulllineq6}
\eea
Here, 
\begin{align}[\bW,W]_\star&=\bW \star W + W \star \bW  \, ,\\ 
[\bW, \Epsilon]_\star&=\bW\star\Epsilon-\Epsilon\star\bW \, , \\
[\bW , C]_\star^\Pi &= \bW \star C - C \star \PibW\, \label{twistedcdefn}
\end{align} 
are wedge star commutators of the respective $hs_2$ valued form fields. (The relative plus sign in the $[\bW,W]_\star$ commutator is due to both $\bW$ and $W$ being one-forms.) $\cD W$ acts as a generalized field strength and is invariant under linearized gauge transformations \eqref{fulllineq3}.  The ``twisted commutator" \eqref{twistedcdefn}
is part of a covariant derivative $\tilde{\cD}$ associated to the ``twisted-adjoint'' representation of $hs_2$, where $\Pi$ is an automorphism of $so(D,1)$ (which extends in the natural way to the entire $hs_2$ algebra) which sends $T_{aD}\rightarrow -T_{aD}$, $T_{ab}\rightarrow T_{ab}$, so that we have
\be \bW_{\Pi}^{AB}=\begin{cases} \bW_{\Pi}^{aD}={1\over L}e^a \, ,\\ \bW_{\Pi}^{ab}=\omega^{ab}\, .\end{cases}
\ee
 Finally, $\mathcal{C}$ in \eqref{fulllineq1} is a so-called ``co-cycle" which depends linearly on $C$, whose precise form we will not need.  
 
 The first equation, \eqref{fulllineq1}, will give equations that determine the field strengths in terms of the gauge fields.  The second equation, \eqref{fulllineq2}, will give  equations of motion satisfied by the field strengths, and will determine the equations of motion for the non-gauge fields.  The final equation, \eqref{fulllineq3}, contains the gauge transformation laws for all the massless and partially massless fields.
 
\subsection{Patterns of Unfolding\label{unfoldsubsection1}}

We begin by first understanding what AdS fields are contained in the components of $W$ and $C$.  As in the original Vasiliev theory, the degrees of freedom appear in an unfolded formulation.  This means there are many more fields in the master fields $W$ and $C$ then there are physical fields corresponding to the spectrum of the theory, and there are many more gauge parameters in the master gauge parameter field $\Epsilon$ then there are gauge parameters for the physical gauge fields.  The extra gauge parameters in $\Epsilon$ are St\"uckelberg gauge symmetries, generalizations of local Lorentz invariance in GR, which can be algebraically gauged away.  The extra fields in $W$ and $C$ are either auxiliary fields whose values are ultimately determined algebraically in terms of the physical fields, or are St\"uckelberg fields corresponding to the extra St\"uckelberg gauge symmetries.
 
The master fields contain $so(\dso)$ tensors transforming as traceless $[r,r]$ tableaux for all $r$, as well as tensors transforming as traceless $[r+2,r]$ for all $r$. The AdS spacetime fields are these tensors reduced down to $\dads$ dimensions. Therefore we must carry out a dimensional reduction of each of the algebra components.  The components present in the reductions are
\be \raisebox{-70pt}{\epsfig{file=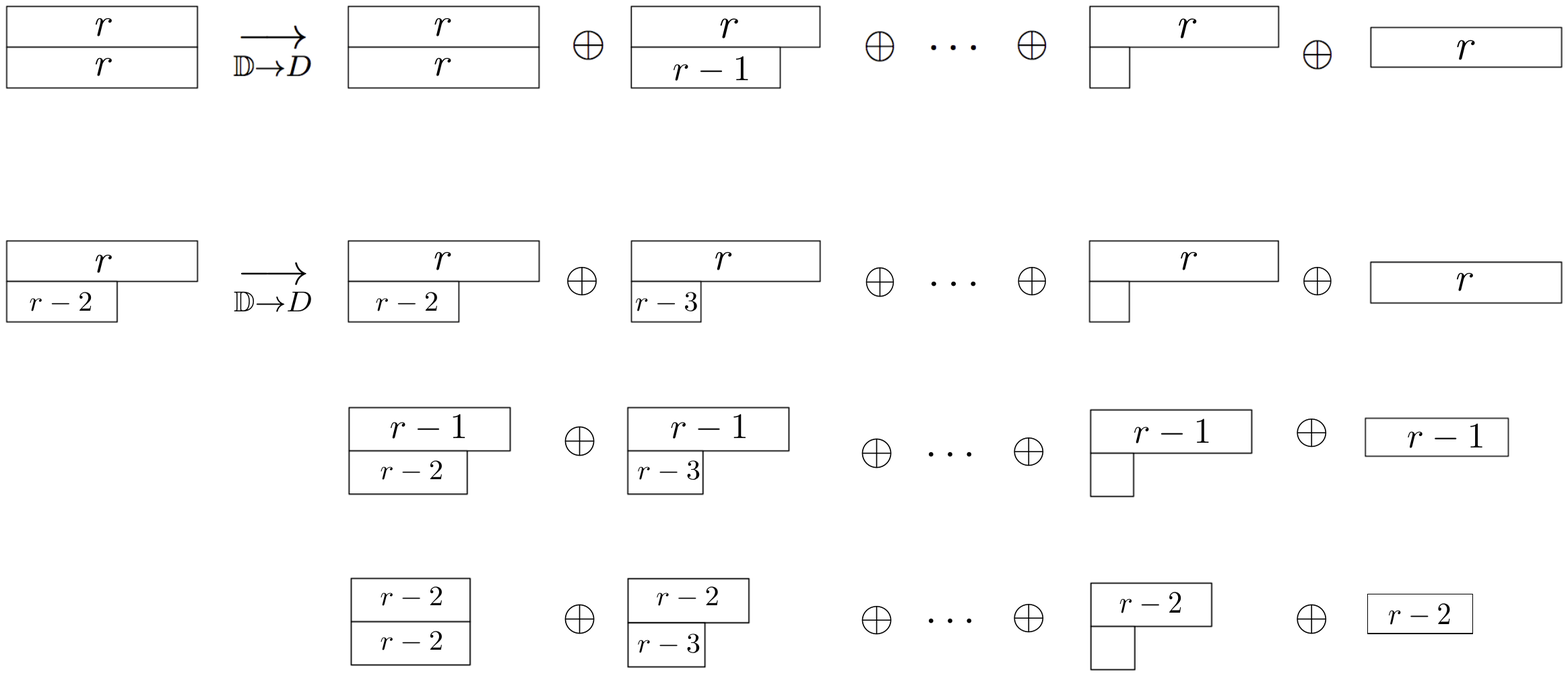,height=4.0in,width=5.5in}}
\ee

We will notate the ${\rm AdS}_D$ tensors coming from the reduction by using the same symbol as the $so(\dso)$ parent tensor, only with a lowercase letter for the name of the tensor as well as lowercase letters for AdS tangent space indices. 
(We abuse notation and keep using $\epsilon$ for the AdS gauge parameters as well, because we don't need to talk about their lower dimensional parts too often.) The number and grouping of AdS indices suffices to uniquely identify each AdS tensor, because any given tableau occurs with multiplicity of at most one in the reduction of a given $so(\dso)$ parent tensor. We use commas to separate indices coming from different columns, so that the number of indices and the placement of the commas uniquely specifies an irrep.

As an example, we list out all symmetry generators up to $O(T^4)$ coming from the $C$ master field in table \ref{tab:fieldsByT}. We color those fields red which do not survive the truncation to the minimal theory with even spins only (as discussed in section \ref{sec:truncation}).  A similar reduction holds for $W$, however, as we discuss in section \ref{sec:truncation}, a different collection of the fields in $W$ survive the truncation to the minimal theory.

{\renewcommand{\arraystretch}{1.3}
\begin{table}[h]
\centering
\begin{tabular}{|rl|rlrlrl|}\hline
\multicolumn{2}{|c|}{$so(\dso)$ tensor} & \multicolumn{6}{c|}{$AdS_{\dads}$ tensors} \\ \hline

\rowcolor{Gray} $\bullet $&$= C_{(0)}$ & $\bullet$&$ = c_{(0)}$& & & &\\

${\tiny \Yvcentermath1 \yng(1,1)}$&$ = C_{(1)}^{AB}$ & ${\color{red} {\tiny \Yvcentermath1 \yng(1,1)} } $&$ {\color{red}= c_{(1)}^{ab}}$ & ${\tiny \Yvcentermath1 \yng(1)} $&$ =c_{(1)}^{a}$ & &\\ 

\rowcolor{Gray} ${\tiny \Yvcentermath1 \yng(2,2)} $&$= C_{(2)}^{AB,CD}$ & $ {\tiny \Yvcentermath1 \yng(2,2)} $&$= c_{(2)}^{ab,cd}$& ${\color{red} {\tiny \Yvcentermath1 \yng(2,1)}}$&$  {\color{red}=c_{(2)}^{ab,c}}$ & ${\tiny \Yvcentermath1 \yng(2)}$&$ = c_{(2)}^{a,b}$ \\

${\tiny \Yvcentermath1 \yng(2)}$&$ = \tilde{C}_{(2)}^{A,B}$ & ${\tiny \Yvcentermath1 \yng(2)} $&$= \tilde{c}_{(2)}^{a,b}$ & ${\color{red} {\tiny \Yvcentermath1 \yng(1)}} $&$ {\color{red}=\tilde{c}_{(2)}^{a}}$ &$\bullet $&$= \tilde{c}_{(2)}$\\ 

\rowcolor{Gray} ${\tiny \Yvcentermath1 \yng(3,3)}$&$ = C_{(3)}^{AB,CD,EF}$ & ${\color{red} {\tiny \Yvcentermath1 \yng(3,3)}}$&${\color{red}=c_{(3)}^{ab,cd,ef}}$ & ${\tiny \Yvcentermath1 \yng(3,2)}$&$=c_{(3)}^{ab,cd,e}$ & ${\color{red} {\tiny \Yvcentermath1 \yng(3,1)}}$&${\color{red}=c_{(3)}^{ab,c,d}}$ \\

\rowcolor{Gray} && ${\tiny \Yvcentermath1 \yng(3)}$&$ = c_{(3)}^{a,b,c}$&&&& \\ 

${\tiny \Yvcentermath1 \yng(3,1)}$&$ = \tilde{C}_{(3)}^{AB,C,D}$ & ${\color{red} {\tiny \Yvcentermath1 \yng(3,1)}} $&$ {\color{red}=\tilde{c}_{(3)}^{ab,c,d}}$ & ${\tiny \Yvcentermath1 \yng(3)} $&$= \tilde{c}_{(3)}^{a,b,c}$ & ${\tiny \Yvcentermath1 \yng(2,1)} $&$= \tilde{c}_{(3)}^{ab,c}$ \\

&&${\color{red} {\tiny \Yvcentermath1 \yng(2)}}$&$  {\color{red}=\tilde{c}_{(3)}^{a,b}}$ & ${\color{red} {\tiny \Yvcentermath1 \yng(1,1)}}$&$  {\color{red}=\tilde{c}_{(3)}^{ab}}$ & ${\tiny \Yvcentermath1 \yng(1)}$&$ = \tilde{c}_{(3)}^a$\\ 

\rowcolor{Gray} ${\tiny \Yvcentermath1 \yng(4,4)}$&$ = C_{(4)}^{AB,CD,EF,GH}$ & ${\tiny \Yvcentermath1 \yng(4,4)} $&$= c_{(4)}^{ab,cd,ef,gh}$ & \color{red} ${\tiny \Yvcentermath1 \yng(4,3)} $&$\color{red}= {c_{(4)}^{ab,cd,ef,g}}$ & &\\

\rowcolor{Gray} &&${\tiny \Yvcentermath1 \yng(4,2)}$&$ = c_{(4)}^{ab,cd,e,f}$&  ${\color{red} {\tiny \Yvcentermath1 \yng(4,1)}}$&$  {\color{red}=c_{(4)}^{ab,c,d,e}}$ & ${\tiny \Yvcentermath1 \yng(4)} $&$= c_{(4)}^{a,b,c,d}$\\ 

${\tiny \Yvcentermath1 \yng(4,2)}$&$ = \tilde{C}_{(4)}^{AB,CD,E,F}$ & ${\tiny \Yvcentermath1 \yng(4,2)} $&$= \tilde{c}_{(4)}^{ab,cd,e,f}$ & ${\color{red} {\tiny \Yvcentermath1 \yng(4,1)}}$&$  {\color{red}=\tilde{c}_{(4)}^{ab,c,d,e}}$ & ${\tiny \Yvcentermath1 \yng(4)}$&$ = \tilde{c}_{(4)}^{a,b,c,d}$ \\

&& ${\color{red} {\tiny \Yvcentermath1 \yng(3,2)}}$&$  {\color{red}=\tilde{c}_{(4)}^{ab,cd,e}}$ & ${\tiny \Yvcentermath1 \yng(3,1)}$&$ = \tilde{c}_{(4)}^{ab,c,d}$ & ${\color{red} {\tiny \Yvcentermath1 \yng(3)}} $&$ {\color{red}=\tilde{c}_{(4)}^{a,b,c}}$ \\

&& ${\tiny \Yvcentermath1 \yng(2,2)} $&$= \tilde{c}_{(4)}^{ab,cd}$ & ${\color{red} {\tiny \Yvcentermath1 \yng(2,1)}}$&$  {\color{red}=\tilde{c}_{(4)}^{ab,c}}$ & ${\tiny \Yvcentermath1 \yng(2)} $&$= \tilde{c}_{(4)}^{a,b}$ \\ \hline
\end{tabular}
\caption{Decomposition of the zero-form $hs_2$-valued master field $C$ into AdS spacetime fields, arranged by powers of $T$, including up to $O(T^4)$. The fields shown in red do not survive the truncation to the minimal theory, as discussed in section \ref{sec:truncation}.}
\label{tab:fieldsByT}
\end{table}
}
In  table \ref{tab:fieldsByIrrep}, we have rearranged the right-hand side of the above table by type of tensor rather than by level of the algebra (note that there are a few tensors which appear which are higher than $O(T^4)$).

{\renewcommand{\arraystretch}{1.3}
\begin{table}[h]
\centering
\begin{tabular}{|c|llll|}\hline
tensor type & \multicolumn{4}{|c|}{$AdS_{\dads}$ tensors} \\ \hline
\rowcolor{Gray}$\bullet$ & $c_{(0)}$ & $\tilde{c}_{(2)}$ & &\\ 
${\tiny \Yvcentermath1 \yng(1)}$ & $c_{(1)}^a$ & ${\color{red} \tilde{c}_{(2)}^a}$ & $\tilde{c}_{(3)}^a$ &\\ 
\rowcolor{Gray}${\tiny \Yvcentermath1 \yng(1,1)}$ & ${\color{red} c_{(1)}^{ab}}$ & ${\color{red} \tilde{c}_{(3)}^{ab}}$ & &\\ 
${\tiny \Yvcentermath1 \yng(2)}$ & $c_{(2)}^{a,b}$ & $\tilde{c}_{(2)}^{a,b}$ & ${\color{red} \tilde{c}_{(3)}^{a,b}}$ & $\tilde{c}_{(4)}^{a,b}$  \\ 
\rowcolor{Gray}${\tiny \Yvcentermath1 \yng(2,1)}$ & ${\color{red} c_{(2)}^{ab,c}}$ & $\tilde{c}_{(3)}^{ab,c}$ & ${\color{red} \tilde{c}_{(4)}^{ab,c}}$ & \\ 
${\tiny \Yvcentermath1 \yng(3)}$ & $c_{(3)}^{a,b,c}$ & $\tilde{c}_{(3)}^{a,b,c}$ & ${\color{red} \tilde{c}_{(4)}^{a,b,c}}$ & $\tilde{c}_{(5)}^{a,b,c}$ \\ 
\rowcolor{Gray}${\tiny \Yvcentermath1 \yng(2,2)}$ & $c_{(2)}^{ab,cd}$ & $\tilde{c}_{(4)}^{ab,cd}$ & &\\ 
${\tiny \Yvcentermath1 \yng(3,1)}$ & ${\color{red} c_{(3)}^{ab,c,d}}$ & ${\color{red} \tilde{c}_{(3)}^{ab,c,d}}$ & $\tilde{c}_{(4)}^{ab,c,d}$ & ${\color{red} \tilde{c}_{(5)}^{ab,c,d}}$\\ 
\rowcolor{Gray}${\tiny \Yvcentermath1 \yng(4)}$ & $c_{(4)}^{a,b,c,d}$ & $\tilde{c}_{(4)}^{a,b,c,d}$ & ${\color{red} \tilde{c}_{(5)}^{a,b,c,d}}$ & $\tilde{c}_{(6)}^{a,b,c,d}$\\ \hline
\end{tabular}
\caption{Decomposition of the zero-form $hs_2$-valued master field $C$ into AdS spacetime fields, arranged by spacetime irreps, including up to four-index tableaux. Again, the fields shown in red do not survive the truncation to the minimal theory.}
\label{tab:fieldsByIrrep}
\end{table}
}

We may do the same for the one-form master field $W$, and the types of fields present are the same, only now each of the fields also carries a space-time one-form index. However, a different collection of fields survives the truncation to the minimal theory. Finally, we may do the same for the zero-form gauge parameter $\Epsilon$.

\begin{figure}[h]
\centering
\includegraphics[width=0.8\textwidth]{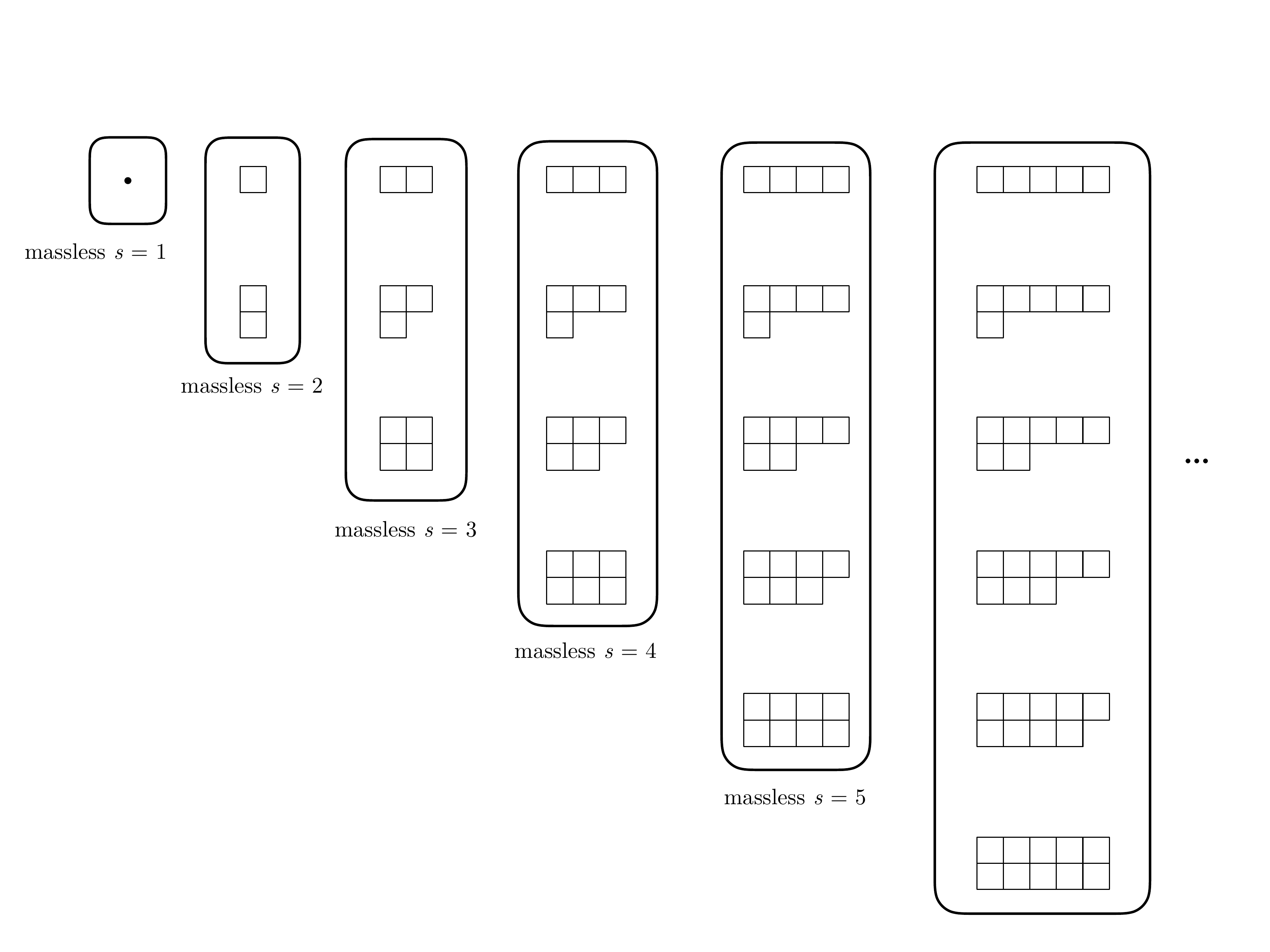}
\caption{\label{fig:unfolding1}The AdS one-forms in the master field $W$ which act as unfolding fields for the massless gauge fields in the theory. These are identical to the usual result in the Vasiliev theory.}
\end{figure}

\begin{figure}[h]
\centering
\includegraphics[width=6in]{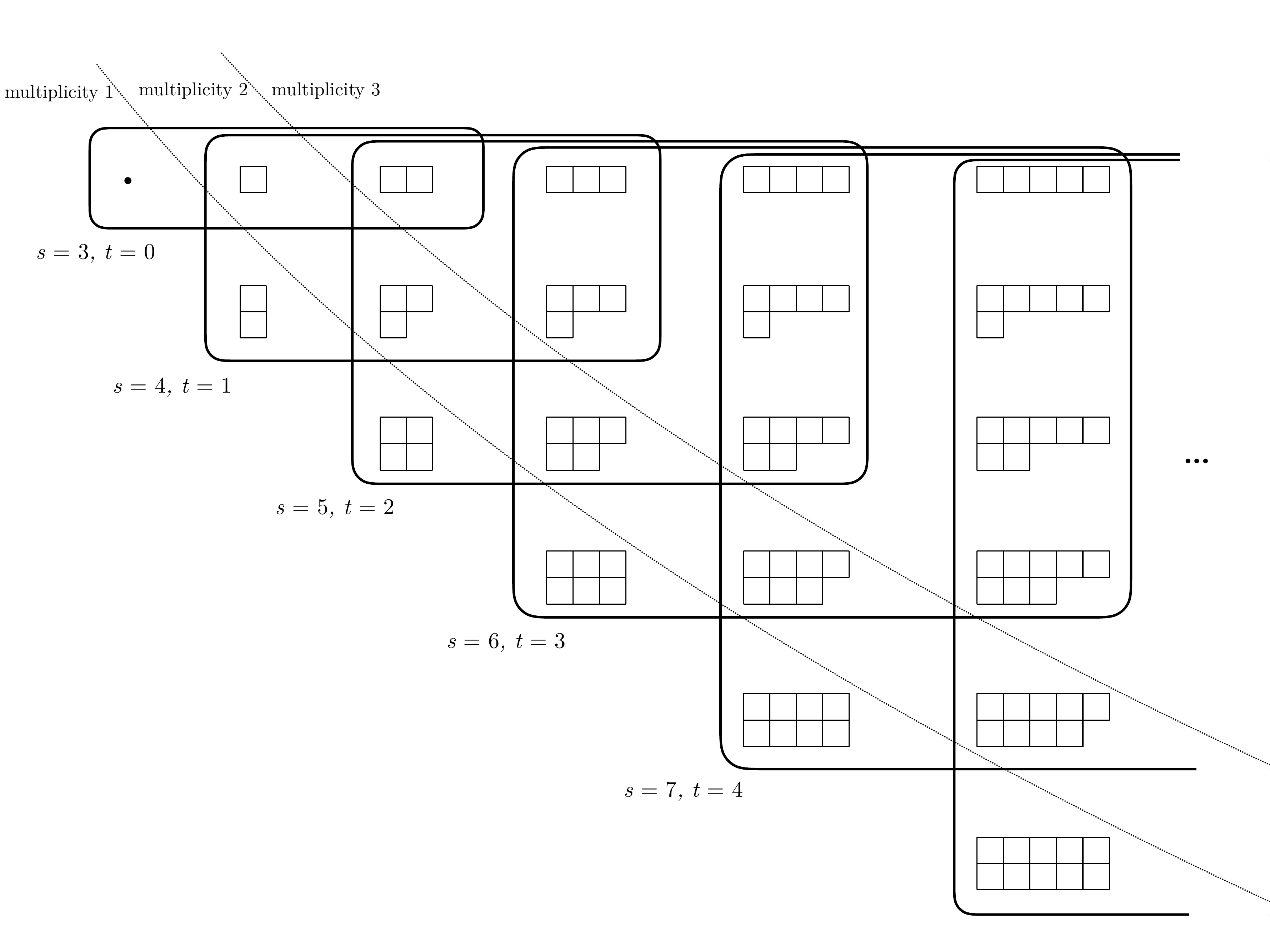}
\caption{\label{fig:unfolding3} Fields in $W$ which correspond to the spin-3 and higher PM fields. The bottom tableau in each column occurs with multiplicity 1, the one to the right of it with multiplicity 2, and all others with multiplicity 3.   We box what fields are needed for a given PM spin.  Note that mixings may occur between tableaux of the same shape.}
\end{figure}

The massless and partially gauge fields as well as their generalized connections are carried by the one-forms $W$.  In figures \eqref{fig:unfolding1} and \eqref{fig:unfolding3} we have arranged the AdS fields coming from $W$ corresponding to the spin of the particle they describe. The fields which are present already in the original $hs$ algebra are in figure \eqref{fig:unfolding1}, and these carry the massless degrees of freedom.  The fields new to $hs_2$ are in figure \eqref{fig:unfolding3}, and these carry the partially massless degrees of freedom.  The gauge parameters are also arranged in a completely analogous fashion.

The massless fields appear in a frame-like formulation just as they do in the original Vasiliev theory.  The symmetric $s-1$ field at the top of each column, when direct producted with the one form index, contains a fully symmetric $s$ index tensor component and its trace,
\be\small
\begin{array}{|c c c c c|}\hline%\vphantom{\biggm|}
&\!\!\!&s-1&&\!\!\!\\
\hline
\end{array}
\ \otimes \
\begin{array}{|c c  c|}\hline%\vphantom{\biggm|}
&\!\!\! &\ \ \ \ \!\!\!\\
\hline
\end{array}
\ \ \ \ \   = {\hspace*{.75cm}}~
\begin{array}{|c c c c c|}\hline%\vphantom{\biggm|}
&~&\ \ \ \ s\ \ \ \ &~&\\
\hline
\end{array}~ \, 
\oplus \ \ \begin{array}{|c c c|}\hline%\vphantom{\biggm|}
&~s-2 &  \\
\hline
\end{array}~ \, ,\label{prod1formdd}
\ee
which together form the $s$ index double traceless symmetric tensor carrying the massless spin $s$ field in the Fronsdal formulation.  All the other one forms in the column below $[s-1] $, are auxiliary fields, St\"uckelberg fields and generalized spin connections.  For the gauge parameters, the $[s-1] $ at the top of each column is the gauge parameter of \eqref{PMgaugesym} corresponding to the massless field.  The remaining fields in the column below are St\"uckelberg gauge parameters corresponding to generalized local Lorentz transformations.

For example, for $s=1$ the standard Maxwell gauge parameter is $\epsilon_{(0)}$ and the massless photon is carried by $w_{(0)\mu}$.  For $s=2$, $\epsilon_{(1)}^a$ is the diffeomorphism gauge symmetry of the graviton, and $\epsilon_{(1)}^{ab}$ is local Lorentz symmetry for the graviton. 
$w_{(0)\mu}^a$ is the vielbein; the symmetric part is the metric and the anti-symmetric part is pure St\"uckelberg and can be gauged away by the local Lorentz transformations.  The field $w_{(0)\mu}^{ab}$ is the spin connection, the gauge field associated to the local Lorentz symmetry, which is auxiliary and is determined algebraically in terms of the metric. 

The partially massless fields appear in the frame-like formulation of \cite{Skvortsov:2006at}.  The symmetric $s-1$ field at the top right corner of each rectangle in \eqref{fig:unfolding3}, when direct producted with the one form index as in \eqref{prod1formdd}, contains a fully symmetric $s$ index tensor component which carries the partially massless spin $s$.  All the other one forms in the rectangle are auxiliary fields, St\"uckelberg fields and generalized spin connections needed for the frame-like description of a spin $s$, depth $t=s-3$ field as described in \cite{Skvortsov:2006at}.   The gauge parameters also arrange as in \eqref{fig:unfolding3};  the field on the top left of each rectangle is the partially massless gauge parameter of \eqref{PMgaugesym}, and the others are all generalized local Lorentz transformations.

For example, consider the $s=3$, $t=0$ partially massless field.  The field $\tilde w_{(2)\mu}^{a,b}$ decomposes into irreducible representations $[3] \oplus [2,1]  \oplus [1]$, the field $\tilde w_{(2)\mu}^{a}$ decomposes into irreducible representations $[2] \oplus [1,1]  \oplus \bullet$, and the field $\tilde W_{(2)\mu}^{}$ is a $[1]$.  The gauge parameters $\tilde\epsilon_{(2)}^{a,b}$ and $\tilde\epsilon_{(2)}^{a}$ are generalized local Lorentz transformations that can be used to gauge away the $[2]$ and one of the $[1]$'s, whereas $\tilde\epsilon_{(2)}$ is the PM gauge parameter.  The $[2,1]$ and $[1,1]$ are gauge fields associated to the generalized local Lorentz symmetries $\tilde\epsilon_{(2)}^{a,b}$ and $\tilde\epsilon_{(2)}^{a}$. They are auxiliary and are determined algebraically.  What remains is a $[1]$ and a $[3] $ which combine into a trace-{\rm full} symmetric rank 3 tensor, and a $[0]$ describing an extra scalar auxiliary.  This is precisely the field content needed to describe a partially massless $s=3$, $t=0$ off shell in the Fronsdal description (see e.g. Appendix  A of \cite{Hinterbichler:2016fgl}, and also \cite{Hallowell:2005np,Gover:2014vxa}).

\begin{figure}[h]
\centering
\includegraphics[width=6in]{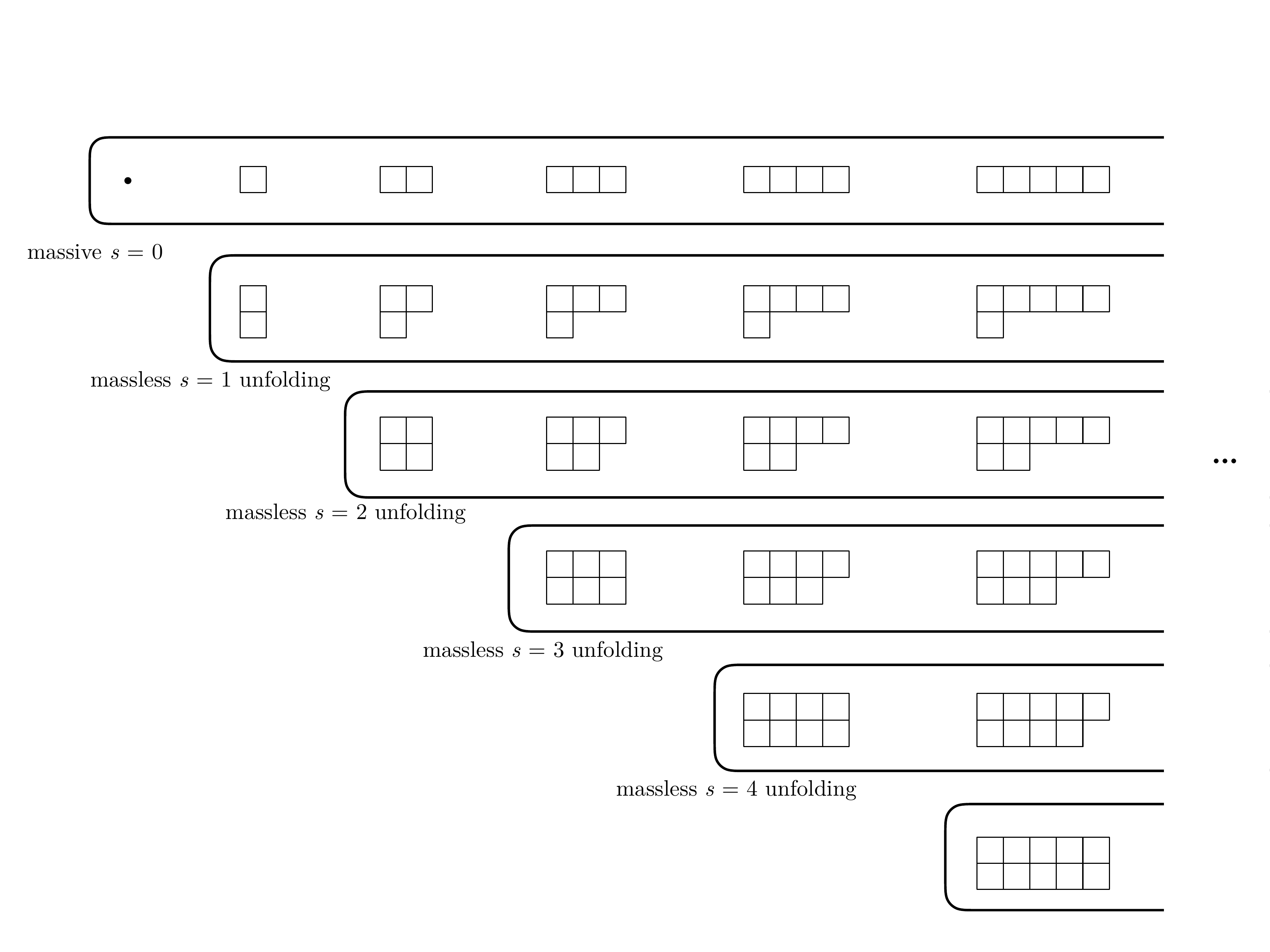}
\caption{\label{fig:unfolding2}The AdS zero-forms in the master field $C$ which act as unfolding fields and field strengths for the massless gauge fields in the theory. These are identical to the usual result in the Vasiliev theory.}
\end{figure}

\begin{figure}[h]
\centering
\includegraphics[width=6in]{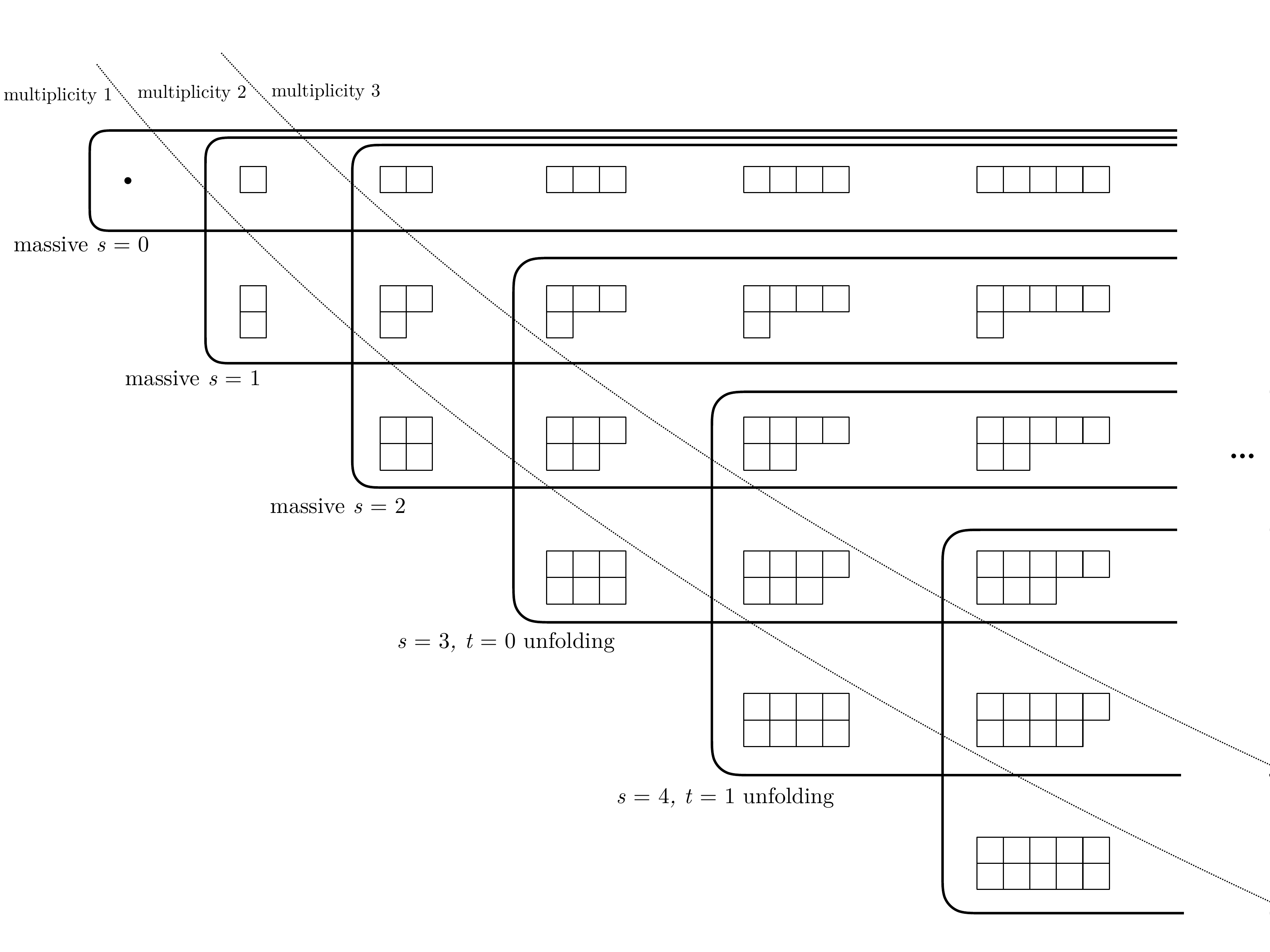}
\caption{\label{fig:unfolding4}We show the unfolding fields in $C$ which describe the three new massive particles, as well as the unfolding fields and field strengths for the PM fields. As with the previous figure, these arise from the new generators in $hs_2$ not present in the $hs$ algebra. The bottom tableau in each column occurs with multiplicity 1, the one to the right of it with multiplicity 2, and all others with multiplicity 3.  We box what fields are needed for a given spin.}
\end{figure}

We now move on to describe the zero-form fields present in $C$.  These are arranged in figures \eqref{fig:unfolding2} and \eqref{fig:unfolding4}.  Figure \eqref{fig:unfolding2} contains the fields present already in the original Vasiliev theory, and figure \eqref{fig:unfolding4} contains the fields new to $hs_2$.  Note that the representations are identical to those present in $W$, but their arrangement by spin is now different, with an infinite number of fields corresponding to each spin.  These fields are the ``unfolding fields," which describe the gauge invariant fields strengths corresponding to the gauge fields, as well as all on-shell non-vanishing derivatives of the field strengths.

Starting with the massless unfolding fields in \eqref{fig:unfolding2}, the field on the right of each row are auxiliary fields which will become the on-shell non-vanishing part of the gauge invariant fields strength of the given spin.  For example, $c_{(1)}^{ab}$ is the standard anti-symmetric Maxwell tensor associated to the massless spin-1, and $c_{(2)}^{ab,cd}$ is the Weyl tensor associated to the massless spin-2.  The tensors to the right are auxiliary fields which will become on-shell non-vanishing derivatives of the field strengths.  In addition to the field strengths associated to spins $s\geq 1$, there is also a row for $s=0$; this parametrizes the massive scalar which is not among the gauge fields and has no associated gauge symmetry.  The scalar field $c_{(0)}$ on the left of the row parametrizes the scalar, and the symmetric traceless tensor fields to the right of it will parametrize all on-shell non-vanishing derivatives of the scalar field.

We now turn to the new fields of $C$ which are not present in the original Vasiliev theory.  These are shown in \eqref{fig:unfolding2}, grouped according to spin.  The rectangles for $s\geq 3$ each contain unfolding fields describing the partially massless fields \cite{Boulanger:2008up,Boulanger:2008kw,Skvortsov:2009nv,Ponomarev:2010st}.  The $[s,s-2]$ tensor in the upper left corner is a field which will be set to the traceless part of the gauge invariant field strength of the partially massless spin $s$. (The partially massless field strengths and their properties are described in \cite{Hinterbichler:2016fgl}.)  The remaining fields in the rectangle will parametrize all on-shell non-vanishing derivatives of the field strength.  In addition to the unfolding fields for spin $s\geq 3$, there are also fields for spins $0,1,2$; these are the new non-gauge fields present in $hs_2$ which are not in the original Vasiliev theory.  One of the fields lying in the upper left corner of the $s=0,1,2$ rectangles parametrizes the massive field, and the remaining fields describe unfolding fields parametrizing its on-shell non-vanishing derivatives.

Throughout this discussion, we have attributed various tensors directly to physical fields and field strengths, however in reality fields of the same representation can mix, and it will in general be a linear combination of fields which describes the desired physical particle or field strength.  We will see examples of this mixing in section \ref{sec:masscomp}.

The structures described in this section have been in line with the predictions of AdS/CFT; the gauge symmetries fix the masses of all but four particles to be precisely what we expected from the CFT dual. However there are four particles in the spectrum which have no linearized gauge symmetry, and so in order to see if those particles' masses are consistent with CFT expectations, we must unfold the equations of motion for those particles and explicitly compute their masses. We turn to that in section \ref{sec:masscomp}.

\subsection{Truncation to the Minimal Theory}
\label{sec:truncation}

Like in the original Vasiliev theory, there is a consistent truncation where we keep only the even-spin fields and their associated unfolding fields.  This truncation is dual to an $O(N)$ $\square^2$ CFT which contains only the even spin single trace primaries.

The truncation is achieved by defining a $\mathbb{Z}_2$ operation, under which the AdS fields are classified as even or odd. The truncated theory keeps only the fields which are even. For the gauge fields of $W$, the ${\rm AdS}$ component fields $W_{(n)}^{a\ldots}$ are even or odd according to whether the level $n$ of the algebra element it comes from is odd or even, respectively. For example, $W_{(0)}$ is level 0, and so we discard it under the truncation, whereas $W_{(1)}^{a}$ and $W_{(1)}^{ab}$ are level 1, and so we keep them. The gauge parameters follow the same rule.

For the fields of $C$, the rule is more complicated, owing to the structure of the twisted-adjoint representation.  Let $m$ denote the number of $D$ indices that appear in the dimensional reduction; in other words, the difference between the number of indices that the embedding space tensor that contains the AdS field of interest has, and the number of indices that the AdS field itself has. For example, consider $c_{(1)}^a$. It has one index, but it descended from the embedding space tensor $C_{(1)}^{AB}$, which had two indices. Therefore, $c_{(1)}^a$ has $m=1$.
The parity of a field $c_{(n)}^{a\ldots}$ is given by $n+m$ mod $2$. Since the embedding space tensors all have an even number of indices, this is equivalent to the following rule: $n$ plus the number of indices in the tensor, mod $2$. To truncate to the minimal theory, we eliminate all the odd fields. To illustrate this rule, we've colored the tensors in tables \ref{tab:fieldsByT} and \ref{tab:fieldsByIrrep} black for the even fields and red for the odd fields.  Only the black fields survive the truncation.

\FloatBarrier

\section{Mass Computations}
\label{sec:masscomp}

In this section, we compute the masses of the four particles (two scalars, a vector and a tensor) in the linearized theory which individually have no gauge symmetries determining their masses.  

From the putative dual CFT, we have expectations for what these masses should be.  In the dual $\square^2$ theory, there are four single-trace primary operators which are not conserved, and so should correspond to the four massive particles in the PM higher-spin theory. The CFT makes predictions for the masses of these particles in AdS via \eqref{m2dimdeadscfte}, since the scaling dimensions are trivial to work out in the CFT. There are two scalars with scaling dimensions $d-4$ and $d-2$, a vector with scaling dimension $d-3$ and a tensor with scaling dimension $d-2$.  Therefore we expect the two scalar masses to be
\be m_0^2L^2= \begin{cases}  -4(D-5)\, , \\  -2(D-3) \, ,\end{cases}\,  \label{scalarmassexpectd}
\ee
 the vector mass to be
\begin{equation}m_1^2L^2 = -2(D-5 )\, ,\label{vectormassexp}\end{equation}
 and the tensor mass to be
\begin{equation}m_2^2L^2 = -2(D-3)\, .\label{tensormassexp}\end{equation}
For all four particles, the minus sign in \eqref{dimeqosc}, \eqref{dimeqo} is necessary, telling us that they should be quantized with alternate boundary conditions.

In this section we work out these masses from the structure of the linearized Vasiliev equations and the bilinear and trilinear forms of the $hs_2$ algebra.
The procedure is the following: all four fields are contained in the $hs_2$ valued scalar field $C$, therefore the masses should be derivable purely from the $C$ equation of motion \eqref{fulllineq2}
\be \tilde{\cD} C= dC+\bW\star C-C\star \PibW=0\, .\label{fulllineq22}\ee

We introduce an $hs_2$-valued zero-form dummy variable $V$ which we'll use to extract independent components of this equation. $V$ has an expansion in terms of $hs_2$ generators mirroring that of $C$ in \eqref{Cexpansion},
\be V=\sum_{r=0}^\infty {1\over 2^r r!} V_{(r)}^{A_1B_1,\ldots,A_rB_r}T_{A_1B_1,\ldots,A_rB_r} 
+\sum_{r=2}^\infty {1\over 2^r r!}\tilde{V}_{(r)}^{A_1B_1,\ldots A_{r-2}B_{r-2}, A_{r-1},A_r} \tilde{T}_{A_1B_1,\ldots,A_{r-2}B_{r-2},A_{r-1},A_r}\, ,\ee
whose components can then be reduced to AdS spacetime components in a manner identical to $C$. 
To project out individual spacetime equations of motion, we star $V$ into \eqref{fulllineq22} and then take a trace using the $hs_2$ trace operation \eqref{tracedefn},
\be \Tr \( V\star dC \)+\Tr \( V\star \bW \star C\)-\Tr \( V\star C\star \PibW\)=0,\label{fulllineq23}\ee
then read off the various components of $V$, that is, we may set all but one AdS spacetime tensor in $V$ to 0 in order to read off a particular spacetime equation, and then iterate over different tensors in $V$ to pick out various equations.  As it turns out, the masses of the four particles of interest can be obtained by turning on tensors in $V$ only up to level 4. 

From here, the computation is reduced to evaluating bilinear and trilinear forms of the $hs_2$ algebra \eqref{bitrideff}, due to the presence of two factor and three factor star products in \eqref{fulllineq23}.  For this we use the techniques described in section \eqref{sec:hs2generalities} and the results of the bilinear and trilinear form calculations compiled in appendix \ref{sec:multilinearforms}.
In a term of a given order in $V$, because of the structure of the star product \eqref{eqn:schematicstarproduct}, and because $\bW$ is only non-vanishing at first level, we only have {a finite number of tensors} from the master field in $C$ contributing; namely, levels of $C$ between the level of $V$ minus one up to the level of $V$ plus one.
Our equations of motion thus become
\bea && {\cal B}_{(0,0)}\( V,dC\)+\left[{\cal T}_{(0,1,1)}\( V, \bW, C\)-{\cal T}_{(0,1,1)}\( V,  C,\PibW\)\right]=0, \label{firsteqform1} \\
&& {\cal B}_{(1,1)}\( V,dC\)+\sum_{i=0}^2 \left[ {\cal T}_{(1,1,i)}\( V, \bW, C\)-{\cal T}_{(1,i,1)}\( V, C, \PibW\)\right]=0,\label{firsteqform2}  \\
&& {\cal B}_{(2,2)}\( V,dC\)+\sum_{i=1}^3 \left[ {\cal T}_{(2,1,i)}\( V, \bW, C\)-{\cal T}_{(2,i,1)}\( V, C, \PibW\)\right]=0, \label{firsteqform3}\\
&& {\cal B}_{(3,3)}\( V,dC\)+\sum_{i=2}^4 \left[ {\cal T}_{(3,1,i)}\( V, \bW, C\)-{\cal T}_{(3,i,1)}\( V, C, \PibW\)\right]=0,  \\
&& {\cal B}_{(4,4)}\( V,dC\)+\sum_{i=3}^5 \left[ {\cal T}_{(4,1,i)}\( V, \bW, C\)-{\cal T}_{(4,i,1)}\( V, C, \PibW\)\right]=0, \\
&& \qquad \qquad \qquad \vdots \nn
\eea
From here, we may plug in our multi-linear forms (expressed in terms of coadjoint orbits) tabulated in appendix \ref{sec:multilinearforms}, and then collect coadjoint orbits into embedding space tensors with equations \eqref{eqn:orbitToTensor1}, \eqref{eqn:orbitToTensor2}.

\subsection{Reduction to $D$ dimensions\label{reductionsectn}}

At this point, we have equations in terms of the $\dso$ dimensional component fields of $C$.  We must then descend down to the AdS component fields described in section \ref{unfoldsubsection1}.  As described there, the embedding space metric decomposes as
\be \eta_{AB}=\begin{cases} \eta_{DD}=-1\, , \\ \eta_{aD}=0 \, , \\ \eta_{ab}=\eta_{ab}\, ,\end{cases}
\ee
where $\eta_{ab}$ is the flat tangent space metric of AdS, and the background one-form decomposes as
\be \bW_{}^{AB}=\begin{cases} \bW_{}^{aD}=-{1\over L}e^a \, ,\\ \bW_{}^{ab}=\omega^{ab}\, ,\end{cases}\, 
\ee
\be \bW_{\Pi}^{AB}=\begin{cases} \bW_{\Pi}^{aD}={1\over L}e^a\, ,\\ \bW_{\Pi}^{ab}=\omega^{ab}\, .\end{cases}
\ee

To compute the masses, we will need the tensors which carry the physical degrees of freedom of the four massive particles, as well as the first level unfolding fields, i.e. the fields which will parametrize first derivatives of the physical fields.  To compute the scalar mass, we will need all even (under the $\mathbb{Z}_2$ truncation discussed in section \ref{sec:truncation}) scalar fields and all the even vector fields.  To compute the vector mass, we will use all odd vector fields and all odd anti-symmetric rank two tensor fields.  To compute the tensor mass, we will use all even tensor fields and even rank 3 mixed symmetry fields.  In total, the fields we need to find are
\bea c_{(0)},\ \tilde{c}_{(2)} &\in& \bullet \ \ \ {\rm (even)} \nn\\
{c}_{(1)}^a,\ \tilde{c}_{(3)}^a &\in& {\Yvcentermath1 \scriptsize\yng(1)} \ \ \ {\rm (even)}\nn\\
 \tilde{c}_{(2)}^a &\in& {\Yvcentermath1 \scriptsize\yng(1)} \ \ \ {\rm (odd)} \nn\\ 
{c}_{(1)}^{ab},\ \tilde{c}_{(3)}^{ab} &\in& {\Yvcentermath1 \scriptsize\yng(1,1)} \ \ \ {\rm (odd)} \nn\\
{c}_{(2)}^{a,b},\ \tilde{c}_{(2)}^{a,b},\ \tilde{c}_{(4)}^{a,b} &\in& {\Yvcentermath1 \scriptsize\yng(2)} \ \ \ {\rm (even)}\nn\\
\tilde{c}_{(3)}^{ab,c} &\in& {\Yvcentermath1 \scriptsize\yng(2,1)} \ \ \ {\rm (even)}\, \ \ \ . \label{necesaryfieldse}
\eea

We must now find the explicit embeddings of the fields \eqref{necesaryfieldse} into their respective $\dso$ tensors.  The scalar component gives a scalar,
\be C_{(0)}=c_{(0)}.\ee
At the first level, we break up
\be C_{(1)}^{AB}=\begin{cases} C_{(1)}^{aD}=c_{(1)}^a \, ,\\ C_{(1)}^{ab}=c_{(1)}^{ab}\, ,\end{cases}
\ee
where ${c}_{(1)}^{ab}\in {\Yvcentermath1 \scriptsize\yng(1,1)} $,  ${c}_{(1)}^{a}\in {\Yvcentermath1 \scriptsize\yng(1)}$.
This realizes the dimensional reduction 
\begin{equation}{\Yvcentermath1 \yng(1,1)} \underset{\dso \rightarrow \dads}{\rightarrow} {\Yvcentermath1 \yng(1,1)} \oplus {\Yvcentermath1 \yng(1)}\, \ \ .\end{equation}

The higher tensors get increasingly trickier to break up.  We get these by knowing what lower dimensional tensors are needing using group theory, then demanding that the $\dso$ tensor is traceless and has the right young symmetry and is traceless, given that the $D$ tensors have these properties. 
At level 2, we have
\be {C}_{(2)}^{AB,CD}=\begin{cases}  {C}_{(2)}^{aD,bD}={c}_{(2)}^{a,b}\, ,\\ {C}_{(2)}^{ab,cD}=0 \, ,%{c}_2^{ab,c} 
\\ {C}_{(2)}^{ab,cd}={1\over D-2}\left(\eta^{ac}c_{(2)}^{b,d}-\eta^{ad}c_{(2)}^{b,c}+\eta^{bd}c_{(2)}^{a,c}-\eta^{bc}c_{(2)}^{a,d}\right)\, ,\end{cases}
\ee
where 
${c}_{(2)}^{a,b}\in {\Yvcentermath1 \scriptsize\yng(2)} $.  This realizes the dimensional reduction,
\begin{equation}{\Yvcentermath1 \yng(2,2)} \underset{\dso \rightarrow \dads}{\supset}  {\Yvcentermath1 \yng(2)}\, .\end{equation}
Next up is the first field new to $hs_2$,
\be \tilde{C}_{(2)}^{A,B}=\begin{cases}  \tilde{C}_{(2)}^{D,D}=\tilde{c}_{(2)}\, ,\\ \tilde{C}_{(2)}^{a,D}=\tilde{c}_{(2)}^a \,,\\ \tilde{C}_{(2)}^{a,b}=\tilde{c}_{(2)}^{a,b}+{1\over D}\tilde{c}_{(2)} \eta^{ab}\, ,\end{cases}
\ee
where $\tilde {c}_{(2)}^{a,b}\in {\Yvcentermath1 \scriptsize\yng(2)} $,  $\tilde {c}_{(2)}^{a}\in {\Yvcentermath1 \scriptsize\yng(1)}$, $\tilde {c}_{(2)}\in {\bullet} $.
This realizes the dimensional reduction for traceless representations:
\begin{equation}{\Yvcentermath1 \yng(2)} \underset{\dso \rightarrow \dads}{\rightarrow} {\Yvcentermath1 \yng(2)} \oplus {\Yvcentermath1 \yng(1)} \oplus {\bullet}\, \ \ \ .\end{equation}

At level 3, we will not need the reduction of $\tilde C_{(3)}^{AB,CD,EF}$, because it contains none of the fields \eqref{necesaryfieldse} in its reduction.
We will however need $\tilde{C}_{(3)}^{AB,C,D}$,
\be \tilde{C}_{(3)}^{AB,C,D}=\begin{cases} \tilde{C}_{(3)}^{aD,D,D}= \tilde{c}_{(3)}^a \, ,\\ \tilde{C}_{(3)}^{ab,D,D}=\tilde{c}_{(3)}^{ab}  \, ,\\  \tilde{C}_{(3)}^{aD,b,D}={1\over 2}\tilde{c}_{(3)}^{ab}  \, ,\\ \tilde{C}_{(3)}^{aD,b,c}=
{2\over 3}\tilde{c}_{(3)}^{a(b,c)}+{D+1\over (D+2)(D-1)}\eta^{b,c}\tilde{c}_{(3)}^a-{2\over (D-1)(D+2)}\eta^{a,(b}\tilde{c}_{(3)}^{c)}\, ,\\ \tilde{C}_{(3)}^{ab,c,D}=\tilde{c}_{(3)}^{ab,c}-{2\over D-1}\eta^{c,[a}\tilde{c}_{(3)}^{b]} \, ,\\ \tilde{C}_{(3)}^{ab,c,d}=  
{1\over D+2}\left[\eta^{c,d}\tilde{c}_{(3)}^{ab}-\eta^{b,(d}\tilde{c}_{(3)}^{c)a}  +\eta^{a,(c}\tilde{c}_{(3)}^{d)b}   \right] \, ,   \end{cases}
\ee
where $\tilde{c}_{(3)}^{a}\in {\Yvcentermath1 \scriptsize\yng(1)} $, $\tilde{c}_{(3)}^{ab}\in {\Yvcentermath1 \scriptsize\yng(1,1)} $,  $\tilde{c}_{(3)}^{ab,c}\in {\Yvcentermath1 \scriptsize\yng(2,1)} $.    
This realizes the dimensional reduction for traceless representations:
\begin{equation}{\Yvcentermath1 \yng(3,1)} \underset{\dso \rightarrow \dads}{\subset}  {\Yvcentermath1 \yng(2,1)}  \oplus {\Yvcentermath1 \yng(1,1)} \oplus {\Yvcentermath1 \yng(1)} \, .\end{equation}

We will not need the reduction of $\tilde C_{(4)}^{AB,CD,EF,GH}$, because it contains no fields of \eqref{necesaryfieldse}
in its reduction, but $\tilde{C}_{(4)}^{AB,CD,E,F}$ contains $\tilde{c}_{(4)}^{a,b} \subset {\Yvcentermath1 \scriptsize\yng(2)}$, and we will need to know how this is embedded,
\be \tilde{C}_{(4)}^{AB,CD,E,F}=\begin{cases} \tilde{C}_{(4)}^{aD,bD,D,D}=\tilde{c}_{(4)}^{a,b} \, ,\\ 
\tilde{C}_{(4)}^{ab,cD,D,D}=0 \, , \\   
\tilde{C}_{(4)}^{aD,bD,c,D}=0  \, ,\\
\tilde{C}_{(4)}^{ab,cd,D,D}=  {1\over D-2}\left(\eta^{ac}\tilde{c}_{(4)}^{b,d}-\eta^{ad}\tilde{c}_{(4)}^{b,c}+\eta^{bd}\tilde{c}_{(4)}^{a,c}-\eta^{bc}\tilde{c}_{(4)}^{a,d}\right)  \, ,\\ 
\tilde{C}_{(4)}^{ab,cD,d,D}=  {1\over D(D-2)}\left(\eta^{ac}\tilde{c}_{(4)}^{b,d}-\eta^{bc}\tilde{c}_{(4)}^{a,d}\right)+{D-1\over D(D-2)}\left(\eta^{bd}\tilde{c}_{(4)}^{a,c}-\eta^{ad}\tilde{c}_{(4)}^{b,c}\right) \, , \\
\tilde{C}_{(4)}^{aD,bD,c,d}=  {4\over D(D-2)(D+4)}\eta^{ab}\tilde{c}_{(4)}^{c,d}+{D^2-2D-4\over D(D-2)(D+4)}\eta^{cd}\tilde{c}_{(4)}^{a,b} \\ \quad\quad\quad\quad\quad +{1\over D^2-2D+8}\left(\eta^{ac}\tilde{c}_4^{b,d}+\eta^{ad}\tilde{c}_{(4)}^{b,c}+\eta^{bd}\tilde{c}_{(4)}^{a,c}+\eta^{bc}\tilde{c}_{(4)}^{a,d}\right)  \, , \\ 
\tilde{C}_{(4)}^{ab,cd,e,D}=0 \, ,\\ 
\tilde{C}_{(4)}^{ab,cD,d,e}=0 \, ,\\
\tilde{C}_{(4)}^{ab,cd,e,f}=-{40\over D(D-2)(D+1)(D+4)} P_{[4,2]}\left[\eta^{ab}\eta^{cd}  \tilde{c}_{(4)}^{e,f}\right] -{20(D^2+3D-2)\over D(D-2)(D+1)(D+4)}P_{[4,2]}\left[\eta^{ab}\eta^{ef}  \tilde{c}_{(4)}^{c,d}\right]  \, .  \end{cases}
\ee
This realizes the dimensional reduction
\begin{equation}{\Yvcentermath1 \yng(4,2)} \underset{\dso \rightarrow \dads}{\supset} {\Yvcentermath1 \yng(2)}\, . \end{equation}
None of the higher levels contain any of the fields \eqref{necesaryfieldse}, so this is all we will need.

The auxiliary master field $V$ must also be reduced from embedding space to AdS tensors. The decompositions look identical to the above, only with $C\rightarrow V$ and $c\rightarrow v$.

\subsection{Extracting AdS Equations of Motion}

Now we get $D$-dimensional equations by plugging in the above and pulling off coefficients of various $v$ components. We are manually setting to zero everything in $C$ which is not one of the fields in \eqref{necesaryfieldse}. Then, we convert all the AdS Lorentz indices to spacetime indices by combining with the background vielbein $e_\mu^{\ a}$.  The background spin connections $\omega_\mu^{\ ab}$ all combine with the differential in \eqref{fulllineq23} to form an $AdS$ covariant spacetime derivative $\nabla_\mu$ (which also provides a nontrivial check on the computations).

\begin{figure}[h]
\centering
\includegraphics[width=7in]{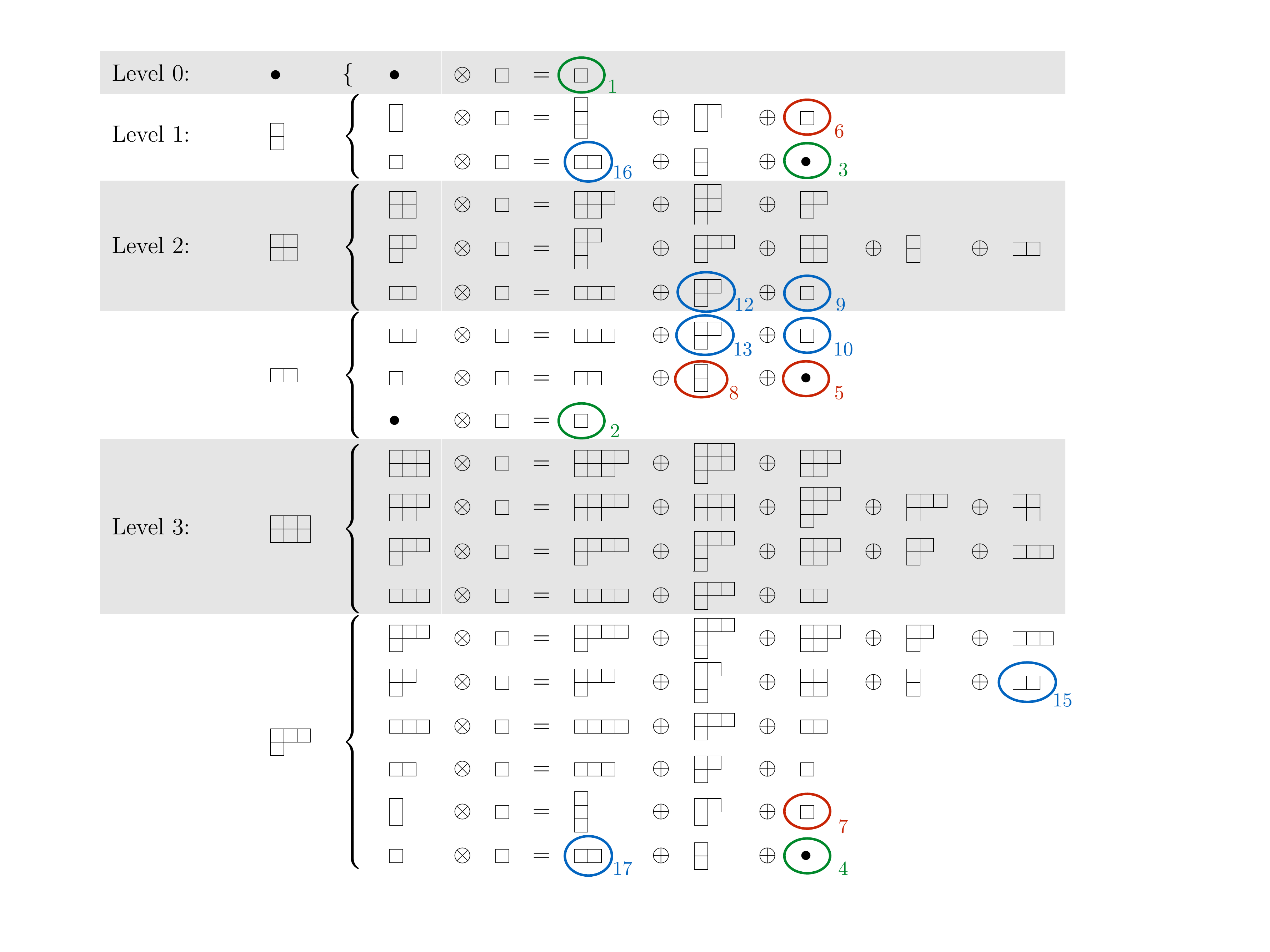}
\caption{\label{fig:massequationstable1} Equations of motion up to level 3.  In the text, the equations are referenced by the numbers appearing by the circles.}
\end{figure}

We'll now go through several examples, starting with the first equation \eqref{firsteqform1}.  Using our bi/tri-linear forms from appendix \ref{sec:multilinearforms} this becomes
\bea &&  V_{(0)}\left(dC_{(0)}-\frac{(D-5) (D+3)}{4 D (D+1)}\(\la \caC \cabW\ra-\la \caPibW \caC\ra\) \right)=0 \,. \label{eq11}
\eea
Converting from coadjoint orbits to tensors using the replacements \eqref{eqn:orbitToTensor1} and reading off the $V_0$ component, we get the equation
 \be   dC_{(0)}+\frac{(D-5) (D+3)}{4 D (D+1)}C^{(1)}_{AB}\(\bW^{AB}-\PidbW^{AB}\)=0\, . \label{eq21} \ee
 Breaking up into AdS components using the reductions in section \ref{reductionsectn}, and converting the exterior derivative to a covariant derivative, this becomes the vector equation 
 \be\nabla_\mu c_{(0)}+\frac{(D-5) (D+3) }{D (D+1) L}{c_{(1)}}_\mu=0\, . \ee
 This is equation number 1 in figure \ref{fig:massequationstable1}, which, as we will discuss below, tabulates all the different AdS equations that arise.

Going on to equation \eqref{firsteqform2}, using the bi/tri-linear forms from appendix \ref{sec:multilinearforms} we get
\bea
&&-\frac{(D-5) (D+3)}{4 D (D+1)}\la \caV \cadC\ra-\frac{(D-5) (D+3)}{4 D (D+1)}\(C_{(0)}\la \caV \cabW\ra-C_{(0)}\la \caPibW \caV\ra\)\nn \\
&&-\frac{(D-5) (D+3)}{4 D (D+1)}\(\la \caV \cabW \caC\ra-\la \caV \caC \caPibW\ra\) 
+ \frac{3 (D-5) (D-3) (D+5) }{32 D (D+1) (D+2)}\( \la \caV \caC\ra \la \cabW \caC\ra-\la \caV \caC\ra \la \caC\caPibW\ra\) \nn \\
&&-\frac{3 (D-5) (D+7) }{8 D (D+1) (D+2)} \(\la \caV \cabW \caC^2\ra - \la \caV \caC^2\caPibW\ra  \)=0\, .  \label{eq12}
\eea
(Note that $\cadC$ is the coadjoint orbit associated to $dC$, not the exterior derivative of the coadjoint orbit $\caC$.) Converting to tensors using \eqref{eqn:orbitToTensor1}, we get the equation
\bea
&&\frac{(D-5) (D+3)}{4 D (D+1)} V^{(1)}_{AB}\bigg[dC_{(1)}^{AB}+C_{(0)}\(\bW^{AB}-\PidbW^{AB}\)-\(\bW^{BC} C_{C}^{{(1)} A}- C_{(1)}^{BC} \bW_{\Pi C}^{\ \ \ A}\) \nn \\
&&  -\frac{6}{(D-1)(D+3)} \(\bW^{BC}\tilde{C}_{C,}^{{(2)} A} - \tilde{C}_{(2)}^{B,C}\bW_{\Pi C}^{\ \ \ A}  \)\bigg]=0.  \label{eq22}
\eea
 Breaking up into AdS components using the reductions in section \ref{reductionsectn}, this leads to two equations, one obtained by reading off the terms proportional to $v_{(1)}^{ab}$,
 \be  {D-5\over 4D(D+1)} \left[\left(D+3\right)\nabla^\rho  c_{(1)}^{\mu\nu} +{12\over (D-1)L}g^{\rho[\mu}\tilde c_{(2)}^{\nu]}+\frac{3 (D-3) (D+5)}{2 (D+2) L}c_{(2)}^{\mu\nu,\rho}\right]=0\, , \label{v2abeq1}\ee
 and one obtained by reading off the terms proportional to $v_{(1)}^{b}$,
 \be -\frac{D-5}{2 D (D+1)}\left[\left(D+3\right)\nabla^\mu c_{(1)}^\nu +g^{\mu\nu}\left({6\over DL}\tilde c_{(2)}-{2(D+3)\over L}c_{(0)}\right) +\frac{3 (D-3) (D+5)}{2 (D+2) L}c_{(2)}^{\mu,\nu}-{6\over (D-1)L}\tilde c_{(2)}^{\mu,\nu}\right]\, . \label{v2abeq2} \ee
 In both cases we combined the exterior derivative with the background spin connection to form a covariant derivative.
\eqref{v2abeq1} has the symmetries of an antisymmetric tensor times the one-form index, which can be split into a totally anti-symmetric part, a traceless mixed symmetry part, and a trace: ${\Yvcentermath1 \scriptsize\yng(1,1)} \otimes  {\Yvcentermath1 \scriptsize\yng(1)} ={\Yvcentermath1 \scriptsize\yng(1,1,1)}  \oplus  {\Yvcentermath1 \scriptsize\yng(2,1)}  \oplus  {\Yvcentermath1 \scriptsize\yng(1)}$.  The trace part, ${\Yvcentermath1 \scriptsize\yng(1)}$, is equation 6 in figure \ref{fig:massequationstable1}.  \eqref{v2abeq2} has the symmetries of a vector times the vector index, which can be split into a symmetric traceless part, an anti-symmetric part, and a trace: ${\Yvcentermath1 \scriptsize\yng(1)} \otimes  {\Yvcentermath1 \scriptsize\yng(1)} ={\Yvcentermath1 \scriptsize\yng(2)}  \oplus  {\Yvcentermath1 \scriptsize\yng(1,1)}  \oplus \bullet $.
The trace part, $\bullet$, is equation 3 in figure \ref{fig:massequationstable1}, and the symmetric traceless part, ${\Yvcentermath1 \scriptsize\yng(2)}$, is equation 16 in figure \ref{fig:massequationstable1}.

We may continue in this way to extract equations.  \eqref{firsteqform3} becomes,  
\bea
&& \frac{3 (D-5) (D-3) (D+5) }{64 D (D+1) (D+2)}\la \caV \cadC\ra ^2-\frac{3 (D-5) (D+7) }{16 D (D+1) (D+2)}\la \caV^2 \cadC^2\ra \nn \\
&& +\frac{3 (D-5) (D-3) (D+5) }{32 D (D+1) (D+2)}\la \caV\caC\ra\( \la \caV\cabW\ra-\la \caV\caPibW\ra\) -\frac{3 (D-5) (D+7)}{8 D (D+1) (D+2)}\(\la \caV^2\cabW\caC\ra-\la \caV^2\caC\caPibW\ra\) \nn\\
&& +\frac{3 (D-5) (D-3) (D+5) }{32 D (D+1) (D+2)} \la \caV\caC\ra \( \la \caV\cabW\caC\ra-\la \caV\caC\caPibW\ra\)+\frac{3 (D-5) (D+7)}{16 D (D+1) (D+2)}\( \la \caV^2\caC^2\cabW\ra+\la \caV^2\caC^2\caPibW\ra\) \nn\\
&& -\frac{3 (D-5) (D-3) (D-1) (D+7) }{128 D (D+1) (D+2) (D+4)} \la \caV\caC\ra ^2 \(\la \cabW\caC\ra -\la \caC\caPibW\ra\) \nn\\ &&+\frac{3 (D-5) (D-3) (D+9) }{16 D
   (D+1) (D+2) (D+4)} \la \caV\caC\ra \( \la \caV\cabW\caC^2\ra-\la \caV\caC^2\caPibW\ra\)  \nn \\ && +\frac{3 (D-5) (D-3) (D+9) }{32 D (D+1) (D+2) (D+4)}\la \caV^2\caC^2\ra \(\la \cabW\caC\ra-\la \caC\caPibW\ra\)=0\, . \label{eq13}
\eea 
The $hs_2$ component $\tilde{V}^{{(2)}}_{A,B}=\caV_{AC}\caV^{C}_{\ B}$ of \eqref{eq13}, which also includes the trace part coming from \eqref{eqn:orbitToTensor1}, becomes
\bea
&&  -\frac{3 (D-5) }{4 D (D+1) (D-1)}\tilde{V}^{{(2)}}_{A,B}\bigg[d\tilde{C}_{(2)}^{A,B} 
 +2\(\bW_{BC}C^C_{\ A}- C_{BC}\bW^C_{\Pi A}\)
- \tilde{C}^{(2)}_{B,C}\(  \bW^C_{\ A}+ \bW^C_{\Pi A}\)\bigg]=0\, . \nn\\ \label{eq23}
\eea 
 Breaking up into AdS components, this leads to three equations, one obtained by reading off the terms proportional to $\tilde v_{(2)}^{a,b}$,
  \be -\frac{3 (D-5)}{D(D-1)  (D+1)}\left[ {1\over 4}\nabla^\rho\tilde c_{(2)}^{\mu,\nu} -\frac{D-3}{(D-1) (D+2)^2 L}g^{\rho(\mu}\tilde c_{(3)}^{\nu)_T}+{1\over L}g^{\rho(\mu} c_{(1)}^{\nu)_T}-{D-3\over 3(D+2)L}\tilde c_{(3)}^{(\mu|\rho|,\nu)} \right]=0\, ,  \label{u2abeq1}\ee
  one obtained by reading off the terms proportional to $\tilde v_{(2)}^{b}$, 
\be \frac{3 (D-5)}{2 D(D-1)  (D+1)}\left[  \nabla^\mu \tilde c_{(2)}^\nu+{2\over L}c_{(1)}^{\mu\nu}+{2(D-3)\over (D+2)L}\tilde c_{(3)}^{\mu,\nu}  +{D-3\over (D+2)L}\tilde c_{(3)}^{\mu\nu}  \right] =0\,, \label{u2abeq2}\ee 
and  one obtained by reading off the terms proportional to $\tilde v_{(2)}^{}$, 
\be   -\frac{3(D-5)  }{D^2(D-1)} \left[\frac{1   }{4  } \nabla^\mu \tilde{c}_{(2)}  +\frac{ (D-1)  }{ (D+1)  L} {c_{(1)}}^\mu +\frac{ 
   (D-3)  }{2 (D+2) L} \tilde{c}^{{(3)}\mu} \right]=0\,.  \label{u2abeq3}\ee 
\eqref{u2abeq1} has the symmetries of a traceless symmetric tensor times the vector index, which can be split into a totally symmetric traceless part, a traceless mixed symmetry part, and a trace: ${\Yvcentermath1 \scriptsize\yng(1,1)} \otimes  {\Yvcentermath1 \scriptsize\yng(1)} ={\Yvcentermath1 \scriptsize\yng(3)}  \oplus  {\Yvcentermath1 \scriptsize\yng(2,1)}  \oplus  {\Yvcentermath1 \scriptsize\yng(1)}$.  The trace part, ${\Yvcentermath1 \scriptsize\yng(1)}$, is equation 10 in figure \ref{fig:massequationstable1}, and the traceless mixed symmetry part, ${\Yvcentermath1 \scriptsize\yng(2,1)}$, is equation 13 in figure \ref{fig:massequationstable1}. \eqref{u2abeq2} has the symmetries of a vector times the vector index, which can be split into a symmetric traceless part, an anti-symmetric part, and a trace: ${\Yvcentermath1 \scriptsize\yng(1)} \otimes  {\Yvcentermath1 \scriptsize\yng(1)} ={\Yvcentermath1 \scriptsize\yng(2)}  \oplus  {\Yvcentermath1 \scriptsize\yng(1,1)}  \oplus \bullet $.  The trace part, $\bullet$, is equation 5 in figure \ref{fig:massequationstable1}, and the symmetric traceless part, ${\Yvcentermath1 \scriptsize\yng(2)} $, is equation 8 in figure \ref{fig:massequationstable1}.  \eqref{u2abeq3} is equation 2 in figure \ref{fig:massequationstable1}.

\begin{figure}[h]
\centering
\includegraphics[width=6in]{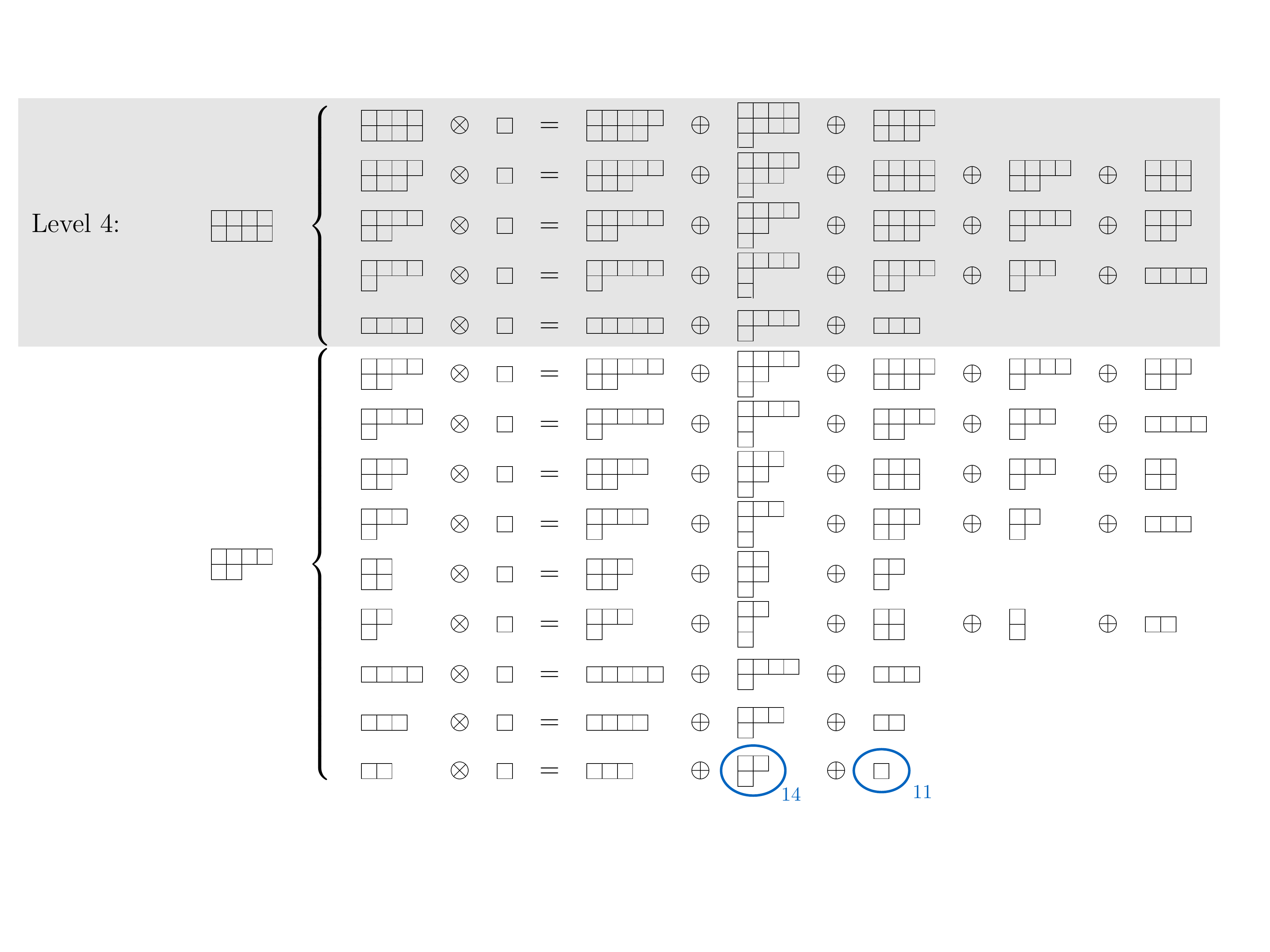}
\caption{\label{fig:massequationstable2} Equations of motion at level 4.  In the text, the equations are referenced by the numbers appearing by the circles.}
\end{figure}

In figures \ref{fig:massequationstable1} and \ref{fig:massequationstable2}, we show graphically all of the equations of motion obtained in this manner up to level 4 in the algebra.  On the left is shown the $\dso$ dimensional component of $V$.  This reduces to the various $D$ dimensional components shown in the next column, which are then multiplied by the one-form index represented by ${\Yvcentermath1 \scriptsize\yng(1)}$.  This product is then broken up into irreducible pieces, each of which becomes a separate spacetime equation of motion.  The circled equations are the ones used in sections \ref{scalarmasssubsection}, \ref{vectormasssubsection}, \ref{tensormasssubsection} to compute masses.  The green circles are the ones used in section \ref{scalarmasssubsection} for the scalar masses, the red circles are the ones used in section \ref{vectormasssubsection} for the vector mass, and the blue circles are the ones used in section \ref{tensormasssubsection} for the tensor mass.  In the text, the equations are referenced by the numbers appearing by the circles.

For each of the scalar, vector and tensor mass computations below, we will see nuances happen in particular dimensions. We will return to these special cases in section \eqref{sec:nuances}.

\subsection{The Scalar Masses\label{scalarmasssubsection}}

The only two scalars in $C$ are $c_{(0)}$ and $\tilde c_{(2)}$, so linear combinations of these must carry the two scalar degrees of freedom.  There are two even vectors $c_{(1)}^a$ and $\tilde c_{(3)}^a$, and so linear combinations of these will be the first unfolding fields.

To find equations for the unfolding fields, we look at even vector equations of the form $d(\bullet)={\Yvcentermath1 \scriptsize\yng(1)}$,

\begin{itemize}
\item Equation 1:
\be \nabla_\mu c_{(0)}+\frac{(D-5) (D+3) }{D (D+1) L}{c_{(1)}}_\mu=0, \label{veq1}\ee
  \item Equation 2:
\be -\frac{3(D-5)  }{D^2(D-1)} \left[\frac{1   }{4  } \nabla_\mu \tilde{c}_{(2)}  +\frac{ (D-1)  }{ (D+1)  L} {c_{(1)}}_\mu +\frac{ 
   (D-3)  }{2 (D+2) L} \tilde{c}_{{(3)}\mu} \right]=0\, . \label{veq2} \ee
\end{itemize}

Consider first $D\not=3,5$.
We see that the two vector fields ${c_{(1)}}_\mu$ and $ \tilde{c}_{{(3)}\mu}$ are auxiliary fields;
we can  use the two vector equations, \eqref{veq1} and \eqref{veq2}, to solve algebraically for the unfolding vector fields $c_{(1)}^\mu$ and $\tilde{c}_{(3)}^\mu$,
\bea && {c_{(1)}}_\mu=-\frac{D (D+1) L }{(D-5) (D+3)} \nabla_\mu c_{(0)}, \label{scalauxsolu1}\\
&& \tilde{c}_{{(3)}\mu}= {L(D+2)\over D-3}\left[ \frac{2D (D-1)}{(D-5)(D+3) } \,. \nabla_\mu c_{(0)} -\frac{1}{2}  \nabla_\mu \tilde{c}_{(2)}\right]. \label{scalauxsolu2}
\eea

The equations of motion then come from even scalar equations of the form $d({\Yvcentermath1 \scriptsize\yng(1)})={\bullet}$:

\begin{itemize}
\item Equation 3:
\be \frac{(D-5) (D+3) }{D+1}\left[-\frac{1 }{2 D } \nabla_\mu c_{(1)}^\mu +\frac{1}{ L}c_{(0)}-\frac{3 
  }{ D (D+3) L}\tilde{c}_{(2)}\right]=0,  \label{seq1} \ee
\item Equation 4:
\be \frac{3(D-5) (D-3) }{2D (D+2)(D-1)}\left[\frac{ D+3 }{2  (D+2) (D+1)} \nabla_\mu \tilde{c}_3^\mu  -\frac{1   }{   L}\tilde{c}_{(2)}\right]=0.   \label{seq2}  \ee
\end{itemize}

Plugging \eqref{scalauxsolu1}, \eqref{scalauxsolu2} into \eqref{seq1}, \eqref{seq2}, we find two closed equations for the two scalars $c_{(0)}$, $\tilde c_{(2)}$,
\be 
\square\left(\begin{array}{c}c_{(0)} \\  \tilde{c}_{(2)} \end{array}\right)= {1\over L^2 (D+1)}\left(\begin{array}{cc}-2(D+3)(D-5) & {6(D-5)\over D} \\ -{8D(D-1)} & -4(D^2-4D+1)\end{array}\right)\left(\begin{array}{c}c_{(0)} \\  \tilde{c}_{(2)} \end{array}\right). \label{scalareqm}
\ee
The eigenvalues of the mass matrix in \eqref{scalareqm} precisely reproduce the scalar masses \eqref{scalarmassexpectd}
\be  m_0^2L^2= \begin{cases}  -4(D-5) \, ,\\  -2(D-3)\,. \end{cases} \label{scalarmassesge}
\ee
We may decouple the scalars by making the change of variables
\be c_{(0)}=c_{(0)}'-{3\over D(D-1)}\tilde c_{(2)}',\ \ \ \tilde c_{(2)}=\tilde c_{(2)}'-{4D\over D-5} c_{(0)}'=0 \,.\label{transs}\ee
in terms of which the equations of motion \eqref{scalareqm} become standard Klein-Gordon equations with the masses \eqref{scalarmassesge},
\be \left(\square+{2(D-3)\over L^2}\right)c_{(0)}'=0,\ \ \ \ \left(\square+{4(D-5)\over L^2}\right)\tilde c_{(2)}'=0\, .  \ee

Note, however, that the Jacobian of the transformation \eqref{transs} is
\be \frac{(D-7) (D+1)}{(D-5) (D-1)},\label{jacobian}\ee
which vanishes when $D=7$.  In this case, the equations \eqref{scalareqm} cannot be decoupled.  In total there are three special cases $D=3,5,7$, which we must treat separately, and which we will come back to in sections \ref{sec:ads3}, \ref{sec:ads5}, \ref{sec:ads7} respectively.

\subsection{The Vector Mass\label{vectormasssubsection}}

The only odd vector field in C is $\tilde{c}_{(2)}^\mu$, so this is expected to carry the massive vector degree of freedom.  Looking for the odd scalar equations of the form $d({\Yvcentermath1 \scriptsize\yng(1)})={\bullet}$, we find
\begin{itemize}
  \item Equation 5:
 \be \frac{3 (D-5)}{2 D(D-1)(D+1)}\nabla_\mu \tilde c_{(2)}^\mu=0\, . \ee
\end{itemize}
For $D\not=5$, this is the transversality constraint \eqref{transverseexp} expected for a massive vector.

There are two odd anti-symmetric tensor fields, $c_{(1)}^{\mu\nu}$, $\tilde{c}_{(3)}^{\mu\nu}$.  The unfolding fields (i.e. Maxwell field strengths) of the massive and massless vectors will each be some linear combination of these fields.  The relevant equations will be vector equations of the form $d\left( \Yvcentermath1 \scriptsize\yng(1,1)\right)={\Yvcentermath1 \scriptsize\yng(1)}$,

\begin{itemize}
\item Equation 6:
\be {D-5\over 2D(D+1)}\left[-{3\over  L}\tilde c_{(2)}^\mu+{D+3\over 2}\nabla_\nu c_{(1)}^{\mu\nu}\right]=0 \, , \label{massiveve1} \ee
  \item Equation 7:
\be -\frac{3 (D-5) (D-3)}{4 D (D+1) (D+2)} \left[{1\over  L}\tilde c_{(2)}^\mu+{D+3\over (D+2)(D+1)}\nabla_\nu \tilde c_{(3)}^{\mu\nu}\right]=0 \, ,\label{massiveve2} \ee
\end{itemize}

\noindent and odd anti-symmetric tensor equations of the form $d\left( \Yvcentermath1 \scriptsize\yng(1)\right)={\Yvcentermath1 \scriptsize\yng(1,1)}$,

\begin{itemize}
\item Equation 8:
\be -\frac{3 (D-5)}{2D (D-1)  (D+1)}\left[{2\over L}c_{(1)}^{\mu\nu}+{D-3\over (D+2)L}\tilde c_{(3)}^{\mu\nu}+\nabla^{[\mu}\tilde c_{(2)}^{\nu]} \right]=0  \label{massivevaux2} \, .\ee
\end{itemize}

Restrict first to $D\not= 3,5$.  We have mixing between the fields strengths of the massive and massless vectors.
\eqref{massivevaux2} tells us which linear combination of $c_{(1)}^{\mu\nu}$, ${\tilde{c}}_{(3)}^{\mu\nu}$ corresponds to the field strength of $\tilde{c}_{(2)}^\mu$.  To find the linear combination corresponding to the massless vector, eliminate $\tilde{c}_{(2)}^\mu$ between \eqref{massiveve1} and \eqref{massiveve2} to obtain
\be -\frac{(D-5) (D-3) (D+3)}{8 D (D+1) (D+2)}\nabla_\nu \left[c_{(1)}^{\mu\nu}+{6\over (D+1)(D+2)}{\tilde{c}}_{(3)}^{\mu\nu}\right]=0. \ee
This fixes a linear combination which satisfies the massless Maxwell equations, so this should be the field strength of the massless vector\footnote{Note that this equation should also follow from the $W$ equation of motion, $\mathcal{D}W = \mathcal{C}(C)$. However, the massless vector field is $W_0$, and so the fact that this is its field strength is a check on a purported form of the co-cycle $\mathcal{C}(C)$. It is nontrivial because it appears at level 0 of the $W$ equation of motion, but nevertheless involves $\tilde{c}_{(3)}$. Presumably this is related to the non-triviality of the factorization procedure in this case \cite{Bekaert:2005vh, Didenko:2012vh}. We thank Evgeny Skvortsov for discussions of this point.}.  

Defining the massless and massive field strength combinations $F_{\mu\nu}$, $G_{\mu\nu}$ respectively, they are related to $c_{(1)}^{\mu\nu}$, ${\tilde{c}}_{(3)}^{\mu\nu}$ by
\bea 
c_{(1)}^{\mu\nu}=F^{\mu\nu}-{6\over (D+1)(D+2)}G^{\mu\nu},\ \ \ {\tilde{c}}_{(3)}^{\mu\nu}=G^{\mu\nu}-{2(D+2)\over D-3}F^{\mu\nu}. \label{fieldvecdefs}
\eea 
Two of the three equations \eqref{massiveve1}, \eqref{massiveve2}, \eqref{massivevaux2} tell us that the massless field strength satisfies the Maxwell equations 
\be \nabla_\nu F^{\mu\nu}=0, \label{masslessmax}\ee
 and that the massive field strength is related to the massive vector by the usual field strength relation
\be  G_{\mu\nu}=-\frac{(D+1) (D+2) L}{2 (D-5) (D+3)}\left[\nabla_\mu \tilde{c}_{{(2)}\nu}-\nabla_\nu \tilde{c}_{{(2)}\mu}\right] \, .\label{massivefsd}
\ee
The final equation then becomes a field equation for the massive vector.  Using the definitions \eqref{fieldvecdefs}, plugging \eqref{massivefsd}, \eqref{masslessmax} into \eqref{massiveve1}, we find
\be \nabla_\nu G^{\nu\mu}+{2(D-5)\over L^2}{\tilde{c}_{(2)}}^\mu=0,\ee
which is the field equation for a massive vector field with mass
\be m^2=-{2(D-5)\over L^2},\ee
matching the expected result \eqref{vectormassexp}.

There are two special cases $D=3,5$, which we must treat separately, and which we will come back to in sections \ref{sec:ads3}, \ref{sec:ads5} respectively.

\subsection{The Tensor Mass\label{tensormasssubsection}}

There are three even symmetric traceless tensor fields in $C$: ${c}_{(2)}^{a,b},\ \tilde{c}_{(2)}^{a,b},\tilde{c}_{(4)}^{a,b}$.  One combination of these will be the massive graviton, the other two combinations will be second level unfolding fields for the two scalars.  To identify the combination corresponding to the massive graviton, we look at the three even vector equations of the form $d\left( \Yvcentermath1 \scriptsize\yng(2)\right)={\Yvcentermath1 \scriptsize\yng(1)}$,

\begin{itemize}
\item Equation 9:
\be \frac{3 (D-5) (D-3) (D-1) (D+5)}{2 D (D+1) (D+2)}\left[ -\frac{D+2}{4 D L}c_{(1)}^\mu+\frac{1}{(D-1) (D+5) L}\tilde c_{(3)}^\mu+\frac{1}{8 (D-2)}  \nabla_\nu c_{(2)}^{\mu,\nu}\right]=0 \, ,\label{equation9lab} \ee
\item Equation 10:
\be \frac{3 (D-5)}{2 D^2 (D+1)} \left[ \frac{D-3}{(D-1) (D+2) L}\tilde c_{(3)}^\mu+\frac{D+2}{L} c_{(1)}^\mu-\frac{D}{2 (D-1)}  \nabla_\nu \tilde c_{(2)}^{\mu,\nu} \right]=0\, ,\label{equation10lab} \ee
\item Equation 11:
\be \frac{15 (D-5) (D-3) (D-1) (D+3)}{16 D^2 (D+1) (D+2) (D+4)}\left[ \frac{1}{L}\tilde c_{(3)}^\mu-\frac{(D-1) (D+2) (D+5)}{2 (D-2) (D+1) (D+3) (D+4)}\nabla_\nu \tilde c_{(4)}^{\mu,\nu} \right]=0\, . \label{equation11lab} \ee
\end{itemize}
First restrict to $D\not=3,5$.  These are then three equations algebraic in the two even unfolding vector fields $c_{(1)}^a$ and $\tilde c_{(3)}^a$, so there is one combination for which the vector fields do not appear and we obtain a constraint equation on a combination of the tensor fields.  This combination is the massive graviton, which we call $h_{\mu,\nu}$,
\bea  && \nabla_\nu h^{\mu,\nu}=0, \label{htransveqn1}\ \ \ \\
&&  h^{\mu,\nu}\equiv \tilde c_{(4)}^{\mu,\nu}+{D(D+1)(D+4)\over 5(D-1)}\left(c_{(2)}^{\mu,\nu}-{D-2\over D-1}\tilde c_{(2)}^{\mu,\nu}\right)\, . \label{hdefeqan}
\eea
Equation \eqref{htransveqn1} is the proper transversality constraint \eqref{transverseexp} for a massive spin-2.

There is one mixed symmetry field which is even, $\tilde c_{(3)}^{ab,c}$, so this should be the first unfolding field for the massive tensor.  The equations from which we should solve for this field are the even mixed symmetry symmetric traceless tensor equations of the form $d\left( \Yvcentermath1 \scriptsize\yng(2)\right)={\Yvcentermath1 \scriptsize\yng(2,1)}$,
\begin{itemize}
\item Equation 12:
\be \frac{(D-5) (D-3) (D-1) (D+5)}{8 D (D-2)  (D+1) (D+2)}\left[ {4\over (D-1)(D+5)L}\tilde c_{(3)}^{\mu\nu,\rho}+\nabla^{[\mu} c_{(2)}^{\nu],\rho}-{1\over D-1}g^{\rho[\mu}\nabla_\lambda c_{(2)}^{\nu],\lambda}\right]=0\, ,\label{equation12lab} \ee
\item Equation 13:
\be -\frac{D-5}{2 D(D-1)  (D+1)}\left[  {D-3\over (D+2)L}\tilde c_{(3)}^{\mu\nu,\rho}+\nabla^{[\mu} \tilde c_{(2)}^{\nu],\rho}-{1\over D-1}g^{\rho[\mu}\nabla_\lambda \tilde c_{(2)}^{\nu],\lambda}\right]=0\, , \label{equation13lab}\ee
\item Equation 14:
\bea -\frac{5 (D-5) (D-3) (D-1)^2 (D+5)}{16 D^2 (D-2)  (D+1)^2 (D+4)^2} \bigg[&&-\frac{8 (D+4)}{(D-1) (D+2) (D+5) L}\tilde c_{(3)}^{\mu\nu,\rho}+\nabla^{[\mu} \tilde c_{(4)}^{\nu],\rho} \nn \\
&&-{1\over D-1}g^{\rho[\mu}\nabla_\lambda \tilde c_{(4)}^{\nu],\lambda} \bigg]=0\, .  \label{equation14lab} \eea
\end{itemize}
There is one linear combination of these three equations for which the combination $h^{\mu,\nu}$ appears under the derivatives.  That combination reads
\be {(D-4)(D+4)\over 5L}\tilde c_{(3)}^{\mu\nu,\rho}= \nabla^{[\mu} h^{\nu],\rho}-{1\over D-1}g^{\rho[\mu}\nabla_\lambda h^{\nu],\lambda} .
\label{tc3solforhe}
\ee
If $D\not=4$, this allows us to solve for $\tilde c_{(3)}^{\mu\nu,\rho}$ in terms of first derivatives of $h^{\mu,\nu}$.  $D=4$ is the dimension in which the CFT predicted that mixings would occur in the tensor, so we now assume $D\not=4$ and return to the case $D=4$ later.

The tensor equation of motion comes from the single even symmetric traceless tensor equation of the form $d\left( \Yvcentermath1 \scriptsize\yng(2,1)\right)={\Yvcentermath1 \scriptsize\yng(2)}$,

\begin{itemize}
\item Equation 15:
\be \frac{(D-5) (D-3)}{D (D+1)^2 (D+2)}\left[ -{D(D+1)\over 2(D-1)L}c_{(2)}^{\mu,\nu}+{D(D+1)(D-2)\over 2(D-1)^2L}\tilde c_{(2)}^{\mu,\nu}-{5\over 2(D+4)L}\tilde c_{(4)}^{\mu,\nu}+\nabla_{\rho}\tilde c_{(3)}^{(\mu|\rho|,\nu)}\right]=0 \, .\label{d21is2eqan}\ee
\end{itemize}
The combination of tensor fields that appears here algebraically is precisely the combination \eqref{hdefeqan} which is the massive graviton, and so \eqref{d21is2eqan} becomes
\be \nabla_\rho \tilde c_{(3)}^{(\mu|\rho|\nu)}-{5\over 2(D+4)L}h^{\mu\nu}=0. \label{dc3tishe}\ee

Solving \eqref{tc3solforhe} for $\tilde c_{(3)}^{\mu\nu,\rho}$  and plugging into \eqref{dc3tishe}, and using the transversality equation \eqref{htransveqn1}, we find a wave equation for $h^{\mu,\nu}$,
\be \square h_{\mu,\nu}+{2(D-2)\over L^2}h_{\mu,\nu}=0\, .\ee
Comparing with \eqref{eomexp}, this is the equation of motion for a massive spin 2 on AdS with mass 
\be m^2=-{2(D-3)\over L^2}\, ,\ee
precisely matching the expected value \eqref{tensormassexp}.

Finally, we can tell which combinations of the even tensor fields ${c}_{(2)}^{a,b},\ \tilde{c}_{(2)}^{a,b},\tilde{c}_{(4)}^{a,b}$ correspond to the unfolding fields of the scalars by looking at the two even symmetric traceless tensor equations of the form $d\left( \Yvcentermath1 \scriptsize\yng(1)\right)={\Yvcentermath1 \scriptsize\yng(2)}$,
\begin{itemize}
\item Equation 16:
\be -\frac{(D-5) (D+3)}{2 D (D+1)}\left[\frac{3 (D-3) (D+5)}{2 (D+2) (D+3) L}c_{(2)}^{\mu,\nu}-\frac{6}{(D-1) (D+3) L}\tilde c_{(2)}^{\mu,\nu}+\nabla^{(\mu} c_{(1)}^{\nu)_T}\right]=0 \, ,\label{spin2auxs1}\ee
\item Equation 17:
\be \small \frac{3 (D-5) (D-3) (D+3)}{4 D (D-1)  (D+1) (D+2)^2}\left[\frac{4 D}{(D+3) L}c_{(2)}^{\mu,\nu}+\frac{4}{(D+3) (D-1) L}\tilde c_{(2)}^{\mu,\nu}+\frac{5 (D-1)}{2 (D+4) L}\tilde c_{(4)}^{\mu,\nu}+\nabla^{(\mu} \tilde c_{(3)}^{\nu)_T}\right]=0 \, .\label{spin2auxs2}\ee
\end{itemize}
Taking linear combinations and using the equations \eqref{scalauxsolu1}, \eqref{scalauxsolu2} for the first level auxiliary fields, we arrive at
\bea && \frac{3 (D-5) (D-3) (D+5)}{2 D (D+1) (D+2) L^2}c_{(2)}^{\mu,\nu}-\frac{6 (D-5)}{D(D-1)  (D+1) L^2} \tilde c_{(2)}^{\mu,\nu}=\nabla^{(\mu}\nabla^{\nu)_T}c_0\, , \nn\\
&&\frac{2 (D-3) (7 D-5)}{(D+1) (D+2) L^2}c_{(2)}^{\mu,\nu}-\frac{8
   (2 D-1)}{(D+1) (D+2) (D-1) L^2} \tilde c_{(2)}^{\mu,\nu}+ \frac{5 (D-3) (D-1)}{(D+2) (D+4) L^2}\tilde c_{(4)}^{\mu,\nu}=\nabla^{(\mu}\nabla^{\nu)_T}\tilde c_{(2)}\, . \nn\\ \label{tensosunfe}
   \eea
The linear combinations on the left hand side are the second level unfolding fields for the two scalars.

There are three special cases $D=3,4,5$, which we must treat separately, and which we will come back to in sections \ref{sec:ads3}, \ref{sec:ads4}, \ref{sec:ads5} respectively.

\section{Nuances in the Spectrum\label{sec:nuances}}

In \cite{Brust:2016gjy}, we saw that peculiarities occur in the spectrum of the dual $\square^2$ CFT in specific low dimensions.  In particular, in $d=2,4$ the spectrum drastically truncates when we choose the propagator to be analytic, resulting in a theory with a finite number of single trace primaries. There were also the ``log CFTs'', but as we'll now show, the AdS theories are the duals of the finite theories rather than the log ones. In $d=3,6$, certain ``extended modules'' arise, which we will see imitated by the AdS theory. In this section, we demonstrate the AdS duals of these nuances in $D=3,4,5,7$.

\subsection{${\rm AdS}_5$\label{sec:ads5}}

In $d=4$, the spectrum of the $\square^2$ CFT completely degenerates when we choose the basic propagator to be $\langle \phi^\dag \phi\rangle = 1$.  The only remaining state is a single scalar primary of dimension zero ($j_0^{(0)}=|\phi|^2$), which has no descendants.  Thus we expect the bulk theory also not have any dynamical fields, with the exception of a single scalar which should allow for only a single mode.

We can see that this indeed happens in ${\rm AdS}_5$: all of the equations we have derived, with the exception of \eqref{veq1}, come with a prefactor $D-5$ and hence degenerate in $D=5$. The origin of this is the truncation of the algebra, as observed in\footnote{Note that the resulting finite dimensional algebras underlying the finite theories discussed here are different from the finite dimensional algebras discussed in \cite{Campoleoni:2010zq,Manvelyan:2013oua}, which occur in the massless $hs$ algebra in certain dimensions where parametrized families of algebras are possible and certain values of the parameters give finite truncations.} \cite{Joung:2015jza}. This truncation may be seen directly by looking at the trilinear form given in the appendix, and noting that every term in it is proportional to $\dso-6$. The only equation which is non-vanishing {\it in the entire theory} is \eqref{veq1}, which leaves only 
\be\nabla_\mu c_{(0)}=0.\label{nondynzeroeqa}\ee

Therefore $c_{(0)}$ is a field that allows only one mode, a constant, consistent with our expectation.  It is the only field in the ${\rm AdS}_5$ theory; every massless, partially massless, and other massive states do not even have equations of motion. This is consistent with the statement that there are no non-trivial bulk dynamics other than a constant solution for $c_0$, dual to a constant two-point function $\langle j_{0}^{(0)}(x)j_0^{(0)}(0)\rangle$ at the boundary.

\subsection{${\rm AdS}_3$}
\label{sec:ads3}

In $d=2$, the spectrum of the $\square^2$ CFT degenerates when we choose the propagator to be analytic, leaving two scalar  primary states of dimensions $0,-2$ and a spin one primary state with dimension $-1$.  These states all have a finite number of descendants, thus we expect the bulk theory to have two scalar fields (one of which allows only a constant mode, as above) and a massive vector field, all allowing only a finite number of modes.

\textbf{Scalars:}

In $D=3$, \eqref{seq2} degenerates and $\tilde{c}_{3\mu}$ decouples from \eqref{veq2}.  Changing to the mass eigenstates using \eqref{transs}, the equations become
\bea
&&\nabla^\mu c_{(0)}+{1\over 6}\nabla^\mu \tilde c_{(2)}+{1\over 3L}c_{(1)}^\mu=0\, ,  \label{veq1d3}\\
&&\nabla^\mu c_{(0)}-{1\over 2} \nabla^\mu\tilde c_{(2)}-{1\over L}c_{(1)}^\mu=0\, ,  \label{veq2d3}\\
&&\nabla_\mu c_{(1)}^\mu+{4\over L}\tilde c_{(2)}=0\, .  \label{veq3d3}
\eea

Eliminating ${c_{(1)}}_\mu$ between \eqref{veq1d3} and \eqref{veq2d3}, we find that $c_{(0)}'$ satisfies,
\be \nabla_{\mu}c_{(0)}'=0.\label{cpd0e}\ee
This is the same equation as \eqref{nondynzeroeqa}, so we identify $c_{(0)}'$ as the bulk dual of the $\Delta=0$ scalar.

Using \eqref{veq1d3} or \eqref{veq2d3} to solve for $c_{(1)}^\mu$, plugging into \eqref{veq3d3}, and using \eqref{cpd0e},
we find a Klein-Gordon equation for $\tilde{c}_{(2)}'$,
\be \left( \square -{8\over L^2}\right)\tilde{c}_{(2)}'=0,\label{c2sd3eom1}\ee
with mass
\be m^2={8\over L^2},\ee
allowing us to identify $\tilde c_{(2)}'$ as dual to the $\Delta=-2$ scalar operator.  Looking at the ordinary Klein-Gordon equation \eqref{c2sd3eom1}, it's not apparent that the field is in a finite-dimensional representation. In order to see that it indeed is, as expected from the CFT dual, we must attempt to quantize this particle and find the allowed modes. We will do this below in section \ref{sec:wavefunctions}.

\textbf{Vectors:}

For $D=3$, \eqref{massiveve2} degenerates, and ${\tilde{c}}_{(3)}^{\mu\nu}$ decouples from \eqref{massivevaux2}, which then tells us that $c_{(1)}^{\mu\nu}$ is the field strength of $\tilde{c}_{(2)}^{\mu}$,
\be c_{(1)}^{\mu\nu}=-{L\over 2}\nabla^{[\mu}\tilde{c}_{(2)}^{\nu]}.\ee
Plugging this into \eqref{massiveve1}, we find 
\be \nabla_\nu G^{\nu\mu}-{4\over L^2}{\tilde{c}_{(2)}}^\mu=0,\ee
which is the field equation for a massive vector field with mass
\be m^2={4\over L^2},\ee
matching the expected result \eqref{vectormassexp} for $D=3$, and allowing us to identify ${\tilde{c}_{(2)}}^\mu$ as the dual of the $\Delta=-1$ vector operator. This will also live in a finite-dimensional representation, though we will not explicitly construct it here.

\textbf{Tensors:}

In $D=3$, all the equations degenerate, with the exception of \eqref{equation10lab}, \eqref{equation13lab}, \eqref{spin2auxs1}, which become, respectively,
\bea &&\nabla_\nu \tilde c_{(2)}^{\mu,\nu}-{20\over 3L}c_{(1)}^\mu=0\, ,\label{3dtense1}\\
&&\nabla^{[\mu}\tilde c_{(2)}^{\nu],\rho}-{1\over 2}g^{\rho[\mu}\nabla_\lambda \tilde c_{(2)}^{\nu],\lambda}=0\,, \label{3dtense2}\\
&&\nabla^{(\mu}c_{(1)}^{\nu)_T}-{1\over 2L}\tilde c_{(2)}^{\mu,\nu}=0\, . \label{3dtense3}
\eea
All the fields except $\tilde c_{(2)}^{\mu,\nu}$ decouple.  $\tilde c_{(2)}^{\mu,\nu}$ is determined algebraically in terms of $c_{(1)}^{\nu}$ by \eqref{3dtense3}, which is given in terms of the scalar $\tilde{c}_{(2)}$ by \eqref{veq3d3}, giving
\be \tilde c_{(2)}^{\mu,\nu}=-L^2 \nabla^{(\mu}\nabla^{\nu)_T}\tilde c_{(2)}\, . \label{d3unfoldc2eq}\ee
Thus $\tilde c_{(2)}^{\mu,\nu}$ is the second level unfolding field for $\tilde c_{(2)}$.  Plugging \eqref{d3unfoldc2eq} into \eqref{3dtense2}, we find that \eqref{3dtense2} is identically satisfied, and plugging \eqref{d3unfoldc2eq} into \eqref{3dtense1}, we find that \eqref{3dtense1} reduces to the gradient of the $\tilde c_{(2)}$ equation of motion \eqref{c2sd3eom1}.  Thus, as expected from the CFT, there is no massive dynamical tensor in $D=3$.

\subsection{${\rm AdS}_7$}
\label{sec:ads7}

In the $\square^2$ CFT in $d=6$, we found in \cite{Brust:2016gjy} that the two scalar Verma modules were linked into one extended module, due to the presence of a particular state, $\square j_0^{(0)}$, becoming both primary and descendant. Consequently, the other linear combination of operators at that scaling dimension and spin, $\tilde{j}_0^{(1)}$, was forced into being an operator which was neither a primary nor a descendant, but was nevertheless in the extended Verma module of $j_0^{(0)}$.  We illustrated this module in figure 4 of \cite{Brust:2016gjy}. We would like to see how the dual of this phenomenon arises in the partially massless higher spin theory.

We saw the first sign of this in the transformation \eqref{transs} to the mass eigenstates, whose Jacobian \eqref{jacobian} vanishes when $D=7$.  (In fact, the scalar equations of motion \eqref{scalareqm} are already a bit unusual in that the mass matrix is not symmetric.)
Plugging in $D=7$ to the equations of motion \eqref{scalareqm}, we have the un-demixable equations
\begin{equation}\square \left(\begin{matrix}c_{(0)} \\ \tilde{c}_{(2)} \end{matrix} \right) = \frac{1}{L^2}\left(\begin{matrix} -5 & \frac{3}{14} \\ -42 & -11\end{matrix}\right) \left(\begin{matrix}c_{(0)} \\ \tilde{c}_{(2)} \end{matrix} \right)\,.\label{scalareqd7e}\end{equation}
The quadratic action which reproduces these equations is (defining $\tilde{c}'_{(2)}=14\tilde{c}_{(2)}$ to canonically normalize the kinetic terms)
\begin{equation}\mathcal{L} = -\frac{1}{2}(\nabla_\mu c_{(0)})^2 + \frac{1}{2} (\nabla_\mu {\tilde{c}'}_{(2)})^2 + \frac{5}{2L^2} c_{(0)}^2 - {3\over L^2} c_{(0)} {\tilde{c}'}_{(2)} -\frac{11}{2L^2} {{\tilde{c}}_{(2)}}^{'2}\, .\label{quadacd7a}\end{equation}

The kinetic terms in \eqref{quadacd7a} are required to have the opposite relative sign; it is not possible to write an action with correct sign kinetic terms that reproduces the equations \eqref{scalareqd7e}.  The internal field space is thus Lorentzian, and the transformations which preserve the kinetic structure are boosts in field space.   We can attempt to do a such a boost in field space to diagonalize the mass terms in \eqref{quadacd7a}, but it cannot be done because the required boost would be infinite.  The mass term in inherently mixed and cannot be unmixed.  Note that this phenomenon cannot happen in the case of normal kinetic terms, where any mass terms can always be diagonalized with a Euclidean rotation in field space.

\begin{figure}[h]
\centering
\includegraphics[width=0.3\textwidth]{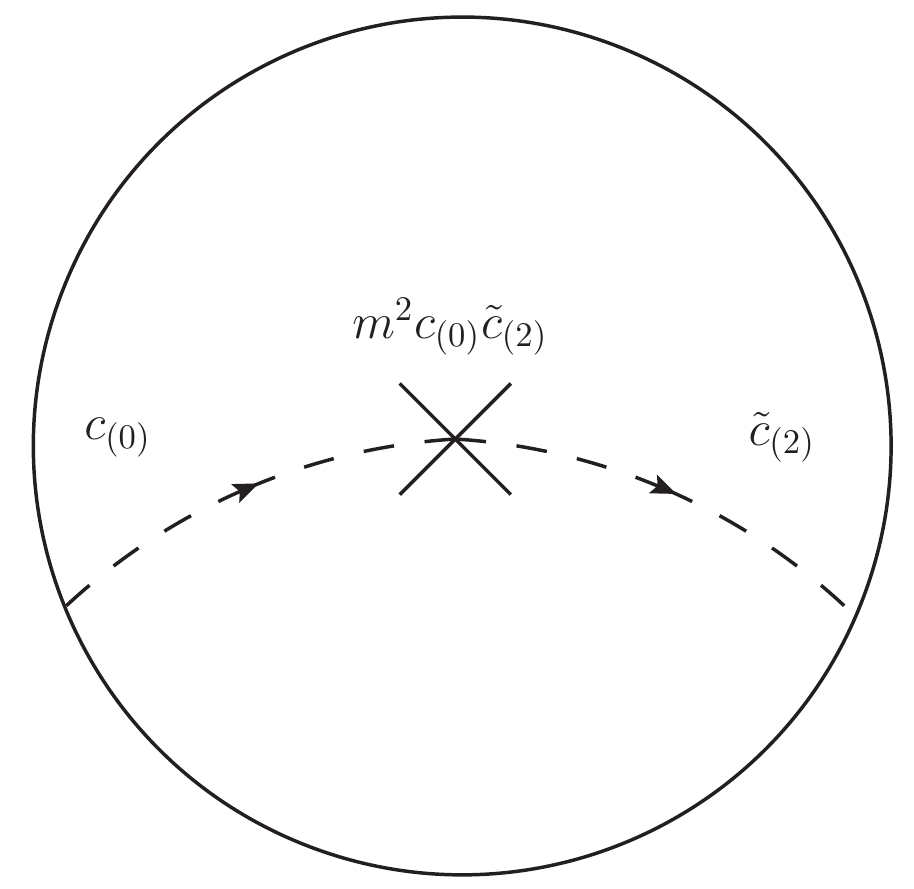}
\caption{\label{fig:7dwitten}A demonstration of how nontrivial off-diagonal correlators can arise in the CFT6, precisely because the mass terms in the 7d AdS action are non-diagonalizable, and we are forced to include Witten diagrams such as the one shown here.}
\end{figure}

This is a field theoretic realization of the spin-0/spin-0 ``extended module" uncovered in \cite{Brust:2016gjy} for the $\square^2$ in $d=6$.  Here we see the ${\rm AdS}_7$ dual of this phenomenon.  The fact that there are mass mixing terms means that when we construct Witten diagrams to evaluate boundary correlators of this theory, we will have diagrams of the form shown in figure \ref{fig:7dwitten}, where we have bulk mixing of the $c_{(0)}$,  ${\tilde{c}}_{(2)}$ degrees of freedom through non-diagonalizable mass insertions.

Although we have not yet done so, it would be very interesting to attempt to quantize \ref{quadacd7a} directly, and directly match the $d=6$ results of \cite{Brust:2016gjy}.

\subsection{${\rm AdS}_4$}
\label{sec:ads4}

The other nuance discussed in \cite{Brust:2016gjy} for the $\square^2$ CFT concerned the $d=3$ mixing of the $j_0^{(0)}$ and $j_2^{(0)}$ modules.  Thus in ${\rm AdS}_4$ we expect there to be irreducible mixing between the massive graviton and the scalar field with mass $m^2={4\over L^2}$.

In $D=4$, we see that due to the vanishing of the left hand side of \eqref{tc3solforhe}, the divergenceless field $h_{\mu\nu}$ will no longer have an auxiliary field and consequently will not satisfy a second order equation expected of the massive graviton.  Thus in $D=4$, the field $h_{\mu\nu}$ does not by itself carry the graviton degrees of freedom, and we will have to use a different strategy to identify the graviton.  

We start by using \eqref{spin2auxs1} and \eqref{spin2auxs2} to eliminate two of the tensor fields in terms of the third tensor and the two scalars $c_{(0)}$, $\tilde c_{(2)}$.  We will choose to eliminate $\tilde c_{(2)}^{\mu,\nu}$ and $\tilde c_{(4)}^{\mu,\nu}$, 
\bea  && \tilde c_{(2)}^{\mu,\nu} = {9\over 8} c_{(2)}^{\mu,\nu}+ 10L^2 \nabla^{(\mu}\nabla^{\nu)_T}c_{(0)}, \nn\\
  &&\tilde c_{(4)}^{\mu,\nu} =- {8\over 3} c_{(2)}^{\mu,\nu}+ {896\over 45}L^2 \nabla^{(\mu}\nabla^{\nu)_T}c_{(0)}+{16\over 5}L^2 \nabla^{(\mu}\nabla^{\nu)_T}\tilde c_{(2)} , \label{eliminatetc2}
  \eea
  where we have replaced the vector fields $c_{(1)}^\mu$ and $\tilde c_{(3)}^\mu$ with their values in terms of the scalars from \eqref{scalauxsolu1} and \eqref{scalauxsolu2},
  \be c_{(1)}^\mu={20\over 7}L\nabla^\mu c_{(0)},\ \ \ \tilde c_{(3)}^\mu=-{144\over 7}L\nabla^\mu c_{(0)}-3L \tilde c_{(2)}\, .\label{vectorinscal4d}\ee
  Having eliminated $\tilde c_{(2)}^{\mu,\nu}$ and $\tilde c_{(4)}^{\mu,\nu}$, our tensor degrees of freedom must now be carried by $c_{(2)}^{\mu,\nu}$.

We now look at the unfolding equations \eqref{equation12lab}, \eqref{equation13lab}, \eqref{equation14lab}, and we find that once we eliminate $\tilde c_{(2)}^{\mu,\nu}$ and $\tilde c_{(2)}^{\mu,\nu}$ using \eqref{eliminatetc2}, all three equations \eqref{equation12lab}, \eqref{equation13lab}, \eqref{equation14lab} reduce to the same equation, which is independent of the scalars and allows us to solve for $\tilde c_{(3)}^{\mu\nu,\rho}$ in terms of first derivatives of $c_{(2)}^{\mu,\nu}$,
\be \tilde c_3^{\mu\nu,\rho} =-{27\over 8L}\left[ \nabla^{[\mu} c_{(2)}^{\nu],\rho}-{1\over 3}g^{\rho[\mu}\nabla_\lambda c_{(2)}^{\nu],\lambda}\right]\, .
\label{21auxd4}\ee

Looking now at the divergence equations \eqref{equation9lab}, \eqref{equation10lab}, \eqref{equation11lab}, we find that upon use of \eqref{vectorinscal4d}, as well as use of the scalar equations of motion \eqref{scalareqm},
\be \square c_{(0)}={14\over 5L^2}c_{(0)}-{3\over 10L^2}\tilde c_{(2)},\ \ \ \square \tilde c_{(2)}=-{96\over 5L^2}c_{(0)}-{4\over 10L^2}\tilde c_{(2)} \label{scalareqm4d}
\ee 
they all reduce to the single equation
\be \nabla_\nu c_{(2)}^{\mu,\nu}={88\over3}\nabla^\mu c_{(0)}+{16\over 9}\nabla^\mu \tilde c_{(2)}\, ,\label{tensordiveq4d}
\ee
which fixes the divergence of $c_{(2)}^{\mu,\nu}$ in terms of the scalars.  

Turning finally to the equation of motion \eqref{d21is2eqan}, using \eqref{21auxd4}, and using \eqref{tensordiveq4d} to eliminate the divergences of $c_{(2)}^{\mu,\nu}$ , we find an equation of motion for $c_{(2)}^{\mu,\nu}$ which is sourced by the scalars
\be \left(\square+{4\over L^2}\right)c_{(2)}^{\mu,\nu}-{928\over 27} \nabla^{(\mu}\nabla^{\nu)_T}c_{(0)}-{8\over 3} \nabla^{(\mu}\nabla^{\nu)_T}\tilde c_{(2)}=0\, . \label{tensord4eq}
\ee
The divergence of \eqref{tensord4eq} vanishes upon use of \eqref{tensordiveq4d} and \eqref{scalareqm4d}, and so provides no new information.

We have now collected all the independent equations of motion, which are the tensor equation \eqref{tensord4eq}, the divergence equation \eqref{tensordiveq4d}, and the scalar equations of motion \eqref{scalareqm4d}.  We can see that one of the scalars decouples by changing to the mass eigenstate scalars \eqref{transs}, 
\be c_{(0)}=c_{(0)}'-{1\over4}\tilde c_{(2)}',\ \ \ \tilde c_{(2)}=\tilde c_{(2)}'+{16} c_{(0)}'=0 \label{transs2}\ee 
and redefining the tensor field as
\be  c_{(2)}^{\mu,\nu}={c'}_{(2)}^{\mu,\nu}-{1040\over 81}L^2  \nabla^{(\mu}\nabla^{\nu)_T}c_{(0)}' \, ,\ee
the equations \eqref{tensord4eq}, \eqref{tensord4eq} and the scalar equations of motion \eqref{scalareqm4d} become
\bea && \left(\square+{4\over L^2}\right){c'}_{(2)}^{\mu,\nu}+{160\over 27} \nabla^{(\mu}\nabla^{\nu)_T}\tilde c_{(2)}'=0 \, , \label{unmixtd4eqn1}\\
&&  \nabla_\nu {c'}_{(2)}^{\mu,\nu}+{50\over 9}\nabla^\mu \tilde c'_{(2)}=0\, , \label{unmixtd4eqn2}\\
&& \left(\square+{2\over L^2}\right)c_{(0)}'=0,\ \ \ \ \left(\square-{4\over L^2}\right)\tilde c_{(2)}'=0 \, .  \label{unmixtd4eqn3}
\eea 
We see that the scalar $c_{(0)}'$ decouples, but the scalar $c_{(2)}'$ remains intrinsically mixed with the tensor.  Note that a transformation ${c'}_{(2)}^{\mu,\nu}\rightarrow {c'}_{(2)}^{\mu,\nu}+\lambda  \nabla^{(\mu}\nabla^{\nu)_T}\tilde c_{(2)}'$, for any constant $\lambda$, leaves the equations invariant, so there is no further de-mixing that can be performed. 
Note also that by comparing with \eqref{eomexp}, we can see that the tensor part of \eqref{unmixtd4eqn1} is that of a graviton with $m^2 = -2$, as expected.

We have found the following Lagrangian formulation of the equations \eqref{unmixtd4eqn1}, \eqref{unmixtd4eqn2}, \eqref{unmixtd4eqn3},
\be {\cal L}={\cal L}_{\rm PM}-{140\over 27}{c'}_{(2)}^{\mu,\nu}\left(\nabla_\mu\nabla_\nu \tilde c_{(2)}'-g_{\mu\nu}\square\tilde c_{(2)}'\right)-{40\over 9L^2} {c'}_{{(2)}\mu}^{\ \ \mu} \tilde c_{(2)}'-{14000\over 243 L^2} \left( \tilde c_{(2)}'\right)^2\, ,\label{lagmixst}
\ee
where now ${c'}_2^{\mu,\nu}$ is a trace-{\it ful} symmetric tensor, and
\bea {\cal L}_{\rm PM}=&& \sqrt{|g|}\bigg[ -{1\over 2} \nabla_\rho {c'}_{(2)}^{\mu,\nu}\nabla^\rho {c'}_{{(2)}\mu,\nu}+ \nabla_\rho {c'}_{(2)}^{\mu,\nu}\nabla^\mu {c'}_{{(2)}\rho,\nu}- \nabla_\nu {c'}_{(2)}^{\mu,\nu}\nabla_\mu {c'}_{{(2)}\rho}^{\ \ \rho}+{1\over 2} \nabla_\mu {c'}_{{(2)}\nu}^{\ \ \nu} \nabla^\mu {c'}_{{(2)}\rho}^{\ \ \rho}  \nn\\
&& -{2\over L} {c'}_{(2)}^{\mu,\nu} {c'}_{{(2)}\mu,\nu} +{1\over 2L^2} {c'}_{{(2)}\mu}^{\ \ \mu} {c'}_{{(2)}\nu}^{\ \ \nu}  \bigg]\, .
\eea
is the standard Fierz-Pauli \cite{Fierz:1939ix} Lagrangian for a partially massless graviton on ${\rm AdS}_4$ (see \cite{Hinterbichler:2011tt,deRham:2014zqa,Schmidt-May:2015vnx} for reviews).  

The equations \eqref{unmixtd4eqn1}, \eqref{unmixtd4eqn2}, \eqref{unmixtd4eqn3} can be derived from \eqref{lagmixst} as follows.  Taking the following combination of the tensor equations of motion, all the higher derivatives and tensor dependence cancels, and we recover the scalar equation of motion,
\be \left(\nabla^\mu\nabla^\nu -{1\over L^2} g^{\mu\nu} \right){\delta {\cal L}\over \delta {c'}_{(2)}^{\mu,\nu}} \propto  \left(\square-{4\over L^2}\right)\tilde c_{(2)}'\, .\label{scalareqlag}\ee
(Note that $\left(\nabla^\mu\nabla^\nu -{1\over L^2} g^{\mu\nu} \right){\delta {\cal L}_{\rm PM}\over \delta {c'}_{(2)}^{\mu,\nu}}$ vanishes identically, due to the Noether identity following from the PM gauge symmetry of ${\cal L}_{\rm PM}$.)  Taking the following combination of the tensor and scalar equations of motion, we recover a constraint telling us that the tensor is traceless,
\be g^{\mu\nu} {\delta {\cal L}\over \delta {c'}_{(2)}^{\mu,\nu}}+{27\over 70}  {\delta {\cal L}\over \delta \tilde c_{(2)}'} \propto {c'}_{{(2)}\mu}^{\ \ \mu}\, . \label{lageqtraces}\ee
The divergence of the tensor equation becomes,
\be \nabla^\nu {\delta {\cal L}\over \delta {c'}_{(2)}^{\mu,\nu}} \propto  \nabla_\nu {c'}_{(2)}^{\mu,\nu}-\nabla_\mu {c'}_{{(2)}\nu}^{\ \ \nu}+{50\over 9}\nabla^\mu \tilde c'_{(2)} \label{divlageq} \ee
which upon use of \eqref{lageqtraces} to set ${c'}_{{(2)}\mu}^{\ \ \mu}=0$ reproduces \eqref{unmixtd4eqn2}.  Finally, the tensor equation ${\delta {\cal L}\over \delta {c'}_{(2)}^{\mu,\nu}}$, after eliminating divergences using \eqref{divlageq}, eliminated traces using \eqref{lageqtraces} and using the scalar equation \eqref{scalareqlag}, reproduces the tensor equation of motion \eqref{unmixtd4eqn1}.

The Lagrangian \eqref{lagmixst} cannot be unmixed into separate Fierz-Pauli and Klein-Gordon Lagrangians for a scalar and a tensor.  It is a field theoretic realization of the spin-0/spin-2 ``extended module" uncovered in \cite{Brust:2016gjy} for the $\square^2$ in $d=3$.  Here we see the ${\rm AdS}_4$ dual of this phenomenon.

\subsection{Wavefunctions in the Finite Theories}
\label{sec:wavefunctions}

In sections \ref{sec:ads5} and \ref{sec:ads3} above, we saw that the theory dramatically truncates in dimensions $D=3,6$, leaving a finite number of fields corresponding to the finite number of primaries in the finite dual CFT's discussed in \cite{Brust:2016gjy}.  Not only do these CFT's have a finite number of primaries, each primary has a finite number of descendants.  In AdS, this should correspond to the fields having a finite number of modes.

In $D=5$ we have already seen that this is the case; the single field remaining is a scalar $c_0$ satisfying an equation of motion $\nabla_\mu c_0=0$, which allows for only one mode, a constant.  This corresponds to the fact that the dual CFT has a single $\Delta=0$ scalar operator which has no descendants.

In $D=3$ there were three fields, two scalars and a vector, corresponding to the three primaries in the dual finite CFT.  One of these was $c_0'$ which satisfied an equation of motion $\nabla_\mu c_0'=0$, allowing only for one constant mode, corresponding to a dual $\Delta=0$ scalar operator with no descendants.  However the other scalar satisfied a full dynamical Klein-Gordon equation \eqref{c2sd3eom1} with $m^2L^2=8$.  The dual CFT tells us that this should correspond to a scalar operator with $\Delta_- = -2$, and so $\Delta = \frac{d}{2}- \sqrt{{d^2\over 4}+L^2m^2}$ with $d=2$, telling us that we should quantize with the alternate boundary conditions.  

The conformal algebra dictates that a scalar operator with $\Delta_- = -2$ have a finite number of descendants, and so a scalar field $m^2L^2=8$ quantized on $\rm{AdS}_3$ should have only a finite number of modes.  The fact that this scalar lives in a finite-dimensional module has been known for some time (to our knowledge it was first uncovered in \cite{Balasubramanian:1998sn}). We review the construction of the wavefunctions here for completeness' sake.

The idea is to construct the ground state wavefunction by solving the Klein-Gordon equation on ${\rm AdS}_3$, and then act with isometries which act as raising operators, adding momentum to the state. We will see that this representation has a ``speed limit'' of sorts; adding too much momentum annihilates the state, spanning a nine-dimensional Verma module, exactly as in the dual finite CFT.

\begin{figure}[h]
\centering
\includegraphics[width=\textwidth]{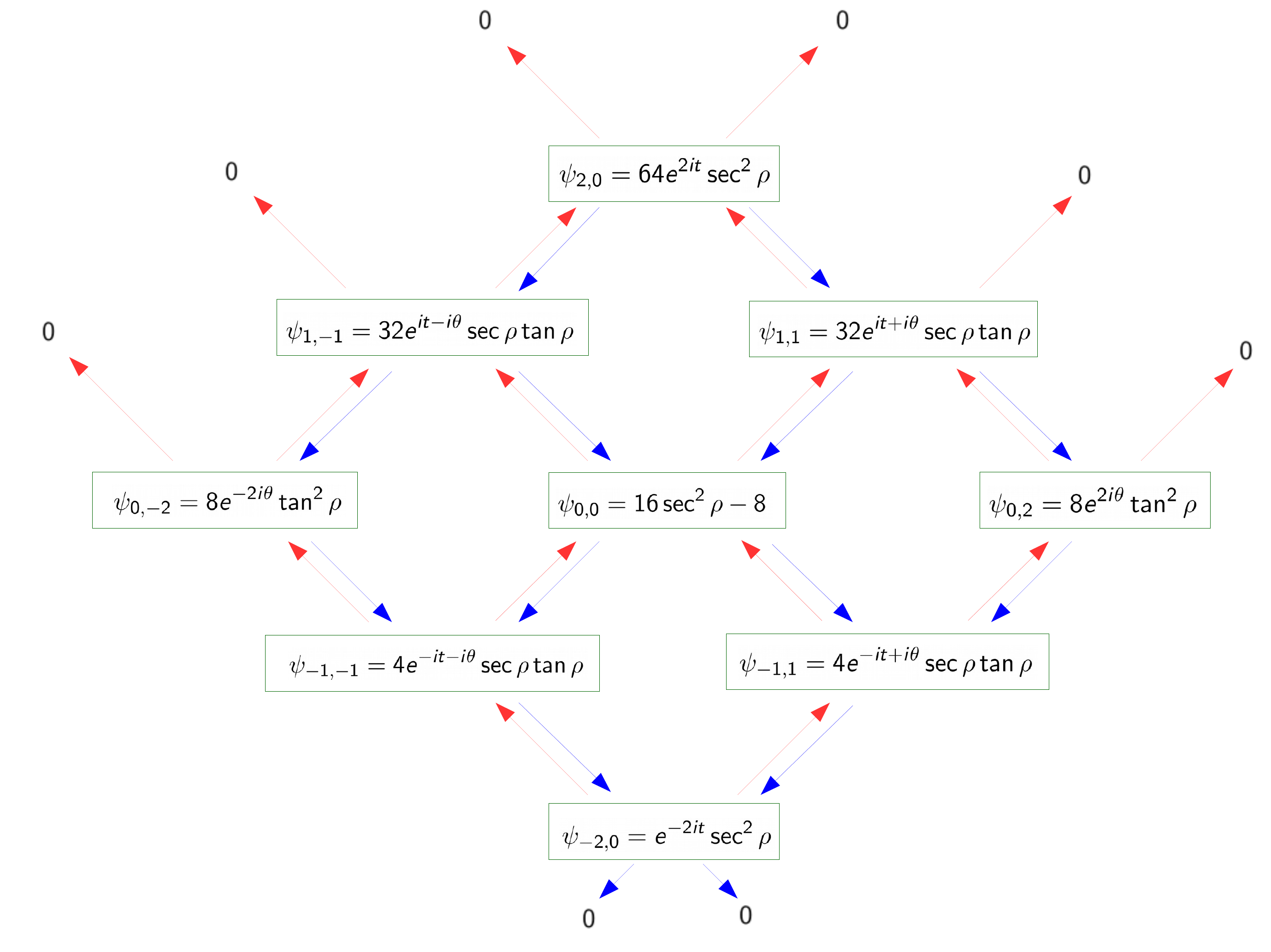}
\caption{\label{fig:ads3}The Verma module and wavefunctions for an alternately quantized scalar with $m^2L^2=8$ on ${\rm AdS}_3$. The scalar lives in a nine-dimensional module, with $\psi_{-2,0}$ being the ground state. We may move up by acting with $P_\pm$ (red arrows) or down by acting with $K_\pm$ (blue arrows), but attempting to act with a third $P_{+}$ or a third $P_-$ annihilates the state.}
\end{figure}

Working in Lorentzian signature and setting $L=1$, we use the ${\rm AdS}_3$ metric
\begin{equation}ds^2=\frac{1}{\cos^2\rho}\left(-dt^2+d\rho^2+\sin^2\rho d\theta^2\right)\, .\end{equation}
We will use the notation $\psi_{\Delta,\ell}$ for a wavefunction dual to a state of scaling dimension $\Delta$ and angular momentum $\ell$, suppressing dependence on the spacetime coordinates.  The ground state wavefunction will be $\psi_{-2,0}$.  The ground state wavefunction will solve the Klein-Gordon equation:
\begin{equation}\square \psi_{-2,0}=\cos^2\rho\left(-\partial_t^2+\partial_\rho^2+\frac{1}{\cos\rho\sin\rho}\partial_\rho+\frac{1}{\sin^2\rho}\partial_\theta^2\right)\psi_{-2,0}=m^2\psi_{-2,0}\, .\end{equation}
The general solutions to this are the wavefunctions
\begin{equation}\psi=c_+e^{i\Delta_+ t}\cos^\Delta_+ \rho + c_- e^{i\Delta_- t} \cos^\Delta_- \rho\, , \ \ \  \Delta_\pm = 4,-2\end{equation}
 but since we are choosing the alternate boundary conditions, we choose the smaller root, and so our ground state wavefunction is
\begin{equation}\psi_{-2,0} =e^{-2it} \cos^{-2}\rho\, .\end{equation}
From here we may move up in the Verma module by acting with isometries which act as raising operators, $P_\pm$, or lowering operators $K_\pm$ (so named because their actions at the boundary match that of the raising and lowering operators of the conformal algebra)
\begin{align}P_{\pm}&=i e^{it\pm i\theta} \left(\sin\rho \partial_t -i \cos\rho \partial_\rho \pm \frac{1}{\sin\rho} \partial_\theta\right)\, ,\nonumber \\
K_{\pm}&=i e^{-it\pm i\theta} \left(\sin\rho \partial_t +i \cos\rho \partial_\rho \mp \frac{1}{\sin\rho} \partial_\theta\right)\, .\end{align}
A straightforward computation shows that the wavefunction vanishes if we act with either $P_+$ or $P_-$ more than twice, and furthermore (as expected) the ground state is annihilated by the lowering operators $K_{\pm}$. Therefore, this $m^2L^2=8$ alternately quantized scalar lives in a finite-dimensional Verma module. We illustrate the structure of the module in figure \ref{fig:ads3}.  There are nine states in total, which matches the expectations from the conformal algebra.

 A similar construction should also go through for the massive vector with mass $m^2L^2={4 }$, quantized with alternate boundary conditions.  There should be a finite number of modes in correspondence with the finite number of descendants of the dual $\Delta=-1$ vector operator.

\FloatBarrier

%%%%%%%%%%%%%%%%%%%%%%%%%%%%%%%%%%%%%%%%%%%%%%%%%%%%%%%%%%%%%%%%%%%%%%%%%%%%%%%%

\section{Conclusions and Future Directions}
\label{sec:conc}

We have presented a construction of a partially massless higher-spin theory which extends the Vasiliev higher-spin theory to include additional partially massless states. The theory is based on a Vasiliev-type gauging of the $hs_2$ algebra, the global symmetry algebra of a free $\square^2$ CFT.  We have worked out the first few dozen terms in the trilinear form of the $hs_2$ algebra, as needed to unfold the $C$ master field equation of motion and work out the masses of the four fully massive particles which do not have any linearized gauge symmetry. We've identified the field content and gauge symmetries of the master fields, demonstrating that they are in agreement with what is expected from a free scalar $\square^2$ CFT.

In certain dimensions, the $\square^2$ spectrum develops various oddities in it \cite{Brust:2016gjy}; in $d=2,4$, there are two different theories; one is the honest $\square^2$ theory with log correlators, and the other is a CFT with a well-defined operator algebra, but only a finite number of single-trace primaries. Furthermore, in $d=3,6$, we find that what would at first appear to be two independent Verma modules are in fact conjoined into a single extended Verma module. This happens between the two scalar single-trace primaries in $d=6$, and between a scalar and a tensor in $d=3$.  In this paper, we explored the AdS duals of all of these phenomena. In $\dso=4,6$, we saw at the level of the trilinear form, in agreement with the observations of \cite{Joung:2015jza}, that the $hs_2$ algebra truncates dramatically into a finite-dimensional algebra. This manifested itself in $D=3,5$ as the truncation of the infinite towers down to a finite number of particles in $D=3$, and down to only a single scalar with a single mode in $D=5$. These are exactly what happens in the finite $d=2,4$ CFTs, supporting the claim that it is truly the finite, rather than the log, theories which are the duals of the PM HS theory. Furthermore, we saw a field-theoretic realization of the extended Verma modules in $D=4,7$ between exactly those particles we would expect from the dual CFT. The module mixing in AdS manifested itself as the non-diagonalizability of the equations of motion and corresponding free actions.

Finally, in the companion paper \cite{Brust:2016xif}, we provide evidence that this theory is sensible at the one-loop level, carrying out the one-loop matching of the coupling of the theory, $G_N^{-1}$ to $N$ of the CFT, with identical findings to what was found for the Vasiliev theory \cite{Giombi:2014iua}.

We believe that all of these checks together constitute significant evidence for the completeness and sensibility of the PM HS theory. Furthermore, as the equations of motion are covariant, they may be formulated in $D=4$ about de Sitter just as easily as about AdS. The dual CFT can be constructed with anti-commuting scalars as well, and our arguments for an AdS/CFT duality lead us to conjecture that, following \cite{Anninos:2011ui}, the dS/Grassmann versions of these theories are dual as well, and constitute a new example of the dS/CFT correspondence. There are many unanswered questions which we hope this new example will help make progress in; one of the most important such questions to address is what about the CFT informs our understanding of the unitarity of the dS theory. We plan to explore this issue in upcoming work.

We also hope that the existence of these extended examples of AdS/CFT and dS/CFT open the door to many exciting future directions, both within and outside of higher-spin holography.  Given a sensible theory of interacting partially massless higher-spin particles, it is worth asking if they may play a role in our own universe. Perhaps in the early universe where massive higher spins may be Hubble scale and possibly partially massless, they might be detectable in future cosmology experiments \cite{Arkani-Hamed:2015bza,Lee:2016vti}.  Such a study would go hand-in-hand with a study of what interactions might be allowed by partially massless higher-spin particles. In principle, the nonlinear Vasiliev theory gauged with the $hs_2$ algebra should produce interaction vertices, though it is not clear a priori whether this will produce all allowable such interactions. In practice, it may be simpler to reconstruct them from correlators of the dual CFT, following some procedure as in \cite{Boulanger:2015ova,Sleight:2016dba,Sleight:2016xqq}.

In particular, we demonstrated that in ${\rm AdS}_3$/${\rm CFT}_2$, the linearized PM HS theory includes just a single propagating scalar (in addition to a scalar with only a zero mode) in the minimal theory, plus an additional single vector in the non-minimal theory, in agreement with CFT predictions. Nevertheless, at finite $N$, the CFT is still exactly solvable, but should now be dual to an {\it interacting non-unitary field theory} (without gravity) on ${\rm AdS}_3$. These two theories would be stable by virtue of the finite number of single-particle states, and the presence of an unbroken $hs_2$ symmetry. Explicitly constructing this theory, and obtaining its action, would be very interesting, and may provide one of the simplest exactly solvable examples of AdS/CFT.

In addition, we believe that this only scratches the surface of non-unitary higher-spin holography. There should be partially massless higher-spin theories dual to the $\square^k$ theories we discussed in \cite{Brust:2016gjy}, with more and more ``Regge trajectories.''  We could also consider the fermionic counterparts, defined by a $\psi^\dagger \slashed{\partial}^k \psi$ action, and also the supersymmetric combination of bosonic and fermionic terms.  Perhaps other interesting field-theoretic mechanisms exist there as well, and perhaps these new additional examples could also be turned into useful examples of dS/CFT, some or all of which will hopefully one day play a role in unlocking the mysteries of quantum gravity in spaces with positive cosmological constants and the higher spin Higgs mechanism.

{\bf Acknowledgements:} We thank Nima Arkani-Hamed, Xavier Bekaert, Frederik Denef, Tudor Dimofte, Davide Gaiotto, Simone Giombi, Bob Holdom, Euihun Joung, Igor Klebanov, Rachel Rosen, Evgeny Skvortsov, and Matt Walters for helpful discussions and comments.  Research at Perimeter Institute is supported by the Government of Canada through Industry Canada and by the Province of Ontario through the Ministry of Economic Development and Innovation.

%%%%%%%%%%%%%%%%%%%%%%%%%%%%%%%%%%%%%%%%%%%%%%%%%%%%%%%%%%%%%%%%%%%%%%%

\appendix

\section{Bilinear and Trilinear Forms in $hs_2$}
\label{sec:multilinearforms}

Here we list the first few results for the $hs_2$ bilinear forms $\cB$ and trilinear forms $\cT$, defined in \eqref{bitrideff}, using the techniques described in section \ref{sec:formsexplanation}. (Note that the bilinear form is known to all orders; it was computed in \cite{Joung:2015jza}). Our results below for the bilinear forms match theirs. Our results for the trilinear forms are new, and include all the trilinear forms necessary for the mass computations of section \eqref{sec:masscomp}

We compute these by directly expanding and evaluating \eqref{eqn:multilinearform}.  The resulting answers may be expressed in terms of powers of the cocycles $\caW_j$ corresponding to each argument appear.   We use notation $\mathcal{M}_{(i_1,\ldots, i_n)}$, where $i_j$ indicates how many powers there are of $\caW_j$ in that term. We work up to fifth level in two of the cocycles, but restrict the third to be at level one (because the background master field $\bW$ which appears in the equations of motion only has support at level one). We use angle brackets to denote matrix traces. Finally, we use a few terms in the text that do not appear in the table below; those are all related to the ones that appear below by the cyclicity of the trace $\tr$ and relabelling.  The results of the computation are listed in table \ref{tab:trilinearForm}. Note that there are potentially multiple terms appearing in a given part of the expansion; this is in one-to-one correspondence with the different tensor structures which emerge.

{\renewcommand{\arraystretch}{1.7}
\begin{longtable}[h]{|c|rl|}\hline \rowcolor{Gray} $\cB_{(1,1)}$& $-\frac{1}{4} \frac{(\dso-6)(\dso+2)}{(\dso-1)\dso}$&$\langle\caW_1\caW_2\rangle $\\

$\cT_{(1,1,1)}$ &$ -\frac{1}{4} \frac{(\dso-6)(\dso+2)}{(\dso-1)\dso} $&$\langle\caW_1\caW_2\caW_3\rangle $\\ 

\rowcolor{Gray} $\cT_{(1,1,2)} $&$ \frac{3}{32} \frac{(\dso-6)(\dso-4)(\dso+4)}{(\dso-1)\dso(\dso+1)}$&$ \langle\caW_1\caW_3)\langle\caW_2\caW_3\rangle$\\

\rowcolor{Gray} & $ - \frac{3}{8} \frac{(\dso-6)(\dso+6)}{(\dso-1)\dso(\dso+1)} $&$\langle\caW_1\caW_2\caW_3\caW_3\rangle $ \\ 

$\cB_{(2,2)}$ & $\frac{3}{64} \frac{(\dso-6)(\dso-4)(\dso+4)}{(\dso-1)\dso(\dso+1)}$&$ \langle\caW_1\caW_2\rangle^2 $ \\ 

& $ - \frac{3}{16} \frac{(\dso-6)(\dso+6)}{(\dso-1)\dso(\dso+1)} $&$\langle\caW_1\caW_1\caW_2\caW_2\rangle$ \\ 

\rowcolor{Gray} $\cT_{(2,1,2)}$ & $\frac{3}{32} \frac{(\dso-6)(\dso-4)(\dso+4)}{(\dso-1)\dso(\dso+1)}$&$ \langle\caW_1\caW_3\rangle\langle\caW_1\caW_2\caW_3\rangle $ \\

\rowcolor{Gray} & $- \frac{3}{16}\frac{(\dso-6)(\dso+6)}{(\dso-1)\dso(\dso+1)}$&$\langle\caW_1\caW_1\caW_2\caW_3\caW_3\rangle$ \\ 

$\cT_{(2,1,3)}$ & $-\frac{3}{128}\frac{(\dso-6)(\dso-4)(\dso-2)(\dso+6)}{(\dso-1)\dso(\dso+1)(\dso+3)}$&$\langle\caW_1\caW_3\rangle^2 \langle\caW_2\caW_3\rangle $ \\

&$ \frac{3}{32} \frac{(\dso-6)(\dso-4)(\dso+8)}{(\dso-1)\dso(\dso+1)(\dso+3)}$&$ \langle\caW_1\caW_1\caW_3\caW_3\rangle\langle\caW_2\caW_3\rangle$ \\ 

& $ \frac{3}{16} \frac{(\dso-6)(\dso-4)(\dso+8)}{(\dso-1)\dso(\dso+1)(\dso+3)}$&$\langle\caW_1\caW_3\rangle\langle\caW_1\caW_2\caW_3\caW_3\rangle$\\

\rowcolor{Gray} $\cB_{(3,3)}$ & $-\frac{1}{128}\frac{(\dso-6)(\dso-4)(\dso-2)(\dso+6)}{(\dso-1)\dso(\dso+1)(\dso+3)}$&$\langle\caW_1\caW_2\rangle^3$ \\

\rowcolor{Gray} & $  \frac{3}{32} \frac{(\dso-6)(\dso-4)(\dso+8)}{(\dso-1)\dso(\dso+1)(\dso+3)} $&$\langle\caW_1\caW_2\rangle\langle\caW_1\caW_1\caW_2\caW_2\rangle$ \\

$\cT_{(3,1,3)}$ & $-\frac{3}{128}\frac{(\dso-6)(\dso-4)(\dso-2)(\dso+6)}{(\dso-1)\dso(\dso+1)(\dso+3)}$&$ \langle\caW_1\caW_3\rangle^2 \langle\caW_1\caW_2\caW_3\rangle $ \\

& $ \frac{3}{64} \frac{(\dso-6)(\dso-4)(\dso+8)}{(\dso-1)\dso(\dso+1)(\dso+3)}$&$ \langle\caW_1\caW_2\caW_3\rangle\langle\caW_1\caW_1\caW_3\caW_3\rangle$\\

& $\frac{15}{128}\frac{(\dso-6)(\dso-4)(\dso+8)}{(\dso-1)\dso(\dso+1)(\dso+3)} $&$ \langle\caW_1\caW_3\rangle\langle\caW_1\caW_1\caW_2\caW_3\caW_3\rangle $ \\

&$ \frac{3}{64}\frac{(\dso-6)(\dso-4)(\dso+8)}{(\dso-1)\dso(\dso+1)(\dso+3)}$&$ \langle\caW_1\caW_1\caW_3\caW_3\caW_1\caW_2\caW_3\rangle$ \\ 

\rowcolor{Gray} $\cT_{(3,1,4)}$ & $\frac{5}{1024}\frac{(\dso-6)(\dso-4)(\dso-2)(\dso+8)}{(\dso-1)(\dso+1)(\dso+3)(\dso+5)}$&$ \langle\caW_1\caW_3\rangle^3 \langle\caW_1\caW_2\caW_3\rangle$ \\

\rowcolor{Gray} & $ - \frac{15}{256} \frac{(\dso-6)(\dso-4)(\dso-2)(\dso+10)}{(\dso-1)\dso(\dso+1)(\dso+3)(\dso+5)} $&$\langle\caW_1\caW_3\rangle\langle\caW_2\caW_3\rangle\langle\caW_1\caW_1\caW_3\caW_3\rangle$ \\

\rowcolor{Gray} & $-\frac{15}{256} \frac{(\dso-6)(\dso-4)(\dso-2)(\dso+10)}{(\dso-1)\dso(\dso+1)(\dso+3)(\dso+5)}$&$ \langle\caW_1\caW_3\rangle^2 \langle\caW_1\caW_2\caW_3\caW_3\rangle $ \\

\rowcolor{Gray} & $ \frac{15}{128}\frac{(\dso-6)(\dso-4)(\dso+12)}{(\dso-1)\dso(\dso+1)(\dso+3)(\dso+5)} $&$\langle\caW_1\caW_1\caW_3\caW_3\rangle\langle\caW_1\caW_2\caW_3\caW_3\rangle$ \\ 

\rowcolor{Gray} & $\frac{15}{128} \frac{(\dso-6)(\dso-4)(\dso+12)}{(\dso-1)\dso(\dso+1)(\dso+3)(\dso+5)} $&$\langle\caW_1\caW_1\caW_3\caW_3\caW_2\caW_1\caW_3\caW_3\rangle$ \\ 

$\cB_{(4,4)}$ & $\frac{5}{4096} \frac{(\dso-6)(\dso-4)(\dso-2)(\dso+8)}{(\dso-1)(\dso+1)(\dso+3)(\dso+5)}$&$ \langle\caW_1\caW_2\rangle^4$ \\

& $ \frac{15}{512} \frac{(\dso-6)(\dso-4)(\dso-2)(\dso+10)}{(\dso-1)\dso(\dso+1)(\dso+3)(\dso+5)} $&$\langle\caW_1\caW_2\rangle^2 \langle\caW_1\caW_1\caW_2\caW_2\rangle$ \\ 

& $ \frac{15}{512} \frac{(\dso-6)(\dso-4)(\dso+12)}{(\dso-1)\dso(\dso+1)(\dso+3)(\dso+5)}$&$ \langle\caW_1\caW_1\caW_2\caW_2\rangle^2 $ \\

& $ \frac{15}{512} \frac{(\dso-6)(\dso-4)(\dso+12)}{(\dso-1)\dso(\dso+1)(\dso+3)(\dso+5)} $&$\langle\caW_1\caW_1\caW_2\caW_2\caW_1\caW_1\caW_2\caW_2\rangle$ \\ 

\rowcolor{Gray} $\cT_{(4,1,4)}$ & $\frac{5}{1024} \frac{(\dso-6)(\dso-4)(\dso-2)(\dso+8)}{(\dso-1)(\dso+1)(\dso+3)(\dso+5)} $&$ \langle\caW_1\caW_3\rangle^3 \langle\caW_1\caW_2\caW_3\rangle $ \\

\rowcolor{Gray} & $- \frac{45}{1024} \frac{(\dso-6)(\dso-4)(\dso-2)(\dso+10)}{(\dso-1)\dso(\dso+1)(\dso+3)(\dso+5)} $&$\langle\caW_1\caW_3\rangle^2\langle\caW_1\caW_1\caW_2\caW_3\caW_3\rangle$\\ 

\rowcolor{Gray} & $\frac{15}{256} \frac{(\dso-6)(\dso-4)(\dso+12)}{(\dso-1)\dso(\dso+1)(\dso+3)(\dso+5)} $&$\langle\caW_1\caW_1\caW_3\caW_3\rangle\langle\caW_1\caW_1\caW_2\caW_3\caW_3\rangle $ \\

\rowcolor{Gray} & $- \frac{15}{512} \frac{(\dso-6)(\dso-4)(\dso-2)(\dso+10)}{(\dso-1)\dso(\dso+1)(\dso+3)(\dso+5)}$&$ \langle\caW_1\caW_3\rangle\langle\caW_1\caW_2\caW_3\rangle\langle\caW_1\caW_1\caW_3\caW_3\rangle$ \\ 

\rowcolor{Gray} & $-\frac{15}{512} \frac{(\dso-6)(\dso-4)(\dso-2)(\dso+10)}{(\dso-1)\dso(\dso+1)(\dso+3)(\dso+5)}$&$ \langle\caW_1\caW_3\rangle\langle\caW_1\caW_1\caW_3\caW_3\caW_1\caW_2\caW_3\rangle $ \\

\rowcolor{Gray} & $ \frac{15}{256}\frac{(\dso-6)(\dso-4)(\dso+12)}{(\dso-1)\dso(\dso+1)(\dso+3)(\dso+5)} $&$\langle\caW_1\caW_1\caW_2\caW_3\caW_3\caW_1\caW_1\caW_3\caW_3\rangle$ \\

$\cT_{(4,1,5)}$ & $-\frac{15}{16384} \frac{(\dso-6)(\dso-4)(\dso-2)(\dso+2)(\dso+10)}{(\dso-1)(\dso+1)(\dso+3)(\dso+5)(\dso+7)}$ & $\langle\caW_1\caW_3\rangle^4 \langle\caW_2\caW_3\rangle$\\ 

& $\frac{45}{2045} \frac{(\dso-6)(\dso-4)(\dso-2)(\dso+12)}{(\dso-1)(\dso+1)(\dso+3)(\dso+5)(\dso+7)}$ & $\langle\caW_1\caW_3\rangle^2 \langle\caW_2\caW_3\rangle\langle\caW_1\caW_1\caW_3\caW_3\rangle$\\

& $\frac{15}{1024} \frac{(\dso-6)(\dso-4)(\dso-2)(\dso+12)}{(\dso-1)(\dso+1)(\dso+3)(\dso+5)(\dso+7)}$ & $\langle\caW_1\caW_3\rangle^3 \langle\caW_1\caW_2\caW_3\caW_3\rangle$\\

& $-\frac{45}{2048} \frac{(\dso-6)(\dso-4)(\dso-2)(\dso+14)}{(\dso-1)\dso(\dso+1)(\dso+3)(\dso+5)(\dso+7)}$&$\langle\caW_2\caW_3\rangle\langle\caW_1\caW_1\caW_3\caW_3\rangle^2$\\

& $-\frac{45}{2048} \frac{(\dso-6)(\dso-4)(\dso-2)(\dso+14)}{(\dso-1)\dso(\dso+1)(\dso+3)(\dso+5)(\dso+7)}$&$\langle\caW_2\caW_3\rangle\langle\caW_1\caW_1\caW_3\caW_3\caW_1\caW_1\caW_3\caW_3\rangle$\\

& $-\frac{45}{512}\frac{(\dso-6)(\dso-4)(\dso-2)(\dso+14)}{(\dso-1)\dso(\dso+1)(\dso+3)(\dso+5)(\dso+7)}$&$\langle\caW_1\caW_3\rangle\langle\caW_1\caW_1\caW_3\caW_3\rangle\langle\caW_1\caW_2\caW_3\caW_3\rangle$\\

& $-\frac{45}{512} \frac{(\dso-6)(\dso-4)(\dso-2)(\dso+14)}{(\dso-1)\dso(\dso+1)(\dso+3)(\dso+5)(\dso+7)}$&$\langle\caW_1\caW_3\rangle\langle\caW_1\caW_1\caW_3\caW_3\caW_2\caW_1\caW_3\caW_3\rangle$\\

\rowcolor{Gray} $\cB_{(5,5)}$ & $-\frac{3}{16384} \frac{(\dso-6)(\dso-4)(\dso-2)(\dso+2)(\dso+10)}{(\dso-1)(\dso+1)(\dso+3)(\dso+5)(\dso+7)}$ & $\langle\caW_1\caW_2\rangle^5$\\ 

\rowcolor{Gray} & $\frac{15}{2048} \frac{(\dso-6)(\dso-4)(\dso-2)(\dso+12)}{(\dso-1)(\dso+1)(\dso+3)(\dso+5)(\dso+7)}$ & $\langle\caW_1\caW_2\rangle^3 \langle\caW_1\caW_1\caW_2\caW_2\rangle$\\ 

\rowcolor{Gray} & $-\frac{45}{2048}\frac{(\dso-6)(\dso-4)(\dso-2)(\dso+14)}{(\dso-1)\dso(\dso+1)(\dso+3)(\dso+5)(\dso+7)}$& $\langle\caW_1\caW_2\rangle\langle\caW_1\caW_1\caW_2\caW_2\rangle^2$\\ 

\rowcolor{Gray} & $-\frac{45}{2048}\frac{(\dso-6)(\dso-4)(\dso-2)(\dso+14)}{(\dso-1)\dso(\dso+1)(\dso+3)(\dso+5)(\dso+7)}$&$\langle\caW_1\caW_2\rangle\langle\caW_1\caW_1\caW_2\caW_2\caW_1\caW_1\caW_2\caW_2\rangle$\\ 

$\cT_{(5,1,5)}$ & $-\frac{15}{16384}\frac{ (\dso-6) (\dso-4) (\dso-2) (\dso+2) (\dso+10) }{ (\dso-1) (\dso+1) (\dso+3) (\dso+5) (\dso+7)}$ & $\langle\caW_1\caW_3\rangle^4\langle\caW_1\caW_2\caW_3\rangle$ \\ 

& $\frac{105}{8192} \frac{ (\dso-6) (\dso-4) (\dso-2) (\dso+12) }{ (\dso-1) (\dso+1) (\dso+3) (\dso+5) (\dso+7)}$ & $\langle\caW_1\caW_3\rangle^3 \langle\caW_1\caW_1\caW_2\caW_3\caW_3\rangle$ \\ 

& $-\frac{15}{1024} \frac{ (\dso-6) (\dso-4) (\dso-2) (\dso+14) }{ (\dso-1) \dso (\dso+1) (\dso+3) (\dso+5) (\dso+7)}$ & $\langle\caW_1\caW_1\caW_3\caW_3\rangle\langle\caW_1\caW_1\caW_3\caW_3\caW_1\caW_2\caW_3\rangle$ \\ 

& $\frac{45}{4096} \frac{ (\dso-6) (\dso-4) (\dso-2) (\dso+12) }{ (\dso-1) (\dso+1) (\dso+3) (\dso+5) (\dso+7)}$ & $\langle\caW_1\caW_3\rangle^2\langle\caW_1\caW_2\caW_3)\langle\caW_1\caW_1\caW_3\caW_3\rangle$\\ 

& $\frac{45}{4096} \frac{ (\dso-6) (\dso-4) (\dso-2) (\dso+12) }{ (\dso-1) (\dso+1) (\dso+3) (\dso+5) (\dso+7)}$ & $\langle\caW_1\caW_3\rangle^2\langle\caW_1\caW_1\caW_3\caW_3\caW_1\caW_2\caW_3\rangle$\\ 

& $-\frac{15}{2048} \frac{ (\dso-6) (\dso-4) (\dso-2) (\dso+14) }{ (\dso-1) \dso (\dso+1) (\dso+3) (\dso+5) (\dso+7)}$ & $\langle\caW_1\caW_2\caW_3\rangle\langle\caW_1\caW_1\caW_3\caW_3\rangle^2$ \\ 

& $-\frac{15}{2048} \frac{ (\dso-6) (\dso-4) (\dso-2) (\dso+14) }{ (\dso-1) \dso (\dso+1) (\dso+3) (\dso+5) (\dso+7)}$ & $\langle\caW_1\caW_2\caW_3\rangle\langle\caW_1\caW_1\caW_3\caW_3\caW_1\caW_1\caW_3\caW_3\rangle$\\ 

& $-\frac{105}{2048} \frac{ (\dso-6) (\dso-4) (\dso-2) (\dso+14) }{ (\dso-1) \dso (\dso+1) (\dso+3) (\dso+5) (\dso+7)}$ & $\langle\caW_1\caW_3\rangle\langle\caW_1\caW_1\caW_3\caW_3\rangle\langle\caW_1\caW_1\caW_2\caW_3\caW_3\rangle$ \\ 

& $-\frac{105}{2048} \frac{ (\dso-6) (\dso-4) (\dso-2) (\dso+14) }{ (\dso-1) \dso (\dso+1) (\dso+3) (\dso+5) (\dso+7)}$ & $\langle\caW_1\caW_3\rangle\langle\caW_1\caW_1\caW_2\caW_3\caW_3\caW_1\caW_1\caW_3\caW_3\rangle$ \\ 

& $-\frac{15}{1024} \frac{ (\dso-6) (\dso-4) (\dso-2) (\dso+14)}{ (\dso-1) \dso (\dso+1) (\dso+3) (\dso+5) (\dso+7)}$ & $\langle\caW_1\caW_1\caW_3\caW_3\caW_1\caW_1\caW_3\caW_3\caW_1\caW_2\caW_3\rangle$\\ \hline
\caption{The first 11 orders of the relevant parts of the bi- and tri-linear forms for the $hs_2$ algebra.}
\label{tab:trilinearForm}
\end{longtable}
}

 \bibliographystyle{utphys}
\addcontentsline{toc}{section}{References}
\bibliography{adsv2}

\end{document}